\documentclass[12pt]{article}

\usepackage{amsmath}
\usepackage{amssymb}

\newcommand{\be}{\begin{equation}}
\newcommand{\ee}{\end{equation}}
\newcommand{\ba}{\begin{eqnarray}}
\newcommand{\ea}{\end{eqnarray}}
\newcommand{\nn}{\nonumber}

\newcommand{\half}{\frac 1 2} 
\newcommand{\z}{\mathfrak{z}}

\newcommand{\X}{\underline{\overline{X}}}
\newcommand{\Y}{Y(u_1,u_2|x_4,\dots, x_1)}
\newcommand{\p}{\mathbf{p}}
\newcommand{\x}{\mathbf{x}}
\newcommand{\y}{\mathbf{y}}
\renewcommand{\H}{\mathcal{H}}
\newcommand{\M}{\mathcal{M}}
\newcommand{\C}{\mathbf{C}}
\renewcommand{\c}{\mathbf{c}} 

\title{D-independent representation of Conformal Field Theories in $D $ dimensions via  transformation to auxiliary Dual Resonance Models.
Scalar amplitudes
  }
\author{G. Mack\\
II. Institut f\"ur Theoretische Physik, Universit\"at Hamburg
}

\date{\today}
\begin{document}

\maketitle 
\begin{abstract}
The Euklidean correlation functions and vacuum expectation values of products
 of $n$ field operators $\phi^{k_i}(x_i)$ of some Lorentz spin $l_i$ and 
dimension $d_i$ are expressed through Mellin amplitudes $M_{k_n,...k_1}(\{ \delta_{ij}\})$ which depend on complex dimensions $\delta_{ij}=\delta_{ji}$, $1\leq i<j\leq n$ subject to linear constraints $\sum_j\delta_{ij}=d_i$. The constraints can be solved in terms of conserved momenta $p_i$ whose squares are given by the field dimensions $d_i$, and related Mandelstam variables $s_{ij}=(p_i+p_j)^2$, viz. $\delta_{ij}=-p_ip_j$. The Mellin amplitudes furnish a universal representation of conformal field theories without explicit reference to D. The costumary principles of quantum field theory plus conformal invariance and operator product expansions (OPE) say that the Mellin amplitudes are amplitudes of dual resonance models with exact duality and a form of factorization which follows from OPE. It suffices to ensure these properties for all 4-point functions, together with positivity of all 2-point functions. A field with spin $l$ and dimension $d$ in the OPE's of scalar field products $\phi^{k_2}\phi^{k_1}$ and of $\phi^{k_3\ast}\phi^{k_4\ast}$  produces simple poles in the 4-point Mellin amplitude $M_{k_4...k_1}$ at $s_{12}==d-l+2n$, $ n=0,1,2...$. Their residues are polynomials of $l$-the order; they depend on $D$. The leading pole determines the satellites $n=1,2,3,..$.
\end{abstract}

\nocite{fradkin:palchik:3,palchik:secondary,palchik:fradkin:secondary,pis'mak:2005me}

\section{Introduction}\label{sec:1} 
There has recently been renewed interest in conformal field theories (CFT)
in more than two space time dimensions $D$. 
 Besides the perennial interest in 
the critical behavior of $D=3$ dimensional models of statistical mechanics,
including the Ising model, 
additional motivation comes from the $AdS/CFT$ correspondence discovered by
 Maldacena 
\cite{Aharony:1999ti,DHoker:2002aw} 
 which relates string theories to supersymmetric 
Yang Mills Theories in $D=4$ dimensions. They are conformal field theories with
the atypical property of admitting an expansion parameter, and are being 
studied vigorously  \cite{Beisert:2004ry}.
 CFT might also be useful in the description of unparticles
\cite{georgi:unparticles}.

Assuming Osterwalder Schrader positivity  \cite{OS:I,OS:II}, Euklidean Green functions, furnished e.g. by the correlation functions of a critical statistical mechanical system, can be analytically continued to Minkowski space and further to the 
$\infty$-sheeted cover $\M_D\simeq S^{D-1}\times \mathbf{R}$ of compactified Minkowski space \cite{mack:luescher:global}. This space admits a conformal invariant global causal ordering \cite{mack:deRiese}.

Conformal theories on compactified Minkowski space, whose fields have
half-integral dimensions, are also being studied \cite{GCI:todorov:gauge,GCI:infinite:algebra,GCI:bilocal}.

 But here I am interested in theories with fields with anomalous dimensions. 

The $AdS/CFT$ correspondence also 
stimulated interest in relations between theories which live on space time 
manifolds of different dimension $D$ (including relations between quantum field theories \cite{ruehl:lifing:AdS}). The known way how this can happen 
is through a holographic principle 
\cite{holography:susskind,holography:rehren}. I hope to pave the way for the study of different possibilities such as dimensional induction. 

In conformal field theory, operator product expansions (OPE)\cite{wilson:OPE,wilson:OPE:Thirring} 
have special properties. As noted first by Ferrara et.al \cite{ferraraEtAl:shadow,ferraraEtAl:conformalAlgebraOPE,ferraraEtAl:NPB:OPE}, the contributions from derivatives of fields are determined by the others; a partial summation is therefore possible. The conformal OPE are partially summed forms of the Wilson OPE in which the contributions from all derivatives of a field are summed. Moreover, the conformal 
operator product expansions converge on the vacuum $\Omega$. 

This was proven in four dimensions \cite{mack:OPE} 
and also in two 
\cite{Luescher:OPE} and is generally true because it has a simple reason: It furnishes the orthogonal decomposition of the Hilbert space $\H$ of physical states in 
irreducible positive energy representation spaces $\H^\chi$  of the conformal 
group \footnote{It was conjectured in \cite{mack:duality} that convergent operator product expansions exist also in the absence of conformal symmetry}. The positive energy representations in 4 dimensions were determined in ref.\cite{mack:irreps}.

A field $\phi$ with Lorentz spin $l$ takes its values in a 
finite dimensional representation space $V^l$ of the Lorentz group. Its 
components $\phi_\alpha$ are obtained by expanding in a basis of $V^l$.
I will mostly use vector notation. 

In a conformal theory, the OPE take the form \cite{mack:OPE}
\footnote{Throughout the paper, I use a special font, $\mathfrak{Q}$,$\mathfrak{V}$, $\mathfrak{M}$, to indicate {\em kinematically determined} quantities.}
\ba
\phi^j(x)\phi^i(y)\Omega &=& \sum_k \sum_a g^{ji}_{k,a}\int d^Dz \phi^k(z)\Omega \label{OPE} \\
&& \qquad \qquad \mathfrak{Q}^a(\chi_k,z;\chi_j, x,\chi_i, y) \nn 
\ea
The $z$-integration is over Minkowski space (or a conformal translate of it in 
the tube $\M_D$; the result is independent of the translation), and summation is over 
nonderivative fields. 
\be
\mathfrak{Q}^a(\chi_k,z;\chi_j , x, \chi_i, y): V^{l_k}\mapsto V^{l_j}\otimes V^{l_i} 
\ee
are {\em kinematically determined} generalized $c$-number functions (amputated
three-point functions, see later). For scalar fields $\phi^j, \phi^i$ there is 
only one of them, so that the index $a$ and the sum over it can be dropped.
The spin and dimension $\chi_k=[l_k,d_k]$ of all the fields and the coupling 
constants $g^{ij}_{k,a}$ are what has to be consistently determined to fix a 
theory, and in addition the normalizations  $\c_k$ of some two point functions which 
have a physical meaning (i.e. for 
currents and the stress energy tensor).

There is the freedom of choosing a normalization factor in $\mathfrak{Q}$. 
The coupling constants 
$g^{ji}_{k}$ depend on this choice. I will later adopt a particular choice of
 normalization; the kernel $\mathfrak{Q}(\chi ...)$ with this normalization
 will be
denoted by $\mathfrak{Q}^u(\tilde{\chi}...)$. 


In any quantum field theory with operator product expansions which converge on the vacuum $\Omega$, all the vacuum expectation values ($n$-point Wightman functions)
\ba 
W_{i_n,...,i_1}(x_n,...,x_1) &=& 
\langle \Omega , \phi^{i_n}(x_n)...\phi^{i_1}(x_1)\Omega 
\rangle \nn \\
&\in&
 V^{l_{i_n}}\otimes ...\otimes V^{l_{i_1}} \label{Wightman}
\ea 
for arbitrary field operators 
$\phi^i(x)$ 
can be constructed from a consistent set of two- and three-point functions
\cite{mack:duality}. 
Consistency requires that all the 4-point functions obtained in this way 
satisfy locality. The four point functions  possess a factorization property as a consequence of the OPE, see section
\ref{sec:positivity} below, and they are positive if all 2-point functions are positive. 

As shown in \cite{mack:duality}, any collection of 4-point functions with these properties defines a consistent 
quantum field theory. In CFT, the two and
 three-point functions 
of fields of given Lorentz spin and dimension are finite linear combinations
 of kinematically determined quantities \cite{migdal:1,migdal:2,mack:OPE}. The difficulty lies in satisfying 
locality. 

All this was spelled out in detail long ago in the paper \cite{mack:duality}
on duality in quantum field theory, and it was 
also observed there that structures much reminiscent of dual resonance models 
emerge.

In particular, locality is a duality property. Here I derive more detailed results. 

In conformal theories, the dynamical content is in functions of anharmonic ratios. Performing a Mellin transformation on these, 
one obtains a {\em Mellin representation}  in which  amplitudes 
$M_{i_n...i_1}^c(\{\delta_{ij}\})$ figure as 
coefficients. It represents the holomorphic function, whose boundary values are the Wightman functions, on their axiomatic domain of analyticity  \cite{PCT}.
  The {\em same} dual amplitude $M^c$ enters the Mellin representation of the Euklidean Green functions, eqs.(\ref{MellinRep},\ref{MellinRep:n}),
 the Wightman functions, eq.(\ref{wightman:iepsilon}), and also 
the time ordered Green functions in Minkowski space, eq.(\ref{mellin:timeordered}). 
The latter two differ only through $i\epsilon$-prescriptions. The momentum space version of the Mellin representation of the time ordered Green function involves a generalized Feynman integral in the sense of Speer \cite{speer}.

The Mellin amplitudes depend on complex dimensions
 $\delta_{ij}=\delta_{ji}$, $1\leq i < j \leq n$ subject to linear constraints
$\sum_i \delta_{ij}=d_j$. In the case $n=4$, there are two independent variables. Using variables $\beta_{ij}$ which differ from $\delta_{ij}$ by a shift, so that $\sum_i \beta_{ij}=0$,  $\beta_{12}$ and $\gamma_{12}=\beta_{23}-\beta_{13}$ may be used.

I conjecture that the Mellin representation is also valid for massive field 
theories. The constraint $\sum_i \delta_{ij}=d_j$ on the variables 
$\delta_{ij}$ of integration  is characteristic of conformal symmetry and  
will have to be relaxed in order to produce the subleading, i.e. less singular,
contributions to the coefficients in the Wilson OPE. In contrast with scale invariant theories, these coefficients no longer scale like single  powers of the 
distance in massive theories.

In this paper I show how
the costumary principles of local relativistic quantum field theory
 plus conformal invariance and OPE, for 4-point functions, 
translate into the statement that the Mellin amplitudes $M^c_{i_4...i_1}$ have properties of dual resonance models. 

Locality and some information from the OPE
translate into exact duality, i.e. crossing symmetry and meromorphy 
in $\delta_{ij}$ with simple poles in single variables $\delta_{ij}$ [for n=4] whose residues are polynomials in the other independent variable(s). 
Duality is a $D$-independent property. 
CFT's in different dimensions $D$ may have the same Mellin amplitude; for more on this see section \ref{sec:dimensionalReduction}. 
The precise relation between fields 
$\phi^k(x)$ of Lorentz spin $l_k$ and dimension $d_k$ which figure in the OPE
and poles in the Mellin amplitudes depends on $D$ and will be explained below. To every field there correspond a leading pole and satellites labeled by $n=1,2,...$. 

I emphasize that the {\em meromorphy} of the dual amplitudes
considered here is an {\em exact} property. In contrast to the dual resonance models of old, there is no
 ``narrow resonance approximation'' \cite{veneziano:erice1970} involved, and no need to consider multiloop dual diagrams to satisfy unitarity \cite{gross:neveu:scherk:schwartz1970}. 
The deeper reason for this is that the conformal two-point functions of fields $\phi^k$ usually have non-vanishing absorptive parts. 

 The interest in this result is in that there exist 
lots of examples of dual amplitudes; Veneziano's Beta-function Ansatz \cite{Veneziano:model} was the first example.

 Given duality, OPE translate into factorization properties of the residues of poles.   Their precise form for scalar 4-point functions is described in section \ref{sec:factorization}. Basically it is  required that the residues of the leading poles factorize and the satellites are determined by them.

By the Osterwalder Schrader (OS) reconstruction theorem, both the spectrum condition and Wightman-positivity (unitarity) is satisfied if the Euklidean Green functions satisfy OS-positivity. Given duality and factorization, OS-positivity will hold if all the 2-point functions of the fields $\phi^k$ are 
OS-positive. This requires that $\chi_k=[l_k,d_k]$ label unitarizable representations of the confomal group, and it exorcizes $(-)$-signs in the factorization properties which signal ghosts.

 Now I turn to the positions of the poles, for 4-point functions of scalar fields $\phi^{i_4},..., \phi^{i_1}$.  I abbreviate $[l_{i_j},d_{i_j}]$ by  $[l_j,d_j]$.

If $p_i$ are Lorentz vectors in any dimension, not necessarily equal to $D$, subject to the conservation law $\sum_{i=1}^n p_i=0$, and with squares given by the dimensions of the external fields, $p_i^2=d_i$, then the linear constraints 
$\sum_i \delta_{ij}=d_j$
can be solved in terms of Mandelstam variables 
\be s_{ij}= (p_i + p_j)^2 \label{Mandelstam} \ee
viz.
\be \delta_{ij}= -p_ip_j = -\frac 1 2 (s_{ij}-d_i-d_j)\label{delta:Mandelstam} \ee

Consider a particular channel $(21)\mapsto(43)$ or, equivalently, the Wightman function $W_{i_4...i_1}$ with ordering of the fields as in eq.(\ref{Wightman}). Fields $\phi^k \neq \mathbf{1}$ with spin $l_k$ and dimension $d_k$ which appear in the OPE of  both 
$\phi^{i_2}\phi^{i_1}$ and of $\phi^{i_3 \ast} \phi^{i_4 \ast}$ will lead to an integer spaced family of poles of $M_{i_4...i_1}$ in $\delta_{12}$. 

Stated in terms of the Mandelstam variables (\ref{delta:Mandelstam}), poles in the channel $(21)\mapsto(43)$ are at positions
\be s_{12}= d_k-l_k +2n,\qquad  n=0,1,2,... \label{poles:Mandelstam}\ee  
{\em independent} of the dimensions $d_1,...,d_4$ of the external fields. This is the key result which permits the interpretation as dual resonance models, with twists $d_k-l_k$ substituting for mass squares. 
(For scalar fields $l_k=0$.) To appreciate this, remember that in a 4-point scattering amplitude, poles in the Mandelstam variable $s_{12}=(p_1+p_2)^2 $ are determined by masses squared of resonances, {\em independent} of the masses of the particles which form the resonances.

To appreciate the role of field dimensions $d_i$, remember that a CFT with anomalous dimensions lives on $\M_D\simeq S^{D-1}\times \mathbf{R}$. There is a conformal Hamiltonian $H$ which translates $\mathbf{R}$, and the radius of the compact space $S^{D-1}$ furnishes a unit of length. The states $\phi^i(x)\Omega$ span an irreducible representation space labeled by $\chi_i=[l_i,d_i]$, and the eigenvalues of $H$ in it are $d_i+n$, $n=0,1,2,...$ \cite{mack:luescher:global,mack:irreps}.  Thus, $d_i$ are energies in appropriate units. 

What kinds of poles do we expect? 
It is expected that the fields which figure in the OPE, with Lorentz spin $s$ and dimension $d_s$ lie on trajectories $d_s=\alpha(s)$ involving fields with different $s$. These trajectories are what corresponds in quantum field theory to Regge-trajectories in S-matrix theory; they determine light cone singularities.  It is expected that these trajectories are approximately, but not exactly, linear with slope $1$. The evidence for this comes from the study of models which possess an expansion parameter. In $\phi^6$-theory in $D=6+\epsilon$ dimensions \cite{mack:shortDistance}, there is a 
fundamental real field of dimension $d=\frac D 2 -1 + \Delta$, $\Delta = \frac 1 {18}\epsilon + \dots$, and a trajectory of traceless symmetric tensor fields of even rank $s \geq 2$  with dimensions 
$d_s=D-2 + s +\sigma_s$ whose anomalous part is given by the formula
\be
\half \sigma_s = \left[ \frac 1 {18} - \frac 2 {3(s+2)(s+1)}\right] \epsilon + \dots
\label{sigma_s}\ee 
so that $d_s - s \mapsto 2d $ as $s\mapsto \infty$.
Callan and Gross showed that this property of the asymptotic behavior of $d_s$ is always true in scalar field theory \cite{Callan:1973pu}. 

 The member of the trajectory with $s=0$ is absent. It is the shadow in the sense of Ferrara et al. \cite{ferraraEtAl:shadow} of the fundamental field. i.e. $d_0=D-d$. The member $s=2$ is the stress energy tensor; it has dimension $d_2=D$. 

The result (\ref{sigma_s}) was obtained by conformal bootstrap. 
Expansions for anomalous dimensions in $\phi^4$-theory in $4-\epsilon$ dimensions and in $N$-vector models are also known \cite{Wilson:1973jj,pismak:1overNexpansion:bootstrap,pismak:wegner:kehrein:epsilon,vasiliev:RG}.

In the maximally supersymmetric Yang Mills theory in 4 dimensions, there is a trajectory whose anomalous part behaves as $\gamma \ln s$ for large $s$. This contribution is known as the cusp anomalous dimension; its coefficient $\gamma$  was recently calculated \cite{bassoEtAl:cusp}. 

Let us now examine what trajectories and satellites in the dual resonance model will correspond to the trajectories of fields $\phi^k$. If $d_k =\alpha(l_k)$ then eq.(\ref{poles:Mandelstam}) produces a trajectory of poles in the dual resonance model whose position in $s_{12}$ is independent of the dimensions of the external fields
\be  s_{12}=  (\alpha(l_k)-l_k) +2n , \label{dynPoles:dualRes}
\ee
$n=0$, and satellites at $n=1,2,3,...$. We see that an exactly linear field trajectory of slope $1$ would produce poles which fall on top of each other, i.e a linear  trajectory of slope $0$, plus satellites. In this paper, I am  interested in fields with anomalous dimensions which lie on approximately linear field trajectories and I assume that there are at most finitely many poles in the Mellin amplitude which fall on top of each other. In a situation like (\ref{sigma_s}), which is expected to occur in Ising-models, there will be  limit points of poles. On the other hand, in the presence of a cusp anomalous part of the dimension, $\alpha(l_k)-l_k$ will go like $\gamma\ln l_k$ and there are no limit points. In either case, the CFT's correspond to a 0-slope limit of dual resonance models with only approximately linear trajectories. 

It is a challenge to find a substitute for the operator formalism of old in dual resonance models \cite{fubini:veneziano:NC.67A} which ensures the appropriate form of factorization. 

An alternative is to appeal to string theory. String theories were an outgrowth of dual resonance models
\cite{frampton:duality:reprint,green:schwarz,freund:26D,heterotic:string}
and they are also 2-dimensional CFT's.
Positivity in the 
CFT is needed to have a probability interpretation and would therefore be 
ensured by a consistent quantization of the string theory, i.e. a positive definite Hilbert space of its states, {\em if} the state spaces can be identified. 
Identification of the states of the narrow resonances 
with states of a 2-dimensional CFT was an important tool in the old studies of dual resonance models.
In the case of the Maldacena AdS/CFT -duality, the identification of the state spaces was established by group theoretical means, showing that the global symmetry and its representations are identical \cite{DHoker:2002aw,GAdsCFT1}.  
The hope is, ultimately, that string theory will help to construct local relativistic quantum field theories.
Exploiting relation (\ref{delta:Mandelstam}), where $p_i$ are Lorentz vectors of some dimension (determined by a string theory) with $p_i^2=d_i$, 
one could try to get $n$-point Mellin amplitudes $M(\{\delta_{ij})$ from string field expectation values by  an Ansatz like
\be
\delta(\sum p_i)M(\{ -p_i p_j\})
=\langle \int dV \ e^{i\sum_i p_{i\mu} X^\mu(\sigma_i,\tau_i)} \rangle\ .
 \label{string:Ansatz}
\ee
  The prototype of such a formula is in Veneziano's 1970 Erice lectures \cite{veneziano:erice1970}.
 $dV$ is an invariant  volume element which integrates over $(n-3)$ of 
the arguments $(\sigma_i,\tau_i)$ of the string field $X^\mu$. The time ordered Green functions of the $D$-dimensional CFT with scalar fields $\phi^i$ of dimension $d_i$ 
are expressed in terms of $M$ by eq.(\ref{mellin:timeordered}), viz.
\ba
&& \langle \Omega, T\left\{\phi^{n}(x_n)...\phi^{1}(x_1)\right\}\Omega\rangle \nn \\ 
&&  =
(2\pi i)^{-n(n-3)/2}\int \int d\delta\ 
M(\{\delta_{ij}\})\nn \\ && \qquad \qquad \prod_{i>j} \Gamma(\delta_{ij})(-\frac 1 2 x_{ij}^2+i\epsilon )^{-\delta_{ij}} . \nn
\ea
Connected parts may have to be taken, see below. 

 To incorporate supersymmetry requires consideration of 
amplitudes with spin. This is postponed. The ultimate desire is to find an {\em explicit} duality transformation between quantum field theory at strong coupling and string theory at weak coupling.

There is a technical complication which affects the Mellin representation.
 It comes from the presence of disconnected parts and leads to an unexpected feature of 
the dual amplitudes.
 It is best stated in terms of the Euklidean Green functions 
$G_{i_n...i_1}(\x_n,...,\x_1)$ 
which are the analytic continuation of the Wightman functions. 

It is the full disconnected Greens function  $G_{i_n...i_1}(\x_n,...,\x_1)$
which satisfies the Osterwalder Schrader axioms, and to which operator product
 expansions apply.

In the dual resonance models of old, there are no disconnected parts, because 
they would correspond to processes $i,j\mapsto \mbox{nothing}$ in some channel,
which is impossible on-shell.

 The full disconnected Green functions  decompose into a sum of 
disconnected parts 
 and the connected Green functions $G^c_{i_n...i_1}$. In case of the 4-point
 functions, the disconnected parts involve only two-point functions, 
\ba G_{i_4...i_1}(\x_n,...,\x_1) &=&
 G_{i_4 i_3}(\x_4,\x_3)G_{i_2 i_1}(\x_2,\x_1)+\nn \\
 &&  G_{i_4 i_2}(\x_4,\x_2)G_{i_3 i_1}(\x_3,\x_1)+ \nn \\
 &&  G_{i_4 i_1}(\x_4,\x_1)G_{i_3 i_2}(\x_3,\x_2)+\nn \\
 &&  G^c_{i_4...i_1}(\x_4,...,\x_1) \ . \label{disc}\ea
There is one disconnected part for each channel $(ji)$, e.g. $(21)\mapsto(43)$;
 it is given by the contribution of the unit operator ${\bf 1}$ 
 to the operator product $\phi^j\phi^i$ and vanishes if there is no such contribution.

For technical reasons, it is necessary to write a Mellin representation 
involving the dual amplitude $M^c$ for the {\em connected} 
Green functions. Now consider the contribution from field $\phi^k$ in the
 operator product expansion of $\phi^{i_2}\phi^{i_1}$.  It furnishes 
contributions to the asymptotic expansion of the full disconnected Green 
function
 $G_{i_4...i_1}(\x_4,...,\x_1)$ 
when $\x_{12}=\x_1 - \x_2\mapsto 0$. These
contributions must match with contributions from the asymptotic expansion 
which obtains from the Mellin representation of the {\em connected} Green 
function, {\em plus} the disconnected parts. If $\phi^k=\mathbf{1}$, then the
 matching contribution is the disconnected part  associated with channel 
$(ji)$. Suppose $\phi^k\neq \mathbf 1$. Then the matching contribution comes 
from 
the integer spaced family of poles (\ref{poles:Mandelstam}) of $M^c$ in $\delta_{12}$.

However, if there are other non-vanishing disconnected parts than the afore mentioned one associated with channel $(ji)$, then they must be cancelled by 
contributions from the Mellin representation of the connected Green function.
 For generic anomalous dimensions they would  have the wrong degrees of 
homogeneity in $x_{21}$ for contributions from OPE. Therefore the dual amplitude for the connected part must have kinematical poles in addition to the dynamical ones. They are fully determined in their positions and residues by the requirement that their contribution cancels the asymptotic expansion of disconnected parts (when they are 
nonzero and incompatible with OPE).    

The Euklidean partial wave expansions suggest that these technical complications could be avoided by an illegitimate interchange 
of sums and integrations, which amounts to writing down a formal Mellin 
representation of the full disconnected Green function. But I do not know how 
to do this in a legitimate way.

This paper deals with scalar amplitudes, mostly 4-point functions. n-point functions will be considered in section \ref{sec:nPointFct}. The strategy of the paper is as follows. It would be difficult to 
find the polynomial residues of the dual amplitudes directly from the match between asymptotic 
expansions from OPE and from the Mellin representation. Therefore I will make a
 detour, using results established in the 70's on partial waves expansions of
Euklidean Green functions (on the Euklidean conformal group $G^E=Spin(1,D+1)$),
and their relation to OPE in Minkowski space \cite{mack:groupth:1,mack:groupth:2,mack:dobrevBook,mack:dobrevClebsch,dobrevEtAl:OPE}. As a check of the result, I will verify that the match between asymptotic expansions obtains for the special case of the contribution of a scalar field to the OPE and the associated family of poles in the Mellin amplitude.

The dynamical information on the 
scalar four point function $G_{i_4 i_3 i_2 i_1} $ within the Euklidean partial wave expansion in 
channel $(21)\mapsto (43) $ resides in a complex function $g^{(43)(21)}(\chi) $
 of 
$\chi=[l,\delta]$, where $l$ is again a (traceless symmetric tensor) representation of the (Euklidean) Lorentz group, and $\delta$  is a complex dimension.
It has the symmetry property
$$g^{(43)(21)}(\chi)=g^{(43)(21)}(\tilde{\chi})$$
where $\tilde{\chi}=[\bar{l}, D-\delta ]$, with $\bar{l}$ the dual 
representation to
$l$. Validity of OPE expansions requires that $g^{(43)(21)}(\chi)$ is a 
meromorphic function of $\delta$ and there is a bijective correspondence between (pairs of) poles of $g^{(43)(21)}(\chi)$ and fields $\phi^k$ in the OPE as follows. A field with Lorentz spin and dimension $\chi_k=[l_k,d_k]$ and  which contributes to 
the OPE (\ref{OPE}) of $\phi^{i_2}\phi^{i_1}$, with coupling constant 
$g^{21}_k$ {\em and} of $\phi^{i_3\ast}\phi^{i_4\ast }$, 
with coupling constant $\bar{g}^{43}_k$ 
 requires a pole of 
$g^{(43)(21)}([l_k,\delta])$ at $\delta=d_k$ with residue  
\be \bar{g}^{43}_kg^{21}_k \c_k \label{factRes}\ee 
%
%
times a kinematical factor. $\mathbf{c}_k$ is given by the normalization of the 2-point function of $\phi^k$. 
I will first derive the Mellin representation of individual Euklidean 
conformal partial waves, using an integration formula of Symanziks
 \cite{symanzik:nstar}, 
 and then deduce the meromorphy properties of the dual 
amplitude, including its residues, from the meromorphy properties of
$ g^{(43)(21)}(\chi)$. The factorization properties of the dual amplitude come 
from the factorized expression (\ref{factRes}) of the residue of  
 $g^{(43)(21)}([l_k,\delta])$. 
\section{Positivity}\label{sec:positivity}
\subsection{Fields and positive energy representations}\label{sec:fields&posERep}
 I consider collections of nonderivative  
field operators $\phi^i(x)$, including the unit operator $\mathbf{1}$, which
are closed under OPE. The conformal transformation law of $\phi^i$ is 
specified  by 
\be \chi_i=[l_i,d_i]\label{chi}\ee
 where $d_i$ are positive real, called the dimension of the field $\phi^i$, and
 $l_i$ specifies an irreducible finite 
dimensional representation of the Lorentz group $M$ by operators $D^{l_i}(m)$
in a vector space $V^{l_i}$.
 I write $l=0$ for the trivial 1-dimensional representation of $M$. 
Scalar fields have $l=0$. 
For the unit operator, $l=0$, $d=0$. Physically allowed values of $\chi$ are 
in bijective correspondence with unitary irreducible positive energy
 representations of the conformal group $G$ in $D$ dimensions. They depend on 
$D$. Fields $\phi^i_\alpha(x)$ with spinor indices $\alpha$ are obtained
by expanding the vector $\phi^i(x)\in V^{l_i}$ in a multispinor basis of
$V^{l_i}$.    

For the sake of transparency, I will write multispinor indices $\alpha$ of 
fields $\phi^i_\alpha (x)$ in this section. 

When anomalous (i.e. not half integral) field dimensions are admitted, CFT
lives on the $\infty$-sheeted simply connected universal covering $\M_D$ of conformally compactified Minkowski space, and $G$ is the infinite 
sheeted universal covering of $SO(2,D)$  \cite{mack:luescher:global}. 
 The Wightman 
functions are determined by their values on Minkowski space, but admit unique analytic continuation to $\M_D$ and
appropriately defined field operators may be regarded to live on $\M_D$. 
State vectors of the special form $\phi^k_\alpha(z)\Omega $ span 
irreducible representation spaces $\H^k$ of $G$. These state vectors are periodic on $\M_D$ so that they can be regarded as sections of vector bundles over 
compactified Minkowski space.
 (In even dimensions this follows from Schur's lemma,
applied to the center of $G$).  In the following, all integrations $\int d^Dz $
are over Minkowski space.
\subsection{Positivity from Operator Product Expansions}\label{sec:WightmanPositivity}
I demand the standard properties of Wightman functions \cite{PCT} - spectrum
 condition, positivity and locality plus conformal invariance, and validity of
 operator product expansions (OPE) (\ref{OPE}) which converge on the
 vacuum $\Omega$.
 
Let us analyze positivity. 
Consider finite sequences $f$ of test functions on Minkowski space, 
$$ f^{\emptyset}, f^{i_1}_{\alpha_1}(x_1),\dots , 
f^{i_1...i_N}_{\alpha_1\dots \alpha_N}(x_1,\dots ,x_N) $$
where $f^\emptyset$ is a constant. For given sequence $f$,
consider the state vector
\ba \Psi (f)&=& 
f^\emptyset\Omega +
\sum_{n=1}^N \int d^Dx_n...\int d^Dx_1 
f^{i_1...i_n}_{\alpha_1...\alpha_n}(x_1...x_n)\nn \\
&&\qquad
\phi^{i_n}_{\alpha_n}(x_n)
 \dots \phi^{i_1}_{\alpha_1}(x_1)\Omega \ . \label{state}\ea
Summation over all repeated indices $\alpha_1,..., i_n$ is understood. 
$\Psi(f)$ is an element of the Hilbert space of physical states, and must 
therefore have positive semidefinite norm,
\be
\langle \Psi (f), \Psi (f)\rangle \geq 0  \label{posNorm}
\ee
Inserting the definition (\ref{state}) one obtains an inequality on Wightman
 functions \cite{PCT} which is known as Wightman positivity. 
It is related to unitarity in the language of time ordered Green functions. 

The analysis of Wightman positivity is much simplified by validity of OPE. 
Consider first sequences with $N=2$. Inserting in definition 
(\ref{state}) the OPE (\ref{OPE}) for 
\mbox{$\phi^{i_2}_{\alpha_2}(x_2)\phi^{i_1}_{\alpha_1}(x_1)\Omega$}
we obtain
\be
\Psi(f)=\sum_k^\prime \int d^Dz h^k_\alpha (z) \phi^k_\alpha(z)\Omega 
+ h^\emptyset \Omega 
\label{stateWithOPE}
\ee
Summation $\sum^\prime $ is over all non-derivative fields {\em other than} the unit operator.
Explicitly, the functions $h^k_\alpha (z)$ are given by
\ba
h^k_\alpha (z)&=&\sum^\prime_a g^{i_2i_1}_{k,a}\int\int d^Dx_2 d^Dx_1
f^{i_1i_2}_{\alpha_1\alpha_2}(x_1,x_2)\nn \\
&& \mathfrak{Q}^a_{\alpha \alpha_2 \alpha_1}
(\chi_k,z;\chi_{i_2}, x_2,\chi_{i_1}, x_1)
+f^k_\alpha(z)\nn  \\ 
h^\emptyset &=& g^{i_2 i_1}_0 \int \int d^Dx_2 d^Dx_1 
f^{i_1i_2}_{\alpha_1\alpha_2}(x_1,x_2) \nn \\ 
&& \mathfrak{Q}_{\emptyset \alpha_2 \alpha_1}
(0,z;\chi_{i_2}, x_2,\chi_{i_1}, x_1)+f^\emptyset \ . \nn 
\ea
Here I write $\chi=0$ for the trivial 1-dimensional representation and
 $\alpha=\emptyset$ when there is no index. The
kernel $\mathfrak{Q}$ in the expression for $h^\emptyset$ is actually a 
two-point function independent of $z$. There is only one such, therefore 
there is no index $a$ to sum over in the formula for $h^\emptyset$. 

Inserting  eq.(\ref{stateWithOPE}), and using that in a conformal theory
$\langle \Omega , \phi^k_\alpha(z) \Omega \rangle =0 $ for all fields 
$\phi^k$ other than the unit operator, we obtain
\ba
\langle \Psi (f), \Psi (f)\rangle &=&
\sum^\prime_k\sum^\prime_{k^\prime} 
\int \int d^Dz d^Dz^\prime \label{normWithOPE}
\\
&&
 \bar{h}^k_\alpha(z) 
\langle \Omega, \phi^{k\ast}_\alpha (z) 
 \phi^{k^\prime}_{\alpha^\prime} (z^\prime)\Omega \rangle   
h^{k^\prime}_{\alpha^\prime}(z^\prime) \nn \\
&&+ \bar{h}^\emptyset h^\emptyset \langle \Omega , \Omega\rangle 
\ea
This is positive semidefinite {\em provided} the two-point functions of all 
the fields are positive. 

Using the OPE repeatedly as described in ref. \cite{mack:duality} it is seen 
that eq.(\ref{stateWithOPE}) generalizes to arbitrary finite sequences $f$, 
and the same is therefore also true of the result (\ref{normWithOPE}).   

Conformal invariant two point functions are determined by conformal symmetry 
up to an overall factor. Taking linear combinations of fields if necessary, 
\be \langle \Omega, \phi^{k\ast}_\alpha (z) 
 \phi^{k^\prime}_{\alpha^\prime} (z^\prime)\Omega \rangle
= \c_k\delta_{k k^\prime}\Delta^{\chi_k}_{\alpha \alpha^\prime}(z,z^\prime)
\label{2pt}\ee
where $\Delta^{\chi_k}(z,z^\prime)$ is kinematically determined.
 If the field transformation law is given by $\chi_k=[l_k,d_k]$ 
which determines a unitarizable representations, then $\Delta^{\chi_k}$ is positive and the two point functions are
either positive or negative semidefinite, depending on whether the 
overall factor $\c_k$ is positive or negative.
Fields whose two-point function is negative are called {\em ghosts}. 

In conclusion, once validity of OPE is assured and if the fields $\phi^k$ have 
spin and dimension $\chi_k=[l_k,d_k]$ which correspond to unitary 
representations of $G$, positivity is assured if none of the fields is a ghost.

The requirement that $\chi_k$ should label unitary positive energy 
representations of $G$  will be called the requirement of {\em no tachyons}. 
Actually it also comes out of the requirement of positivity. But it is 
convenient to divide positivity into two requirements:
 Absence of tachyons which gives
D-dependent restrictions on the allowed positions of poles of Mellin 
amplitudes, and absence of ghosts which could be inherited from a 
2-dimensional theory. 
\subsection{Factorization properties from Operator Product Expansions}
\label{sec:FacFRomOPE}
Consider two pairs of fields, labeled by $j,i$ and $l,m$.
Assume that fields are labeled in such a way that 
$\phi^{k\ast}=\phi^{\bar k}$.

Inserting the OPE (\ref{OPE}) for $\phi^j\phi^i\Omega$ and for $\phi^l\phi^m\Omega$, and using the formula (\ref{2pt}) for the conformal invariant two-point functions, the OPE for the four-point Wightman functions takes the form
\ba
&&
 W_{\bar{m},\bar{l},j,i}(x_4,...,x_1) =
\langle \Omega , \phi^{\bar{m}}(x_4)\phi^{\bar{l}}(x_3)\phi^j(x_2)\phi^i(x_1)\Omega \rangle\nn \\
&& = \sum_k \sum_{a,b}\c_k \bar{g}^{lm}_{k,b}g^{ji}_{k,a}
\mathfrak{W}^{ba}_{\chi_k,\bar{m},\bar{l},j,i}(x_4,...,x_1) \label{OPE:4ptfct}
\ea
etc., where $\mathfrak{W}$ are kinematically determined quantities,
\ba
&& \mathfrak{W}^{ba}_{\chi_k,\bar{m},\bar{l},j,i}(x_4,...,x_1)=\int\int dz\ dz^\prime   \label{Wkin}\\  && \!\!\!\!\!\!\!
\bar{\mathfrak{Q}}^b(\chi_k,z; \chi_l,x_3,\chi_m,x_4)
\Delta^{\chi_k}(z,z^\prime)
\mathfrak{Q}^a(\chi_k,z^\prime; \chi_j,x_2,\chi_i,x_1) \nn 
\ea
Suppose that a particular field $\phi^k$ contributes to both 
$\phi^j\phi^i\Omega$ and $\phi^l\phi^m\Omega$. Then it will contribute to all 
of the following Wightman functions and its contributions are related as 
follows.

Assume for transparency that there is only one invariant 3-point function 
so that the label $a$ on $\mathfrak{Q}^a$ in eq.(\ref{OPE}) is redundant. 
Then the contribution of the field $\phi^k $ to the  4-point Wightman functions is
\ba 
W_{\bar{m},\bar{l},j,i}(x_4,...,x_1)&=&
 \c_k \bar{g}^{lm}_kg^{ji}_k  \mathfrak{W}_{\chi_k ,\bar{m},\bar{l},j,i}(x_4,...,x_1)\nn \\ && \quad +\dots ,\nn \\
W_{\bar{i},\bar{j},j,i}(x_4,...,x_1)&=&
 \c_k \bar{g}^{ji}_kg^{ji}_k \mathfrak{W}_{\chi_k ,\bar{i},\bar{j},j,i}(x_4,...,x_1)\nn \\ && \quad +\dots , \nn \\
W_{\bar{m},\bar{l},l,m}(x_4,...,x_1)&=&
 \c_k \bar{g}^{lm}_kg^{lm}_k \mathfrak{W}_{\chi_k ,\bar{m},\bar{l},l,m}(x_4,...,x_1)\nn \\ && \quad +\dots . \nn
\ea
The real constants $\c_k$ are determined by the two-point functions (\ref{2pt}). They are positive in a theory without ghosts. 

We see that the coefficients of the kinematically determined factors $\mathfrak{W}$ factorize in a product of coupling constants in a way which is familiar from dual resonance models. These factorization properties are at the origin of the factorization properties of the residues of the poles in the Mellin amplitude, and they assure positivity under the above mentioned conditions.

  For future use we note that the kinematically determined factors $\mathfrak{W}$ can be written in another way.

Consider three-point functions.
There is a finite number of kinematically determined conformal invariant 
3-point functions $\mathfrak{V}^a$ , so that
\ba
&& \langle \Omega , \phi^{k}(x_3)\phi^j(x_2)\phi^i(x_1)\Omega \rangle
\label{3pt}\\
&& 
= \sum_a \tilde{g}^{ji}_{k,a}\mathfrak{V}^a(\chi_k,x_3,\chi_j,x_2,\chi_i,x_1) 
\nn 
\ea
with dynamically determined coupling constants $\tilde{g}^{ji}_{k,a}$ 
which will be related to $g^{ji}_{k,a}$ below.

Inserting the OPE in the 3-point function and using (\ref{2pt}) gives
\ba
\langle \Omega , \phi^{\bar{k}}(x_3)\phi^j(x_2)\phi^i(x_1)\Omega \rangle
& = & \c_k \sum_a \int dz \Delta^{\chi_k}(x_3,z)\label{amputV}
\\ && g^{ji}_{k,a} 
\mathfrak{Q}^a(\chi_k,z;\chi_j,x_2,\chi_i,x_1) \nn
\ea
Exploiting the freedom of taking linear combinations of the kinematically 
determined quantities we
 may demand the following {\em amputation identity}
\ba
&& \mathfrak{V}^a(\chi_{\bar{k}},x_3,\chi_j,x_2,\chi_i,x_1)= \label{QvsV}\\
&& \int dz.
\Delta^{\chi_k}(x_3,z) 
\mathfrak{Q}^a(\chi_k,z;\chi_j,x_2,\chi_i,x_1) \nn
\ea.
and 
\be \tilde{g}^{ji}_{k,a}=\c_k g^{ji}_{k,a} \ee
This shows that $\mathfrak{Q}$ is an amputated Wightman 3-point function. It is uniquely determined by this and by the requirement that its partial Fourier
 transform in $z$ is a holomorphic function of $p$.

Inserting eq.(\ref{amputV} simplifies the defining formula (\ref{Wkin}) of 
$\mathfrak{W}$.
 The resulting formula will be used in section 
\ref{sec:comparison} for four-point functions of real scalar fields; the labels $a,b$ are redundant in this case, and $\bar{l}=l$ etc..
\ba
&& \mathfrak{W}^{be}_{\bar{m},\bar{l},j,i}(x_4,...,x_1) = \label{QV} \\
&& \int dz. \bar{\mathfrak{Q}}^b(\chi_k,z;\chi_l,x_3,\chi_m,x_4)
\mathfrak{V}^a(\chi_k,z,\chi_j,x_2,\chi_i,x_1)\nn \\
&=&  {\mathfrak{Q}}^b(\chi_k,p;\chi_m,x_4,\chi_l,x_3)
\mathfrak{V}^a(\chi_k,z,\chi_j,x_2,\chi_i,x_1)|_{z=0}\ . \nn
\ea.
In the last formula, $\mathfrak{Q}(\chi_k,p;...)$ 
is the partial Fourier transform of 
${\mathfrak{Q}}$ with respect to the variable $z$; it has the reality property
(\ref{Q:reality}).
  $p$ is to act as a differential operator  with respect to $z$, $p=-i \nabla_z$.
As noted before, indices $a$, $b$ are redundant when scalar four-point 
amplitudes are expanded. 
%
\section{Universal Mellin representation}\label{sec:mellin}
\subsection{Euklidean Green functions}
Wightman functions admit analytic continuation to Euklidean points $x$, 
defining Euklidean Green functions 
\be G_{i_n...i_1}(\x_n,...,\x_1)\in V^{l_n}\otimes..\otimes V^{l_1}.\ee
Euklidean points $x$ have imaginary time components $x^0$ and real space 
components $x^i$, $i=1,...,D-1$. They can also be specified by real coordinates
$\x=(x^1,...,x^D)$, $x^D=ix^0$. {\em Whenever $x$ and $\x$ will appear in the same formula, they are understood to be related in this way.} In particular,
$$ -x^2=\x^2> 0 $$ in Euklidean space.

By Weyls unitary trick, the finite dimensional representation spaces
$V^l$ of the Lorentz group are identified with representation spaces of
its Euklidean brother $Spin(D)$.  The Euklidean Green functions so defined 
include disconnected parts, and satisfy Oterwalder Schrader 
positivity as a substitute for spectrum condition and positivity 
of Wightman functions. Locality becomes (crossing) symmetry,
\be G_{i_{\pi n}...i_{\pi 1}}(\x_{\pi n},..., \x_{\pi 1}) =
\pm \ \hat{\pi}\ G_{i_n...i_1}(\x_n,...,x_1) \ee
for arbitrary permutations $\pi$ of $1...n$. $\pm$ depends on how many Fermi fields are interchanged, and 
$$\hat{\pi}: V^{l_n}\otimes..\otimes V^{l_1}
\mapsto V^{l_{\pi n}}\otimes..\otimes V^{l_{\pi 1}}$$ 
is the map which permutes multispinor indices. It is redundant for scalar Green functions.

The Euklidean Green functions decompose into connected Greens functions plus 
possibly disconnected parts as in eq.(\ref{disc}).
The above symmetry property is shared by the connected Greens functions.

The Eukidean Green functions are invariant under the Euklidean conformal
group $Spin(D+1,1)$ which is isomorphic to the Lorentz group in $D+2$
dimensions.  
\subsection{Mellin representation of scalar 4-point functions}
\label{sec:MellinRep4}
Let us first consider Euklidean 4-point Green functions of four scalar fields of dimensions $d_1,...,d_4$ including in particular
the Greens function  $G_{0000}(\x_4,...,\x_1)\in \C$ 
of a single scalar field $\phi^0$ of dimension $d_0$. For the latter, the
 symmetry property simplifies to 
\be
 G_{0000}(\x_{\pi 4},..., \x_{\pi 1}) = G_{0000}(\x_4,...,\x_1) .
\label{EuklScalar4pt}
\ee
The arbitrariness in a scalar conformal invariant 4-point functions resides in 
an arbitrary function $F(\omega) $ of anharmonic ratios. They are defined as 
follows, with $x_{ij}=x_i -x_j$. 
\be \omega_1=\frac {x_{12}^2 x^2_{34}}{x_{13}^2 x_{24}^2}, \quad
\omega_2=\frac {x_{14}^2 x_{23}^2}{x_{12}^2 x_{34}^2}, \quad
\omega_3=\frac {x_{13}^2 x_{24}^2}{x_{14}^2 x_{23}^2}, \label{def:omegas}
\ee
and satisfy 
\be \omega_1\omega_2\omega_3=1 \ . \label{harmConstraint}\ee
For non-coinciding Euklidean points, $$x_{ij}^2 = -|\x_{ij}|^2 < 0$$ 
and all the anharmonic ratios are positive. We write $\omega=(\omega_1,\omega_2,\omega_3)$ for triples of anharmonic ratios subject to (\ref{harmConstraint}).
Permutations $\pi$ of arguments $\x_1...\x_4$ act on such triples $\omega$ as 
follows. Consider transpositions (ij) of two arguments $i$ and $j$.
\ba
(12)\mbox{ or }(34):&& \omega_1\mapsto\omega_2^{-1}, \quad
\omega_2\mapsto\omega_1^{-1}, \quad
\omega_3\mapsto\omega_3^{-1},\nonumber \\
(13)\mbox{ or }(24):&& \omega_1\mapsto\omega_3^{-1}, \quad
\omega_2\mapsto\omega_2^{-1}, \quad
\omega_3\mapsto\omega_1^{-1}, \nonumber\\
(14)\mbox{ or }(23):&& \omega_1\mapsto\omega_1^{-1}, \quad
\omega_2\mapsto\omega_3^{-1}, \quad
\omega_3\mapsto\omega_2^{-1},\nn 
\ea 
Note that some even permutations of $x_i$ act trivially on $(\omega_1,\omega_2, \omega_3)$,
\be
\pi \omega= \omega \mbox{ for }
 \pi = (12)(34) \mbox{ or } (13)(24) \mbox{ or } (14)(23) \ .\label{trivPermut}
\ee
The other even permutations of $x_i$ act as cyclic permutations of $(\omega_1,\omega_2, \omega_3)$.

Consider complex exponents $\delta_{ij}=\delta_{ji}$ attached to unordered 
pairs $(ij)$ of distinct  $i,j=1...4$ subject to the constraint
\be \sum_i \delta_{ij} = d_j \label{DeltaConstraint} \ee
There is a unique real solution $\{\delta^0_{ij}=\delta^0_{ij}(d_1,...,d_n)\}$
 of this system of equations which satisfies the condition 
\be
\delta^0_{\pi i \pi j} (d_{\pi 1},...,d_{\pi n})= \delta^0_{ij}(d_1,...,d_n)
\label{delta^0}
\ee
For $n=4$ it is given by
\be \delta_{12}^0 = \frac 1 6 (2d_1+2d_2-d_3-d_4) \mbox{ etc.} \ee
Given the dimensions $d=\{d_i\}$ of the four scalar fields, $i=1...4$ 
define the
{\em fundamental scalar 4-point function}
\be f(x_1,...,x_4|d)=\prod_{ij}(-\frac 1 2 x_{ij}^2)^{-\delta_{ij}^0} 
\label{fund}\ee  
The most general conformal invariant Euklidean scalar 4-point Green function has the form
\be 
G_{i_4...i_1}(\x_1,...,\x_4) = f(x_1,...,x_4|d) F_{i_4...i_1}(\omega)\label{genScalar}
\ee
and locality reads
\be
F_{i_{\pi 4}...i_{\pi 1}}(\pi \omega )= F_{i_4...i_1}(\omega)
\label{localityInOmega}\ee
In particular, $F_{0000}(\pi \omega)=F_{0000}(\omega)$, and these relations 
depend neither on $D$ nor on the dimensions $d_i$ of the scalar fields.

We also see that locality condition (\ref{EuklScalar4pt}) is automatically 
fulfilled for those permutations $\pi$, given by eq.(\ref{trivPermut}),
for which $\pi \omega=\omega$.

In general, the full disconnected Green function is the sum of disconnected parts and a connected Green function $G^c$, as in eq.(\ref{disc}).
This translates into sums of terms of the form (\ref{genScalar}) in the obvious way. For the amplitude $G_{0000}$ of four equal fields of dimension $d_0$
 with two-point function 
$$ G_{00}(\x_2,\x_1)= N_{d_0} (-\frac 1 2 x_{12}^2)^{-d_0}, $$
the decomposition reads
\ba
F_{0000}(\omega)&=& N_{d_0}^2 \left( \left(\frac{\omega_2}{\omega_1}\right)^{d_0/3} + \left(\frac{\omega_1}{\omega_3}\right)^{d_0/3} + \left(\frac{\omega_3}{\omega_2}\right)^{d_0/3}\right) \nn \\
&& +F^c_{0000}(\omega) \ . 
\ea
Note that the sum of disconnected parts fulfills locality constraint 
(\ref{localityInOmega}) as it must be. 
In a generalized free field theory, $F^c_{i_4...i_1}=0$.

Now we subject $F^c(\omega)$ to Mellin transformation. Given the dimensions 
of fields, OPE yield information 
about the strength of the singularities of the Euklidean Green functions at 
coinciding arguments, and this translates into information on the strength of 
the singularities of $F(\omega)$ when some $\omega_a \mapsto 0$ or $\infty$. 
If these singularities are not too strong, and if some further boundedness
properties hold,
 a Mellin expansion exists with 
paths that  
can be taken along the imaginary axis %
\footnote{It is known that an expansion of this form exists order by order in a skeleton graph expansion, for Greens functions of fundamental fields. It can be derived 
using Symanzik's integration formula (\ref{nStar})}.
It is convenient to write the expansion in the 
following form. 

The general solution of eq.(\ref{DeltaConstraint}) is of the form
\be
\delta_{ij} = \delta^0_{ij}+\beta_{ij}
\ee
where $\beta=\{\beta_{ij}\}$ is a solution of the homogeneous equation, whence
\ba &&\beta_{34}=\beta_{12}, \quad \beta_{24}=\beta_{13},\quad
 \beta_{14}=\beta_{23}, \label{beta:sol1}\\
&& \beta_{12}+\beta_{13}+\beta_{23}=0 \ . \label{beta:sol2}
\ea
It depends on two independent parameters, e.g. $\beta_{12} $ and $\beta_{13}$.
Abbreviating $(i_4...i_1)=a$ to specify a quadruple of scalar fields,
 the Mellin expansion reads
\be
F^c_a(\omega)=(2\pi i)^{-2} \int_{-i\infty}^{i \infty}\int_{-i\infty}^{i \infty}
d^2\beta \ \hat{M}^c_a(\beta)
\omega_1^{-s_1}  \omega_2^{-s_2}\omega_3^{-s_3},  \label{mellin:F}
\ee
where $s_i$ are determined by $\beta$ up to a common additive constant, 
which is irrelevant by (\ref{harmConstraint}), and
\be
s_2-s_1= -\beta_{12} , \quad s_3-s_2= -\beta_{23}, \quad s_1-s_3=-\beta_{13}
\ee
 $d^2\beta$
is integration over two independent variables, e.g. $d^2\beta=d\beta_{12}d\beta_{13}$.

$\hat{M}^c_a $ will be called the {\em reduced Mellin amplitude}. There is an obvious action of permutations $\pi $ of the four space time points on $\beta$, 
viz.
\be
(\pi \beta)_{ij}= \beta_{\pi^{-1} i \pi^{-1} j} \label{piBeta} 
\ee
and locality reads
\be \hat{M}_{\pi a}(\pi \beta) = \hat{M}_a(\beta) \ee
There is neither an explicit reference to $D$ nor to the dimensions of the scalar fields. 

Defining, for $\delta_{ij} = \delta_{ij}^0+\beta_{ij}$ 
the {\em Mellin amplitude} $M^c$ by
\be M^c_a(\{\delta_{ij}\})= \hat{M}^c_a(\{\beta_{ij}\})\prod_{ij} 
\Gamma(\delta_{ij})^{-1}, \label{reduce}\ee  
the connected scalar 4-point function can be written as
\ba
 G^c_a(\x_4,...,\x_1)&=& (2\pi i)^{-2}\int \int d\delta\ 
 \qquad M^c_a(\{\delta_{ij}\})\nn \\ && \qquad 
 \prod_{ij} \Gamma(\delta_{ij})(-\frac 1 2 x_{ij}^2)^{-\delta_{ij}}, \label{MellinRep}
\ea
with $a=(i_4...i_1)$, $d\delta = d^2\beta = d\beta_{12}d\beta_{13}$
 and integration over imaginary
 $\beta_{12}, \beta_{13}$. In contrast with the reduced Mellin amplitude, 
$M^c$ depends on the dimensions of the scalar fields. 

The explicit $\Gamma$-functions decrease exponentially in the imaginary
 directions. They are included in order that the growth conditions on the 
Mellin amplitude, which ensure the axiomatic analyticity in coordinate space,
 take a simple form; cf. section \ref{sec:spectrum}.  

It will be shown in section \ref{sec:mellin:poles} that OPE in some channel,
 let us say
$(21)\mapsto (43)$, determine the poles of $M^c$ in $\delta_{12}$ (and $\delta_{34}$) and their residues. But there are also zeroes at non-positive integers $\delta_{12}$ and $\delta_{34}$ so that the reduced Mellin amplitude $\hat{M}^c$, which is related to $M^c$ by eq.(\ref{reduce}), does not have 
extra poles at non-positive integers $\delta_{12}$ and $\delta_{34}$. 
The result suggests to regard the factor $M^c$ in the Mellin representation (\ref{MellinRep}) as an operator which creates the integrand of (\ref{MellinRep}) from the raw material
 $\prod \Gamma(\delta_{ij})(-\half x_{ij}^2)^{-\delta_{ij}}$ 
by shifting poles and modifying residues.
\subsection{Scalar $n$-point functions for $n>4$}\label{sec:nPointFct}
The generalization of the Mellin representation to $n>4$ is straightforward. 
Given the field dimensions $d_i$, $i=1...n$, consider again solutions 
$\delta_{ij}=\delta_{ji}$, $j\neq i$ of the 
system of equations
\be \sum_i \delta_{ij}=d_j\, \qquad j=1\dots n \nn
\ee
The solution space is $\frac 1 2 n(n-3)$ dimensional \cite{symanzik:nstar}.
 There 
is a unique solution $\delta^0_{ij}$ of this system of equations with the property (\ref{delta^0}), and the general solution is 
$\delta_{ij}=\delta^0_{ij}+\beta_{ij}$, where $\beta_{ij}$ is the general 
solution of the homogeneous equation; $\frac 1 2 n(n-3) $ of these variables 
are independent. We may choose $\beta_{ij}$ with $2\leq i < j <n$, excepting 
$\delta_{23}$ as the independent ones \cite{symanzik:nstar}. The Mellin 
representation of the connected Green function reads
\ba
 G^c_a(\x_n,...,\x_1) &=& (2\pi i)^{-n(n-3)/2}\int \int d\delta\ 
M^c_a(\{\delta_{ij}\})\nn \\ && \qquad \qquad
 \prod_{ij} \Gamma(\delta_{ij})(-\frac 1 2 x_{ij}^2)^{-\delta_{ij}},
\label{MellinRep:n}\ea
with $a=(i_n,...,i_1)$ and $d\delta = d^{n(n-3)/2}\beta$. A reduced 
Mellin amplitude $\hat{M}^c(\beta) $ can be defined by 
eq. (\ref{reduce}).

The $n$-point functions and their Mellin amplitudes
can be constructed from two- and three-point functions, and these in turn are 
furnished by a collection of factorizing local four-point functions. The
meromorphy and factorization properties of the n-point Mellin amplitudes
follow from this. The coordinate space integrations
which are involved in constructing $n$-point functions from  lower ones
can be converted into operations on the Mellin amplitudes by using Symanzik's integration formula with spin.

Abbreviating $\delta=\{ \delta_{ij}\}$, retaining $ a= (i_n,...,i_1)$, 
and defining 
$(\pi\delta)_{ij} = \delta_{\pi^{-1}i \pi^{-1}j}$, locality reads
\be M^c_{\pi a}(\pi \delta) = M_a^c(\delta) \label{Mn:locality} \ee
\subsection{Spectrum condition}\label{sec:spectrum} 
Given suitable growth conditions on $M^c_a$,
the right hand sides of the Mellin representations 
(\ref{MellinRep},\ref{MellinRep:n}) 
 also defines the analytic continuation of the 
Euklidean Green function throughout the axiomatic analyticity domain, known as the permuted extended tube, and also 
its boundary values, the Wightman functions, when the appropriate 
$i\epsilon$-prescriptions are understood which are appropriate for 
a particular ordering of the fields. 


The $i\epsilon$-prescriptions assure validity of the spectrum condition, and follow from it. 

Consider the Wightman function which is defined by a particular 
$i\epsilon$-prescription, and its Fourier transform
\ba 
&&\langle \Omega, \phi^{i_n}(x_n)...\phi^{i_1}(x_1)\Omega\rangle 
= \nn \\ && (2\pi)^{-nD} \int ... \int 
\left(\prod_{i=1}^n dp_i e^{-ip_ix_i}\right)W_a(p_n,...,p_1) \nn \\
&&= \
(2\pi i)^{-n(n-3)/2}\int \int d\delta\ 
M^c_a(\{\delta_{ij}\})\nn \\
&&\qquad \qquad \prod_{i>j} \Gamma(\delta_{ij})(-\frac 1 2 x_{ij}^2+i\epsilon x^0_{ij})^{-\delta_{ij}},
\label{wightman:iepsilon}
\ea
with the abbreviation $a=(i_n,...,i_1)$. 

The spectrum condition says that all physical states must have $D$-momentum 
$p$ in the closed forward light-cone $V_+$.
 This is equivalent \cite{PCT} to the 
requirement that $W(p_n,...,p_1)$ vanishes unless
\be
q_k \equiv -\sum_{i=1}^k p_i \in V_+ \label{def:q}
\ee
for all $k=1...n-1$.
Now insert the integral representation of the factors
\ba
\Gamma(\delta)(-\half x^2+i\epsilon x^0)^{-\delta} 
&=& \rho_D(\delta) \int_{V_+}dp e^{-ipx} 
\left( \frac {p^2}{2}\right)^{-h+\delta}\ ,
\label{propagator:momentum} \ea
with
\be
\rho_D(\delta) = 2^{-2\delta +1} \pi^{-h+1} \Gamma(\delta)[\Gamma(2h-\delta)
\Gamma(\delta-h+1)]^{-1} \ ,\nn 
\ee
$h=\half D$. Define variables of integration $p_{ij}$ for $1\leq j < i \leq n$, one for each unordered pair $(ij)$, and set $p_{ij}=-p_{ji}$ for $i<j$, and $p_{ii}=0$.
Inserting the integral representation (\ref{def:q}) for each $x$-dependent
 factor in expression (\ref{wightman:iepsilon}), one can read off the Fourier transform
\ba && (2\pi)^{nD}W(p_n,...,p_1) = \label{Mellin:W:FT} \\
&& \int_{V_+} ... \int_{V_+}
 \prod_{i>j} dp_{ij} \prod_k \delta(p_k-\sum_{m=1}^n p_{km}) 
\tilde{M}^c(\{ p_{ij}\})  \nn
\ea
with
\ba
\tilde{M}(\{ p_{ij}\})&=&
(2\pi i)^{-n(n-3)/2}\int \int d\delta\ 
M^c(\{\delta_{ij}\}) \nn \\ && \quad \prod_{i>j}  \rho_D(\delta_{ij}) 
\left( \frac {p_{ij}^2}{2}\right)^{-h+\delta_{ij}} \label{M:p:wightman}
\ea
From the $\delta$-functions it follows that 
\be -\sum_{i=1}^k p_i = \sum_{j=k+1}^n\sum_{i=1}^k p_{ji} \ \in \ V_+ \nn \ee
throughout the domain of integration. Therefore the spectrum condition is 
fulfilled. 

Conversely, it follows \cite{PCT} from the spectrum condition
that the Wightman function, considered as a function of differences of coordinates, is holomorphic in the half planes $Im\ x_{ij}^0 < 0$, for $i>j$.
The $i\epsilon $ prescription shifts infinitesimally into this half plane
since $-\half(x^0-i\epsilon)^2 = -\half (x^0)^2 + i\epsilon x^0$. 

Concerning the growth conditions, we do not need to worry about them once we have Osterwalder Schrader positivity as a consequence of OPE, see section \ref{sec:OS}. Since the other Euklidean axioms are fulfilled by construction, the 
Osterwalder Schrader reconstruction theorem can then be applied
 \cite{OS:I,OS:II}.
It will guarantee validity of the spectrum condition and analyticity in coordinate space in the axiomatic domain. 

It is nevertheless instructive to have a look. Suppose that some $x_{ij}$ in expression (\ref{wightman:iepsilon}) is positive time-like. Then 
$$
(-\half x_{ij}^2+i\epsilon x_{ij}^0)^{-\delta_{ij}}
= (\half |x_{ij}^2|)^{-\delta_{ij}} e^{(\pi - \epsilon)\Im m \ \delta_{ij}}\cdot phase
$$
The second factor increases exponentially when the imaginary part of 
$\delta_{ij}$ becomes large. 
The $\Gamma$-functions in expression (\ref{wightman:iepsilon}) decay 
exponentially in the imaginary directions. 
One finds in a case by case study for
 $n=4$ that the $\Gamma$-functions  
 always keep the integrand bounded at large imaginary parts of $\delta_{ij}$, provided $M^c$ does not grow exponentially. 
\subsection{Time ordered Green functions in Minkowski space}\label{sec:timeordered}
The time ordered Green functions are vacuum expectation values of time ordered products of fields. Time ordered means that the fields with later times as arguments stand further to the left. According to this definition, they can be expressed in terms of the Wightman functions. For Bose fields
\be 
\langle \Omega, T\left\{\phi^{i_n}(x_n)...\phi^{i_1}(x_1)\right\}\Omega\rangle
= 
\langle \Omega, \phi^{i_{\pi n}}(x_{\pi n})...\phi^{i_{\pi 1}}(x_{\pi 1})\Omega\rangle
\ee
where $\pi $ is that permutation of $1,...,n$ which assures that
$$ x^0_{\pi n}\geq ... \geq x^0_{\pi 2} \geq x^0_{\pi 1} \ . $$
Inserting the Mellin representation of the Wightman function and using the 
locality property (\ref{Mn:locality}) of the Mellin amplitude, it follows that 
the time ordered Green function in Minkowski space possesses the Mellin representation 
\ba
&& \langle \Omega, T\left\{\phi^{i_n}(x_n)...\phi^{i_1}(x_1)\right\}\Omega\rangle \label{mellin:timeordered} \\
&&  =
(2\pi i)^{-n(n-3)/2}\int \int d\delta\ 
M^c_{i_n...i_1}(\{\delta_{ij}\})\nn \\ && \qquad \qquad \prod_{i>j} \Gamma(\delta_{ij})(-\frac 1 2 x_{ij}^2+i\epsilon )^{-\delta_{ij}} . \nn
\ea
It differs from the Wightman functions only through the $i\epsilon$-prescriptions.

 $\Gamma(\delta)(-\half x^2 + i\epsilon)^{-\delta}$ is a generalized Feynman propagator in the sense of Speer \cite{speer}. Therefore the $x$-dependent factor is the inverse Fourier-transform of a generalized Feynman integral. In momentum space, the result looks like expression (\ref{Mellin:W:FT}), (\ref{M:p:wightman}) except that the 
factor $\rho_D(\delta)$ is different, $p_{ij}$-integrations are over all of 
momentum space, and $p^2 $ is to be read as $p^2+i\epsilon$, as is appropriate for Feynman propagators. 

\section{Relations to theories in lower dimensions}\label{sec:lowerDimension}
\subsection{Dimensional reduction: AdS-orbits \& their boundary}\label{sec:dimensionalReduction}
The conformal group in $D$ dimensions, $G=\mbox{universal covering of SO(2,D)}$
, acts transitively on the tube $\M_D\simeq S^{D-1}\times \mathbf{R}$ (the $\infty$-sheeted cover of conformally compactified Minkowski space).
Under the subgroup which covers $SO(2,D-1)$, $\mathcal{M}_D$ decomposes into orbits as follows 
\be \mathcal{M}_D=\M_{D-1}\cup AdS_{D}\cup AdS_D \ee  
There are two copies of Anti-de Sitter space in $D$ dimensions - or rather, its simply connected covering. The first factor is their common boundary. Points on the space $\M_D$ can be coordinatized by $(e,\tau)$, $\-\infty < \tau < \infty$, and $e$ a unit $D$-vector with components $(e_1,...,e_{D-1},e_{D+1})$  The orbit $\M_{D-1}$ consists of points with $e_{D-1}=0$, the equator of the sphere $S^{D-1}$. 

This is most easily seen by using Dirac's manifestly covariant formalism \cite{Dirac36}. In place of points $x$ of (conformally compactified) Minkowski space it uses rays of light-like $D+2$-vectors $\xi = (\xi_0,\xi_1...\xi_{D-1}, \xi_{D+1}, \xi_{D+2})$. Light-like means 
\be (\xi_0)^2 - (\xi_1)^2 - ... - (\xi_{D-1})^2 -(\xi_{D+1})^2 + (\xi_{D+2})^2 = 0 \label{cone}\ee
 Vectors $\xi $ and $\xi^\prime = \lambda \xi$, $\lambda > 0$ are on the same ray.

Elements of the group $SO(2,D)$ act as pseudo-rotations on the $D+2$-vectors 
$\xi$. The subgroup $SO(2,D-1)$ leaves $\xi_{D-1}$ invariant. There are therefore two invariant domains: rays with $\xi_{D-1}=0$ and rays with $\xi_{D-1}\neq 0$. The covering of the first domain is $\M_{D-1}$. 
 For the second domain we may use homogeneity to scale $\xi $ to $r\xi$, $r>0$ such that either $\xi_{D-1}=1$ or $\xi_{D-1}=-1.$ In either case Eq.(\ref{cone}) then tells  us that we are dealing with Anti-de Sitter space.

The relation between the rays and the coordinates $(\tau, e)$ is as follows \cite{mack:luescher:global}:
\be \xi_0= r \cos \tau, \ \xi_{D+2}=r \sin \tau,\  \xi_k = r e_k \ee
for $(k=1,...,D-1,D+1)$. 
Therefore, $\xi_{D-1}=0$ means $e_{D-1}=0$ as claimed, and $\xi_{D-1} > 0 $ means $e_{D-1}>0$ while $\xi_{D-1}<0$ means $e_{D-1}<0$. 

For future use I also give the relation between Minkowski space coordinates $x_\mu$ and $\xi$. Set 

$\kappa = \xi_{D+1}+\xi_{D+2}$. Then 
$ x_\mu = \xi_\mu / \kappa , \quad (\mu = 0...D-1).$

Abbreviating $\{ i_n,...,i_1 \}=a$ as before,
the Mellin representation of a conformally invariant scalar $n$-point Wightman  function can be written in manifestly covariant form as follows. 
\ba && W_a(\xi^n,...,\xi^1) \equiv 
\kappa_1^{-d_1}... \kappa_n^{-d_n}  W_a(x_n,...,x_1)\\ && = 
(2\pi i)^{-n(n-3)/2}  \int d\delta M_a(\{\delta_{ij}\})
\prod_{i>j} ( \xi_i\xi_j)^{-\delta_{ij}} \  . \nn 
\ea 
We see that the restriction of the amplitude  $W_a$ to $\M_{D-1}\times ... \times \M_{D-1}$,
(i. e to points with $\xi^i_{D-1}=0$) exists and is given by identically the same Mellin representation with the same Mellin amplitude $M_a$, with the understanding that  $\xi^i_{D-1}=0$. Conversely, this restriction suffices to determine the Mellin amplitude $M_a$, and the Mellin amplitude yields the 
amplitude $W_a$ on all of  $\M_D\times ... \times \M_D$. This is why I call the Mellin representation a {\em $D$-independent representation}.

Wightman positivity as discussed in section \ref{sec:WightmanPositivity} can be formulated for scalar amplitudes $W_a(\xi_n,...\xi_1)$ by using sequences of test functions 
\be f^\emptyset, f^{i_1}(\xi^1)... f^{i_1...i_N}(\xi^1,...,\xi^N) \ee
which are homogeneous functions of $\xi_k$ of degree $-D+d_k$. The existence of the restriction implies that test functions are permissible which are concentrated on the hyper-surface $\xi^k_{D-1}=0$, i.e. 
\be f^{i_1...i_k}(\xi^1,...,\xi^k) \propto \delta(\xi^1_{D-1})...\delta(\xi^k_{D-1}) \ee
Thus, Wightman positivity of the restricted Wightman functions in $D-1$ dimensions follows from Wightman positivity of the original Wightman functions. Therefore {\em a unitary CFT in $D$ dimensions restricts to a unitary CFT in $D-1$ dimensions}.

The operator content of $D-1$-dimensional CFT's which are obtainable  by dimensional reduction has special properties. Together 
with every field $\phi^i$ of dimension $d_i$ there is a tower of fields $\phi^i_{,n}$, $n=1,2,3,...$ of dimensions $d_i+n$. They arise because derivatives of fields in $D$ dimensions need not be derivatives in $D-1$ dimensions. In the manifestly covariant formalism, 
\be \phi^i_{,n} = D_{D-1}...D_{D-1}\phi^i|_{\xi_{D-1}=0} \ee
($n$ factors $D_{D-1}$). $D_A$ is the $SO(2,D)$ covariant interior differential operator on the $D+2$-dimensional cone $\xi^2=0$ \cite{todorov:bargmann}. The missing generators $J_{AB}$, $B=D-1$ of $SO(2,D)$ transform $\phi^i\leftrightarrow D_{D-1}\phi^i$, leaving $2$- and $3$-point functions invariant.
\subsection{Dual amplitudes from 2-dimensional conformal field theories}\label{sec:dualFrom2d}
In two dimensions, some things are different. $S^1\times {\bf R}$ is not simply connected, and is therefore not the universal covering of compactified 
Minkowski space. Spin needs not be half integral, and more general types of 
statistics, known as braid statistics, are possible \cite{frs:CMP}.
There exist chiral theories, in which fields depend only on 
one of the light-cone variables $x^{\pm}=x^0\pm x^1$; they are effectively 
1-dimensional. All this gives rise to possibilities which do not 
exist in higher dimensions, such as anyons \cite{wilczek:anyons}.

Let us examine sufficient conditions, such that a universal Mellin representation as discussed in section \ref{sec:mellin} exists and  furnishes a dual amplitude. It will turn out that invariance under space reflections and conventional 
locality, i.e. Bose/Fermi-statistics are sufficient.

In two dimensionsional Minkowski space $x^2=x^+x^-$. As a result, the anharmonic ratios factorize,
\ba \omega_1 &=&\omega_1^+\omega_1^-, \quad 
\omega_2=\omega_2^+\omega_2^-, \quad
\omega_3=\omega_3^+\omega_3^-, \label{omega:factorize}\\
 \omega_1^\pm&=&\frac {x_{12}^\pm  x^\pm_{34}}{x_{13}^\pm x_{24}^\pm}, \quad
\omega_2^\pm=\frac {x_{14}^\pm x_{23}^\pm}{x_{12}^\pm x_{34}^\pm}, \quad
\omega_3^\pm=\frac {x_{13}^\pm x_{24}^\pm}{x_{14}^\pm x_{23}^\pm}.
 \label{def:omegas:pm}
\ea
They share the property that $\omega_1\omega_2\omega_3=1$. 
But it follows from the definition (\ref{def:omegas:pm}) that the three quantities $\omega^+_i$ are actually all dependent. Take $\omega^+_1$ as the independent one and call it $x$, temporarily. Then
\be
\omega^+_1= x , \quad (\omega^+_3)^{-1}= 1-x, \quad \omega^+_2 = (1-x)/x , \label{omega+:expr}
\ee
and similarly for $\omega_i^-$. 

Space reflection takes $x^1\mapsto -x^1$, therefore $x^+ \leftrightarrow x^-$ and $\omega_i^+\leftrightarrow \omega_i^-$. 
Let us now use the above equations to  determine the two independent quantities $\omega_1^+$ and $\omega_1^-$ in terms of two independent anharmonic rations $\omega_1$ and $\omega_2$ of the 
two-dimensional theory. There are two solutions which go into each other by interchanging $\omega_1^+ \leftrightarrow \omega_1^-$. This is expected since 
space reflections leave all $\omega_i$ invariant, but interchange $\omega_1^+$ and $\omega_1^-$. In other words, $\omega^+$ and $\omega^-$ depend on a 
variable $\sigma=\pm 1$ in addition to $\omega_1$, $\omega_2$. Explicitly,
\ba \omega^+ &=& -b+ \sigma\sqrt{b^2 - \omega_1},\nn  \\
\omega^- &=& -b- \sigma\sqrt{b^2 - \omega_1}, \nn \\
b&=& -\half\left( \omega_1(\omega_2-1)-1\right) \ . \label{omega:pm:solution}
\ea
Retain the notation $\omega$ for triples $(\omega_1,\omega_2,\omega_3)$ subject to $\omega_1\omega_2\omega_3=1$.  
Conformal 4-point Green functions $G_{i_4...i_1}$ are products of kinematical factors times 
functions of $\omega_1^+$ and $\omega_1^-$, that is functions 
$F_{i_4...i_1}(\omega,\sigma)$.

Under space reflections, $\sigma\mapsto -\sigma$, while $\omega$ is invariant. Suppose now that the 2-dimensional theory is space reflection invariant and 
obeys conventional locality. Then $F_{i_4...i_1}(\omega , \sigma)$ is independent of $\sigma$, and obeys the crossing relation  (\ref{localityInOmega}).

Now we may Mellin transform $F_{i_4...i_1}(\omega)$ as in eq.(\ref{mellin:F}) and this yields a dual amplitude $\hat{M}^c(\beta)$. 

In terms of the quantities $\omega_1^\pm$, the factor
\ba \omega_1^{-s_1}\omega_2^{-s_2}\omega_3^{-s_3}
=&&  (\omega_1^+)^{-\beta_{12}}(1-\omega_1^+)^{-\beta_{23}} \nn \\
&& (\omega_1^-)^{-\beta_{12}}(1-\omega_1^-)^{-\beta_{23}}  \ . \ea
Space reflection invariance is not a necessary condition. Consider chiral theories whose fields depend only on one variable; let us say $x^+$. 
 The 4-point amplitudes will be equal to  kinematical 
factors times functions $F_{i_4...i_1}(\omega^+)$ of anharmonic ratios. 
There is now only 
one independent anharmonic ratio, and Mellin transformation is of no help. 
But suppose that the construction of $F$ yields an integral representation
\ba F_a(\omega^+)&=&(2\pi i)^{-2}\int \int d\beta_{12} d\beta_{23}
\hat{M}_a(\{\beta_{ij}\})\nn \\
&&\qquad  (\omega_1^+)^{-\beta_{12}}(1-\omega_1^+)^{-\beta_{23}}.
\label{chiral:rep}
\ea
Then this can serve as a substitute for a Mellin transformation.
Considering $\hat{M}_a$ as a function of variables 
$\beta_{ij}=\beta_{ji}$, $i\neq j$ 
subject to the constraints (\ref{beta:sol1},\ref{beta:sol2}) will serve to 
define the action 
$\beta\mapsto \pi \beta$ of permutations $\pi$ on $\beta=\{\beta_{ij}\}$.
Suppose that 
$F_a(\omega )$, $a=i_4,...,i_1$  is invariant under the fractional linear 
transformations of 
$\omega_1^+$ which amount to permutations $\pi$ of $\omega_i^+$ as defined by 
eq.(\ref{omega+:expr}), invariance meaning that 
$F_{\pi a}(\pi \omega^+)=F_a(\omega^+)$. Suppose 
that this is implemented by $\hat{M}_{\pi a}(\pi \beta)=\hat{M}_a(\beta)$.
 And suppose further that 
$\hat{M}_a(\{\beta_{ij}\})$ has poles in individual variables, with residues 
that are polynomials in the other(s). Then $\hat{M}_a$ furnishes 
a dual amplitude. 

When $F_a$ come from a dual resonance model, then the supposition regarding 
poles and residues is true by construction.

\section{Euklidean partial wave expansion of scalar 4-point functions}
\label{EuklideanPWE}
\subsection{Fundamental fields}
Fields with dimension $d<\frac D 2$ will be called {\em fundamental fields}. 
By the restrictions on $\chi$, they must be either scalar or spinor fields, 
in $D=4$ dimensions. A theory may have fundamental fields or not.
This does not affect duality. But fundamental fields play a 
special role in the Euklidean partial wave expansions, requiring exorcision
of a ghost - the shadow pole of Ferrara Gatto, Grillo and Parisi
\cite{ferraraEtAl:shadow}. Absence of this ghost  is equivalent to validity of renormalized 
Schwinger Dyson equations for all $n$-point Green functions 
in Lagrangean field theory, see the end of subsection \ref{EuklPWEsub}. 
\subsection{Euklidean partial wave expansions} \label{EuklPWEsub}
Typically, the scalar Euklidean Greens functions are correlation functions of a statistical mechanical system, e.g. 
\ba
&& G_{i_4i_3i_2i_1}(\x_4,...,\x_1)=\nn \\
&& \quad <\varphi^{Ei_4} (\x_4),\varphi^{Ei_3} (\x_3),\varphi^{Ei_2} (\x_2),\varphi^{Ei_1} (\x_1)> . \nn \ea
where the (commuting) Euklidean fields $\varphi^E$ are variables of the statistical 
mechanical system.
Suppose that we replace the two fields on the right, $\varphi^{E2}(\x_2)$ and 
$\varphi^{E1}(\x_1)$ by their conformally transformed brothers, transformed by one and the same element $g\in G^E$. The result will be a function
 $\hat{G}_{i_4i_3i_2i_1}$ of $g$. It follows from the transitivity of the action of 
the Euklidean conformal group on 
triples of distinct points, that $G_{0000}$ is determined for all
 $\x_1,...,\x_4$ if we know this function $\hat{G}$ of $g$  for some standard choice of $\hat{\x}_1,...,\hat{\x}_4$, with $\hat{\x}_1\neq \hat{\x}_2, \hat{\x}_3\neq \hat{\x}_4$. One may now perform a partial wave expansion (i.e. expansion in representation functions of $G^E$ of $\hat{G}(g) $ on $G^E$) for this standard choice of $\hat{\x}_4,...,\hat{\x}_1$. For scalar amplitudes, such partial wave expansions have been extensively studied in the 70's. 
The program was initiated by the author \cite{mack:groupth:1,mack:groupth:2,mack:OS} and completed by Todorov and his collaborators \cite{dobrevEtAl:OPE}.
Normalization factors are 
computed in \cite{mack:dobrevClebsch}; a review in book form is 
\cite{mack:dobrevBook}.
 The result, rewritten for $G_{i_4i_3i_2i_1}$ is as follows. 

It is convenient to introduce the notation
\be h=\frac D 2 \label{h} \ee
The representations which enter are labeled 
\footnote{The representations which we here label by $\chi=[l,h+c]$ are 
labeled by $\chi=[l,c]$ in the literature on Euklidean partial wave expansions} 
by $\chi=[l,\delta=h+c]$ 
just like 
the representations of $G$, where $l$ is now interpreted as a representation 
of the Euklidean Lorentz group $M^E=Spin(D)$ in a vector space $V^l$ (this makes no difference by Weyl's unitary trick). But the dimension $\delta$ is now either of the form 
$\delta=h+c$, $c$ imaginary 
({\em principal series of representations})
 or $c$ real in some interval ({\em complementary series}). There is also a
{\em discrete series} if $D$ is odd, but it does not contribute to scalar amplitudes. 
The representations $\chi = [l, h+c ] $ and
 $\tilde{\chi}=[\bar{l},h-c ] $ of the principal or complementary series are equivalent, where $\bar{l} $ is the dual representation to $l$; for symmetric tensor representations $\bar{l} \simeq l$. 
 I write $l=0$ for the trivial representation. 

The unitary representations of $G^E$ are special cases of elementary representations. They are constructed as induced 
representations and are labeled by $\chi=[l,h+c]$ with complex $c$. 

Suppose the scalar Euklidean fields $\varphi^{Ei_1},...$ have real dimensions $d_1,...,d_4$ so that they transform according to representations $\chi_1=[0,d_1],...\chi_4=[0,d_4]$. Given $\chi=[l,\delta]$, a conformal invariant three point function 
\mbox{$\mathfrak{V}(\x_3, \chi, \x_2,\chi_2, \x_1, \chi_1)\in V^l$} 
exists only if $l$ is a traceless symmetric tensor representation of $M^E$; in this case it is unique up to normalization. 

There exists a unique canonically normalized conformal invariant 2-point function $ \Delta^\chi (\x,\y)$. The three point functions can be normalized in such a way that 
\ba
&& \int d^D\y   \Delta^\chi (\x,\y)
  \mathfrak{V}(\y , \tilde{\chi}, \x_2,\chi_2, \x_1, \chi_1)\nn \\ && =
  \mathfrak{V}(\x, \chi, \x_2,\chi_2, \x_1, \chi_1) \label{Norm1Vertex}
\ea
Validity of this amputation identity for arbitrary $\chi$ requires that 
\be
\Delta^{\tilde{\chi}} = \left( \Delta^\chi\right)^{-1}  \ , 
\label{Delta:inverse}
\ee
the inverse being meant in the convolution sense. 

Referring to 4-point functions only for now, a {\em channel} is a division of
the four arguments of a 4-point function labeled by $4,3,2,1$ into two groups
 of
two, viz. $(43)(21)$ or $(42)(31)$ or $(41)(32)$. I will sometimes write $(21)$ in place of $(43)(21)$ for short. 

There is no way of doing partial wave expansions of 4-point Wightman functions
in positive energy representations in different channels without commuting
the fields. But for Euklidean Green functions this can be done, because 
they are symmetric in their arguments - the classical statistical mechanical 
observables $\varphi^{Ei_k}(\x_k
)$ commute - in the bosonic case. 
The above mentioned expansion is the partial wave expansion in the channel
(43)(21).

The Euklidean partial wave expansion of the {\em connected} 
scalar 4-point function of  four 
hermitean scalar field $\phi^{i_4},...,\phi^{i_1}$ with spin and dimension
$\chi_k=[0,d_k], k=4,...,1$, in a theory in which no fundamental field 
appears in the
 OPE of both $\phi^{i_2}\phi^{i_1}$ and $\phi^{i_3}\phi^{i_4}$  reads as 
follows
\ba
&& G^{c}_{i_4i_3i_2i_1}(\x_4,\x_3,\x_2,\x_1) =
 \int d\chi g_{i_4i_3i_2i_1}^{(43)(21)}(\chi)
\label{scalarEpwe}\\
&&  \int d^D\x  
\langle \mathfrak{V}(\x,\tilde{\chi} ,\x_4,\chi_4,\x_3, \chi_3),
 \mathfrak{V}(\x,\chi ,\x_2,\chi_2,\x_1, \chi_1)\rangle
\nn  
\ea
The superscript $(43)(21)$ on $g(\chi)$ indicates the channel in which 
we expand. 
$\langle , \rangle$ is contraction of tensor indices.

The $\chi$-integration is over the part of the principal series 
\mbox{$\chi=[l,\frac D 2 + c]$}, $c$ imaginary, where $l$ is a completely 
symmetric 
traceless tensor representation, with Plancherel measure $d\chi$. Thus
\cite{dobrevEtAl:OPE} 
\ba \int d\chi ...
 &=& \frac 1 {2\pi i} \sum_l \int_{-i\infty}^{\infty} dc\ \rho_l(c)... \nn \\
\rho_l(c) &=& \frac{\Gamma(l+h)}{2 (2\pi)^h l!} 
\frac {\Gamma(h-1+c)\Gamma(h-1-c)}{\Gamma(c)\Gamma(-c)}\nn \\ && 
[(h+l-1)^2-c^2]
\label{PlancherelMeas}
\ea
In even dimensions $D$, the Plancherel weight $\rho_l(c) $ is a polynomial in
 $c$.
Expansion (\ref{scalarEpwe}) can be written in a more symmetrical form by
 inserting eq.(\ref{Norm1Vertex}), but is more 
convenient for our purpose as it stands.

The contribution from a single $\chi$ is called the partial wave. Since all 
the other factors are kinematically determined, I will also refer to 
$g^{(43)(21)}$ as the partial wave. It depends on normalization conventions
including (\ref{Norm1Vertex}). 

\begin{figure}[h]
\setlength{\unitlength}{0.65mm}
\begin{center}
\begin{picture}(100,100)(-2,0)
\thinlines
\put(0,50){
\put(38,25){\line(1,0){4}}
\thicklines
\put(40,0){\vector(0,1){50}}
\put(32,25){\circle*{3}}
\put(30,15){$c_f$}
\put(48,25){\circle{3}}
\put(32,25){\circle{9}}
\put(36.4,24.5){\vector(0,-1){1}}
\put(48,25){\circle{9}}
\put(43.6,24.5){\vector(0,-1){1}}

\put(62,25){\circle*{3}}
\put(74,25){\circle*{3}}
\put(84,25){\circle*{3}}

\put(18,25){\circle{3}}
\put(6,25){\circle{3}}
\put(-4,25){\circle{3}}

\put(0,3){a)} 
}

\put(-100,0){
\thinlines
\put(138,25){\line(1,0){4}}
\put(140,23){\line(0,1){4}}
\thicklines
\put(132,25){\circle*{3}}
\put(130,15){$c_f$}
\put(139,15){$0$}

\put(132,25){\circle{9}}
\put(136.4,24.5){\vector(0,-1){1}}
\put(148,25){\circle{3}}

\put(162,25){\circle*{3}}
\put(162,25){\circle{9}}
\put(166,24.5){\vector(0,-1){1}}

\put(174,25){\circle*{3}}
\put(174,25){\circle{9}}
\put(178,24.5){\vector(0,-1){1}}

\put(184,25){\circle*{3}}
\put(184,25){\circle{9}}
\put(188,24.5){\vector(0,-1){1}}

\put(118,25){\circle{3}}
\put(106,25){\circle{3}}
\put(96,25){\circle{3}}

\put(100,3){b)} 
}
\end{picture}
\caption{Path of the $c$-integration in the complex $c$-plane in the derivation of OPE, for fixed $l$ in the presence of a fundamental field with Lorentz spin $l$ and dimension $d_f=h+c_f$, $c_f<0$: a) before and b) after the shift of the path of integration. To each field $\phi^k$ in the OPE of Lorentz spin $l$ and dimension $d_k=h+c_k$
 there corresponds a pole  $\bullet$ at $c=c_k$ in
 $g_{\dots}^{(43)(21)}([l,c])$, and a shadow pole $\circ$ at $c=-c_k$. The contributions in a) from the the circles  around the poles at $c=\pm c_f$ represent the Born term.
\label{fig:paths}}
\end{center}
\end{figure}
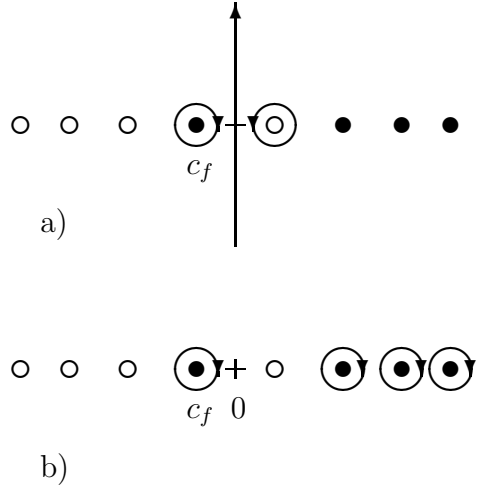

If the OPE of both $\phi^{i_2}\phi^{i_1}$ and $\phi^{i_3}\phi^{i_4}$ 
 contains a fundamental 
scalar field of dimension
 $d_f$ then a Born term must be added to the right hand side of 
eq.(\ref{scalarEpwe}). It has the same form as the integrand
of the $\chi$-integration, 
with $\chi_f=[0,d_f]$
substituted for $\chi$, and a product of coupling constants substituting for
$g_{i_4i_3i_2i_1}^{(43)(21}(\chi)$. It can be included by altering the path of the $c$-integration, see section \ref{sec:Lagrangean} 
below and figure \ref{fig:paths} a).

Let us now consider situations in which there exist non-vanishing disconnected 
parts other than a contribution $G_{i_4i_3}(\x_4,\x_3)G_{i_2i_1}(\x_2,\x_1)$ 
[which may be interpreted as the contribution of the unit operator in the 
OPE]. 
This will be so in the case of four equal hermitean scalar fields 
$\phi^{i_1}=\phi^{i_2}=\phi^{i_3}=\phi^{i_4}=\phi^0$ with dimension $d_0$. 
Set $\chi_0=[0,d_0]$.

There exists a choice of normalization factors such that the coefficient in
 the  Euklidean partial wave expansion of the sum
$ G_{00}(\x_4,\x_2)G_{00}(\x_3,\x_1)+G_{00}(\x_4,\x_1)G_{00}(\x_3,\x_2)$ of 
disconnected parts 
is identically equal to $1$. This will be the case if the normalization factors in eq.(\ref{Vscalar:withN}) for $\mathfrak{V}$ satisfy eq.(\ref{ProdNorm}).

Adopting these normalization conventions, the expansion of the full Green function will be
\ba
&&G^{4}_{0000}(\x_4,\x_3,\x_2,\x_1) \label{scalarEpweDisc}\\
&=& G_{00}(\x_4, \x_3)G_{00}(\x_2,\x_1) + \nn \\
&+&\int d\chi [1+g_{0000}^{(43)(21)}(\chi)] \int d^D\x  \nn \\
&& \quad \langle \mathfrak{V}(\x,\tilde{\chi} ,\x_4,\chi_0,\x_3, \chi_0),
 \mathfrak{V}(\x,\chi ,\x_2,\chi_0,\x_1, \chi_0)\rangle 
\nn \\
&+& \mbox{ possibly a Born term } \nn
\ea

The representation (\ref{scalarEpweDisc}) is also valid for a generalized free field theory, with
 $g^{(43)(21)}(\chi)=0$. 
\subsection{Explicit form of the 3-point function}
In our later computations, we will need the explicit form of the three point function $\mathfrak{V}(\x,\chi ,\x_2,\chi_2,\x_1, \chi_1)$ .

It is convenient for our purposes to use a particular realization of the 
representation space $V^l$ of traceless symmetric tensors of rank $l$. With 
every such tensor $t_{\mu_1,...,\mu_{l}}$ one associates a homogeneous 
polynomial in a complex $D$-vector $\z$ with $\z^2=0$,
$$ t(\z )= \sum t_{\mu_1,...,\mu_{l}}\z^{\mu_1}...\z^{\mu_l}, $$
and one regards $V^l$ as a space of such homogeneous polynomials. 
[Here and everywhere, I simply write $l$ for the rank of a symmetric tensor 
representation $l$].

In this realization, the vertex function $\mathfrak{V}$
 become functions of $\z$.

I use notations for the normalization factors which are in accord with the 
literature \cite{dobrevEtAl:OPE}. Let $d_i=h+c_i$ and set 
$c^+=\half (c_1+c_2)$, $c_-=\half(c_1-c_2)$ and 
$\chi = [l,\delta]=[l,h+c]$. Then
 \ba
\mathfrak{V}(\x_0, \chi, \z, \x_2, \chi_2, \x_1, \chi_1)
&=&N_l(c_+,c_-,c) \label{Vscalar:withN} \\
&&  \mathfrak{V}^u(\x_0, \chi, \z, \x_2, \chi_2, \x_1, \chi_1) 
\nn 
\ea
with
\ba
&& \mathfrak{V}^u(\x_0, \chi, \z, \x_2, \chi_2, \x_1, \chi_1)
=  (2\pi)^{-h} (\lambda\cdot \z)^l \label{Vscalar:u}\\ 
&&
 \left( \frac {2}{\x_{12}^2}\right)^{\frac 1 2 (h-c+l)+c_+} 
 \left( \frac { 4}{\x_{10}^2 \x_{20}^2}\right)^{\frac 1 2 (h+c-l)} 
\left(\frac {\x_{20}^2}{\x_{10}^2}\right)^{c_-}\ ,\nn \\
&& \lambda = 2\left( \frac {\x_{10}}{\x_{10}^2} - \frac {\x_{20}} {\x_{20}^2}
 \right).
\label{lambda}
\ea
The right hand side defines the analytic continuation throughout the axiomatic 
analyticity domain. In Euklidean space $-x^2=\x^2 > 0$. In Minkowski space,
$i\epsilon$-prescriptions have to be put , which depend on the ordering of the fields, similarly as in section \ref{sec:spectrum}.

There exists a canonical choice of the normalization factors $N_l(c_+,0,c)$ in
the  expression (\ref{Vscalar:withN}) for the 3-point function . They 
have been determined in the literature from the requirement that appropriate orthonormality properties of the expansion functions, and amputation identities such as eq.(\ref{Norm1Vertex}) hold
\cite{mack:dobrevClebsch}. The result gives the product
\ba
&& N_l(c_+,0,c)N_l(c_+,0,-c)=  \label{ProdNorm}\\
&& \frac {\Gamma(\half(h-c+l)+c_+)\Gamma(\half(h+c+l)+c_+)}
{\Gamma(\half(h-c+l)-c_+)\Gamma(\half(h+c+l)-c_+)}\nn
\ea
One sees that this is a meromorphic function of $c$. 
%
\subsection{Note on the use of normalization conventions}
Let us finally comment on normalization factors in the general case
of not necessarily equal dimensions $d_4,d_3,d_2,d_1$ of the 
external scalar fields. 

Disconnected parts other than the first term in the Euklidean partial wave
 expansion (\ref{scalarEpweDisc}) can only appear when either $d_1=d_3$ and 
$d_2=d_4$ or $d_1=d_4$ and $d_2=d_3$. When neither is the case, it is 
less convenient to use the canonical normalization factors, because they have 
branch cuts such that the product
 $N_l(c^{12}_+,c^{12}_-,c)N_l(c^{34}_+,c^{34}_-,-c)$ is in general 
not meromorphic. One
may choose $N_l\equiv 1 $ instead, at the cost of introducing (meromorphic)
factors in amputation identities like eq.(\ref{Norm1Vertex}). 
The orthogonality relations remain unchanged.
 In any case, the quantity 
\be N_l(c^{12}_+,c^{12}_-,c)N_l(c^{34}_+,c^{34}_-,-c)g_{i_4i_3i_2i_1}^{(43)(21)}([l,h+c]) \ee
is independent of normalization conventions,
and validity of OPE requires meromorphy of this product.

 In the case of equal dimensions $d_4,...,d_1$ the canonical product 
(\ref{ProdNorm}) of normalization factors is meromorphic, so that also 
$ g^{(43)(21)}([l,h+c])$ is meromorphic. 
%
\subsection{OPE from Euklidean partial wave expansions}\label{sec:OPE:PWE}
From eq.(\ref{scalarEpweDisc}), operator product expansions
(\ref{OPE:4ptfct}) for the 4-point
 function have been derived, 
 assuming meromorphy properties of partial waves $g^{(43)(21)}([l,c]) $.
 A complete derivation was given in ref. \cite{dobrevEtAl:OPE}
for even dimension $D$ and under some restrictions on the field dimensions
$d_1,...,d_4$. It was stated that the result can be generalized. I will
 briefly review the result and slightly generalize it
 to emphasize factorization.

 The first step exploits the equivalence of representations $\chi=[l,h+c]$ and 
$\tilde{\chi}=[l,h-c]$ of the Euklidean conformal group to split the three point function into Clebsch-Gordan kernels of the second kind in a way  modeled after the split of Legendre functions $P_l$ into Legendre functions of the second kind,  $Q_l$,
\ba
&& \mathfrak{V}(\x_0,\tilde{\chi}, \x_2, \chi_2, \x_1, \chi_1)
= \  \label{kernel2ndKind}  \\ && \frac {\pi}{\sin(l+c)}
[\mathfrak{Q}(\x,\tilde{\chi} ,\x_2,\chi_2,\x_1, \chi_1)-  \nn \\
&& \qquad \qquad - \int d^D\y \Delta^{\tilde{\chi}}(\x,\y )\mathfrak{Q}(\y,\chi,\x_2,\chi_1,\x_1, \chi_2)] \ .
\nn
\ea
such that the partial Fourier transform 
$\mathfrak{Q}(p,\tilde{\chi} ,\x_2,\chi_2,\x_1, \chi_1)$
is an entire holomorphic function of $p$.

As a consequence of this split, the integrand of the Euklidean 
partial wave expansion becomes a sum of two terms. Upon a change of variables 
$c\mapsto -c$, the second term becomes equal to the first, producing a factor 
of $2$.

The result can be rewritten in terms of the partial Euklidean Fourier 
transforms as 
\ba
&& G^{c}_{i_4i_3i_2i_1}(\x_4,\x_3,\x_2,\x_1) = \int d\chi 
\frac {2\pi}{\sin \pi(l+c)} 
g_{i_4i_3i_2i_1}^{(43)(21)}(\chi) \nn
 \\ 
&&  \int (d\p) \langle \mathfrak{Q}(-p,\tilde{\chi} ,\x_4,\chi_4,\x_3, \chi_3),
 \mathfrak{V}(p,\chi ,\x_2,\chi_2,\x_1, \chi_1)\rangle
\nn  
\ea
Integration is over Euklidean momenta, i.e. over imaginary $p^0=ip^D$, with
$(d\p)=(2\pi)^{-D}dp^1...dp^D$. 

\begin{figure}[h]
\setlength{\unitlength}{0.75mm}
\begin{center}

\begin{picture}(100,100)(-2,0)
\put(0,50){
\put(0,0){a)}
\thinlines
\put(0,25){\line(1,0){80}}
\thicklines
\put(40,0){\vector(0,1){50}}
\put(40,0){\vector(0,1){15}}
\put(40,25){\circle*{1}}
\put(60,40){$p^0$}
}

\put(-100,0){
\thinlines 
\put(100,0){b)}

\put(140,25){\circle*{1}}
\put(100,25){\line(1,0){80}}
\thicklines
\put(180,27){\vector(-1,0){40}}
\put(140,23){\vector(1,0){40}}
\put(165,27){\vector(-1,0){2}}
\put(149,23){\vector(1,0){2}}
\put(140,25){\oval(4,4)[l]}
\put(140,16){$0$}
\put(160,40){$p^0$}
}
\end{picture}
\caption{Deformation of the path of integration in the complex energy plane. 
a) Integration over Euklidean space involves integration over imaginary 
$p^0$. b) The deformed path  \label{deformation:EnergyPath}
}
\end{center}
\end{figure}
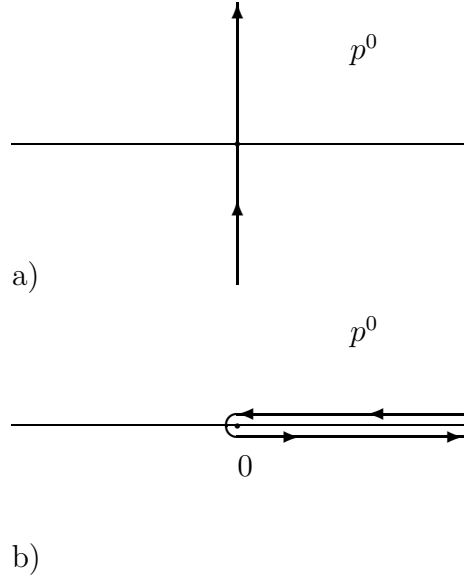

Let us now restrict attention to (Euklidean) arguments  $\x_i$ where the 
Euklidean time 
components satisfy inequalities
\be x^D_1 >0, \ x^D_2 > 0\ , \  x_3^D < 0, \ x^D_4 < 0 \ .\ee
Inserting the split (\ref{kernel2ndKind}) with $\chi $ substituted for $\tilde{\chi}$ also for the remaining factor $\mathfrak{V}$, a sum of two terms 
appears, 
but  the term involving the entire function 
$\mathfrak{Q}(-p,\tilde{\chi}...)\mathfrak{Q}(p,\chi,...)$  of $p$ will make 
zero contribution after the following deformation of the path of the
 $p$-integration (figure \ref{deformation:EnergyPath}. The path of the $p^0$-integration is deformed as shown in 
figure \ref{deformation:EnergyPath}. The closure of the path is allowed, because the integrand has exponential falloff as $\Re e\ p^0 \mapsto \infty$.

Let $\Delta^\chi(p) $ be the function of $p$ which is the analytic continuation of the Euklidean Fourier transform of $\Delta^\chi$, with a cut 
along the positive $p^0$-axis. Writing the discontinuity as 
$$ \frac {i}{\pi}\sin\pi(l-c)\Delta_+^\chi(p)\ , $$
then $\Delta^\chi_+(p)$ is a Wightman function which is positive when 
$\chi$ labels a positive energy representation of the conformal group. 

%
In the light of the remarks of the last subsection, I will extract normalization factors, 
\ba
 \mathfrak{Q}(\x,\tilde{\chi} ,\x_2,\chi_2,\x_1, \chi_1)
&=& 
N_l(c_+^{12},c_-^{12},-c) \nn \\
&& \mathfrak{Q}^u(\x,\tilde{\chi} ,\x_2,\chi_2,\x_1, \chi_1)\nn
\ea
An explicit formula for $\mathfrak{Q}^u$ is given the appendix, eq.(\ref{Q}).


The notation for the normalization factor has been chosen to be in agreement with the literature \cite{dobrevEtAl:OPE}; $c_\pm^{21}=\half (c_2\pm c_1)$ when 
$d_1=h+c_1$, $d_2=h+c_2$. 

Extracting normalization factors, the result after the deformation of the path of integration reads
\ba
&& G^{c}_{i_4i_3i_2i_1}(\x_4,\x_3,\x_2,\x_1) = \int d\chi 
\frac {-2\pi}{\sin \pi(l+c)} \nn \\ 
&&\qquad  N_l(c^{34}_+,c^{34}_-,-c)N_l(c^{12}_+,c_-^{12},c)
g_{i_4i_3i_2i_1}^{(43)(21)}(\chi) \nn
 \\ 
&& \qquad  \int_{V_+} (dp) \langle \mathfrak{Q}^u(-p,\tilde{\chi} ,\x_4,\chi_4,\x_3, \chi_3), \nn \\
&& \qquad \qquad \qquad
 \Delta^\chi_+(p) \mathfrak{Q}^u(p,\tilde{\chi} ,\x_2,\chi_2,\x_1, \chi_1)\rangle
\nn  
\ea
The integration now runs over the forward light cone $V_+$ in Minkowskian 
momentum space, $(dp)=(2\pi)^{-D}dp^0...dp^{D-1}$.

 Thereafter, the path of the $c$-integration is closed in the half plane 
$Re \ c>0$ as shown in figure \ref{fig:paths}, assuming growth conditions on partial wave amplitudes $g(\chi)$.
Remember that $d\chi$ includes a factor $\rho_l(c)$, the Plancherel weight. 
 Assume meromorphy of 
\be  \rho_l(c)\ N_l(c^{34}_+,c_-^{34},-c)N_l(c^{12}_+,c_-^{12},c)\
g_{i_4i_3i_2i_1}^{(43)(21)}(\chi)
\label{poles}
\ee
in $c$ with simple poles at real positions $c=c_k > 0$ for $l=l_k$. 
Every such pole will make a discrete contribution which is in accord with the presence of a field $\phi^k$ in the OPE with  spin and dimension 
$l_k,d_k=h+c_k$, {\em provided} the residue has factorization properties as detailed below. The absence of contributions from other poles, notably from 
the poles of $\pi/\sin\pi(l+c)$ in even dimensions $D$ is a subtle issue in the derivation of OPE. It was discussed in detail in \cite{dobrevEtAl:OPE} 
 and will concern us again later. In the present 
paper, I assume the validity of OPE  and conclude that the partial waves
$g(\chi)$ must have properties such that there are no other contributions, 
cf. section  \ref{footnote:poles}. 

Suppose expression (\ref{poles}) has simple poles at $\chi=\chi_k=[l_k,h+c_k]$
with residues which factorize as follows
 \ba && 
\frac {2\pi}{\sin\pi(c_k+l)}  res_{c=c_k} \nn \\
&&  \left[ \rho_l(c) N_l(c^{34}_+,c_-^{34},-c)
N_l(c^{12}_+,c_-^{12},c) \ g^{(43)(21)}([l,h+c])\right] \nn 
\\
&&= \bar{g}^{i_3i_4}_k g^{i_2i_1}_k \ .\label{coupling:const}
\ea

Then  we obtain as our final result
\ba &&
G_{i_4i_3i_2i_1}(\x_4 ... \x_1) = \sum_k \bar{g}^{i_3i_4}_k g^{i_2i_1}_k
\int_{V_+} (dp)
\label{E:OPE:pSpace} \\
&& \! \! \! \! \! \! 
\langle \mathfrak{Q}^u(-p,\tilde{\chi}_k,\x_2, \chi_2, \x_1, \chi_1)
\Delta^\chi_+(p)
\mathfrak{Q^u}(p , \tilde{\chi}_k, \x_4, \chi_4, \x_3, \chi_3)\rangle
\nn \ea
This admits analytic continuation to Minkowski space, and satisfies 
Osterwalder Schrader positivity as shown below.

Consider now also the situation with disconnected parts. 
Eq.(\ref{E:OPE:pSpace})
is valid as it stands for $G_{0000}$ except that the contribution 
$G_{00}(\x_4\x_3)G_{00}(\x_2\x_1)$ of the unit operator to the OPE must be added on the right hand side. But in  the definition (\ref{coupling:const}) 
of coupling constants, $[1+g_{0000}^{(43)(21)}(\chi )]$ is to be substituted
 for $g_{i_4i_3i_2i_1}^{(43)(21)}(\chi)$, and normalization conventions 
(\ref{ProdNorm}) are in force. 

In conclusion, there is a bijective correspondence between poles at
 $\chi=\chi_k = [l_k,h+c_k] ,\ c_k > 0$ of expression 
 (\ref{poles}) and contributions of fields $\phi^k $ with 
conformal transformation law specified by $\chi_k$ to the OPE for the 4-point function, and the residues must have appropriate factorization properties. 

If there is a fundamental field $\phi^f$ transforming according to
\mbox{ $\chi_f=[0,d_f]$} with $d_f < D/2$ in the OPE then the above statement
 must be qualified as 
follows: Among the poles of expression (\ref{poles}), there is a pole at 
$\chi=\chi_f$ whose contribution must be included, but the contribution of 
its 
``shadow'' at $\chi=\tilde{\chi_f}=[l,D-d_f]$ must be omitted, as shown in figure \ref{fig:paths} b).

This has a deep reason which was extensively discussed in the literature
\cite{mack:groupth:1,mack:groupth:2,dobrevEtAl:OPE}, and will be briefly reviewed in section 
\ref{sec:Lagrangean} below. 
\subsection{Osterwalder Schrader positivity}\label{sec:OS}
In section \ref{sec:positivity} it was pointed out that validity of OPE
implies positivity of Wightman functions. This statement has a Euklidean 
version. It involves Osterwalder Schrader (OS) positivity, also known as 
reflection positivity \cite{glimm:jaffe}. This result is useful
for us because it saves us from having to worry about validity of the spectrum
condition. Validity of the spectrum condition follows from 
OS- positivity and other Euklidean axioms by the Osterwalder Schrader 
reconstruction theorem \cite{OS:I,OS:II}. The other Euklidean axioms include
 crossing 
symmetry and are fulfilled by construction in the present approach. 

Similarly as for Wightman functions, it suffices to have OS-positivity of all 
4-point functions, for all fields, including those of higher spin. Here I only consider scalar 4-point functions. 
  Moreover, OS-positivity implies 
convergence of conformal OPE on the vacuum because these expansions amount to an orthogonal decomposition of a vector in a Hilbert space.

I pause for some explanatory remarks.
In the Euklidean approach the Hilbert space $\H$ is the completion of a space
 of sequences of test functions 
$f$
with support in half spaces, and the scalar product is furnished by 
OS-positivity. $\H$ carries a contractive representation of a sub-semigroup
$S$ of the Euklidean conformal group $G^E$. $S$ leaves half spaces invariant.  Its contractive representations possess an 
analytic continuation to  unitary positive energy representations of $G$.
This was proven in  \cite{mack:luescher:global},
 and generalized in \cite{luescher:phd}; in mathematics a similar method was later introduced by Gel'fand and Gindikin \cite{GelfandGindikin:CM-78}.
 $S$ possesses a polar decomposition. Such semigroups are nowadays known in mathematics as 
Ol'shanskii-semigroups \cite{Lawson:semigroups}. It would be interesting to reformulate axiomatic analyticity as holomorphy on a quadruple of semigroups $S$ 
which are sewn together along parts of their boundaries.

OS positivity of 4-point functions is the following statement. Write 
$\theta $ for the Euklidean time reflection $x^D \mapsto -x^D$. 
Consider finite sequences $f=\{ f_{i_2i_1}(\x_2,\x_1)\}$ of test functions
which vanish with all their derivatives unless the Euklidean time components 
satisfy  $x_1^D > 0$ and $x_2^D>0$. Then 
\ba && A_f \equiv \sum_{i_4,i_3,i_2,i_1} \int ... \int d\x_4...d\x_1 \label{A:f}\\
&& \qquad \bar{f}_{i_3i_4}(\theta \x_3 , \theta \x_4)
 G_{i_4...i_1}(\x_4,...,\x_1)
f_{i_2i_1}(\x_2,\x_1) \geq 0 \nn
\ea
Let us verify that this follows from OPE (\ref{E:OPE:pSpace}) of Euklidean 
4-point functions, given positivity of the Wightman 2-point functions 
$\Delta_+^{\chi_k}(p) $ of all the fields $\phi^k$ in the OPE. 

For Euklidean points $x=(x^0,...,x^{D-1})$, $x^0$ is imaginary, and Euklidean 
time reflection $\theta$ is equivalent to complex conjugation 
$x\mapsto \bar{x}$.  For Minkowski
momenta $p$, including in particular all $p\in V_+$, the 
analytic continuation  (\ref{Q})
of the 
$\mathfrak{Q}^u$
kernels has the following reality property.
\be 
\mathfrak{Q}^u(-p,\tilde{\chi},\bar{x}_4,\chi_4,\bar{x}_3,\chi_3)
= \overline{\mathfrak{Q}^u( p,\tilde{\chi},x_3,\chi_3,x_4,\chi_4)} \ . 
\label{Q:reality}\ee
Insert eq. (\ref{E:OPE:pSpace}) for $G_{i_4...i_1}$ in expression (\ref{A:f})
for $A_f$. Define
\ba
v_k(p)&=&\sum_{i_2,i_1}g_k^{i_2i_1}\int \int d\x_2\ d\x_1 
\nn \\ 
&& \qquad  f_{i_2i_1}(\x_2,\x_1)
\mathfrak{Q}^u(p,\tilde{\chi}, \x_2,\chi_2,\x_1,\chi_1)\nn 
\ea
Then 
\be A_f = \sum_k \int_{V_+} (dp)\ \langle \bar{v}_k(p),\Delta_+^{\chi_k}(p) v_k(p) 
\rangle \geq 0
 \ .\ee
Remember that $\langle , \rangle$ stands for the bilinear form which is given 
by contraction of tensor indices; there is no complex conjugation involved in 
it.

\section{Lessons from Lagrangean field theory}\label{sec:Lagrangean}
\subsection{Renormalized Schwinger Dyson equations}\label{sec:Lagrangean:SD} 
Lagrangean field theories are theories of fundamental fields. 
For simplicity, assume that there is one such field $\phi^0$.
In Lagrangean 
field theory there exists an  infinite set of renormalized 
Schwinger Dyson equations which the collection of all connected n-point
 Green functions of $\phi^0$ 
must satisfy; it involves no coupling constants
\cite{symanzik:hercegnovi,fradkin:1,fradkin:2}. 
The equations \cite{johnson:SDequations} for $\phi^4$-theory 
could be simplified by introducing a second fundamental field $:(\phi^0)^2:$. 

 Renormalized perturbation theory is the iterative solution of all these equations, together with renormalization conditions which introduce the renormalized coupling constants. 
When only the equations for the higher n-point functions are solved by iteration, the result is known as a skeleton graph expansions. 
This leaves the renormalized Schwinger Dyson equations for two- and three-point functions to be solved; they are 
also known as bootstrap equations in the context of conformal field theory 
\cite{migdal:1,migdal:2,parisi:peliti:bootstrap,mack:todorov:UV}. 

As an aside: It has been known for some time that gravity has a polynomial 
action in suitable variables \cite{witten:GR}. In the presence of a
 cosmological term it can be brought to a $F^2$-form, where $F$
is a de Sitter (or anti de Sitter) field strength \cite{mack:pruestel:GR}.
Although they are not in the literature, as far as I know, renormalized 
Schwinger Dyson equations could be written down, after introducing suitable 
composites as auxiliary fields, similarly to $:(\phi^0)^2:$ above, assuming 
that an appropriate gauge fixing can be found. But these equations do not 
admit an iterative solution. This is in agreement with the conviction that 
gravity is perturbatively non-renormalizable. The possibility of a conformal 
invariant short distance behaviour is being discussed \cite{reuter:UVfixpoint}.
 
Let us return to conformal field theory. 
Suppose that the fundamental field $\phi^0$ has spin $0$ and dimension $d_0$. 

The [totality of] renormalized Schwinger Dyson equations have
  been solved by Euklidean conformal partial wave 
expansions \cite{mack:groupth:1,mack:groupth:2,dobrevEtAl:OPE}, and this leads to relations between the square of a coupling constant which appears in a Born term, and the residue of a pole of
 $g^{(43)(21)}(\chi)=g^{(43)(21)}(\tilde{\chi})$ at $\chi=\chi_0=[0,d_0]$.

 As a consequence, a Born term can be included in the Euklidean partial wave expansion by altering the path of the $c$-integration as shown in figure \ref{fig:paths} a). 

When the path of the $c$-integration is shifted (after a split of the integrand) in the derivation of OPE from the Euklidean partial wave expansion 
(section \ref{sec:OPE:PWE}) there results a cancellation 
of the contribution of the shadow pole as shown in figure \ref{fig:paths} b).
This cancellation  was first proposed by Ferrara, Gatto, Grillo and Parisi
\cite{ferraraEtAl:shadow}, see also \cite{ferraraEtAl:OPE,ferraraEtAl:conformalAlgebraOPE}.

 The cancellation is illustrated at the example of  
$\phi^3$-theory in $D=6+\epsilon$ dimensions by the result eq.(\ref{sigma_s})
and remarks following it.

\subsection{Consequences of Bethe Salpeter equations}\label{footnote:poles}
Now consider composite fields. Lagrangean field theory teaches that 
all three point Green functions $V$ with one composite and two fundamental 
fields should satisfy 
the Bethe Salpeter equation. In short hand, it reads $BV=V$, where $B$ is the
 Bethe Salpeter kernel. It has the same form as the Schwinger Dyson equation 
for the 3-point function in case of a fundamental field. 

In conformal theories, $V$ is a sum of kinematically 
determined three-point functions $\mathfrak{V}^a$ times coupling constants.
Currents and stress tensor have been examined in \cite{mack:symanzik} and 
\cite{erdmenger:osborn:currentsEnergy}.

Consider fields $\phi_k$, with spin and dimension $\chi_k=[l_k,d_k]$.
Assuming normalization convention (\ref{ProdNorm}), validity of the 
Bethe Salpeter equation amounts to the requirement 
$b(\chi_k)=1$ on the Euklidean partial wave $b$ of the Bethe Salpeter kernel
 $B$.

 If $\phi^0$ is the fundamental field of a Lagrangean field theory,
this can be used to give an argument  that the physical fields $\phi_k$ which appear in the OPE of $\phi^0\phi^0\Omega $ are in bijective correspondence with pairs of poles of  $g(\chi)\equiv g_{0000}^{(43)(21)}(\chi)$ at $\chi=\chi_k$ and $\chi=\tilde{\chi}_k$ .

 This result  implies that the OPE contains no contributions other than from 
poles of 
$g(\chi)$. In particular, the poles from the normalization factor 
(\ref{ProdNorm}) in $N_l(...)N_l(...)[1+g(\chi)]$ must all be cancelled by 
zeroes of 
$[1+g(\chi)]$. 

The argument is based on the 
Bethe Salpeter equation for the 4-point function. It is an equation for the 4-point Green function $G$ which is 1-particle irreducible in the given channel, i.e. without the above mentioned Born term.
It reads, in shorthand, $G=B+BG$. 
Upon Euklidean partial wave expansion, this translates into 
$g(\chi)=b(\chi)+b(\chi)g(\chi)$, or 
$g(\chi)=b(\chi)[1-b(\chi)]^{-1}$. Therefore, $g(\chi)$ has a pole at 
$\chi=\chi_k$. Because $g(\chi)=g(\tilde{\chi})$, it also has a pole at 
$\chi=\tilde{\chi}_k.$ .


\section{Mellin amplitudes of individual  Euklidean partial waves}
Here I derive the Mellin representation of an individual Euklidean partial wave
$I^\chi$, $\chi=[l,h+c]$,  which enters the Euklidean partial wave expansion of the 4-point function of four scalar fields $\phi^{i_1},...,\phi^{i_4}$ of dimension $d_1,...,d_4$. Without loss of generality I assume that the scalar fields are real and that the normalization constants in the 3-point functions are chosen real for real $c$. 

To make contact with the formulae in earlier work, I introduce the notations,
for $(ij)=(12),(34)$
\ba \half (d_i + d_j) &=& h + c_+^{ij} = h+c_+^{ji}\nn \\
     \half (d_i - d_j) &=&   c_-^{ij} = -c_-^{ji}\nn \label{c:+-}
\ea
By definition
 \ba && I^{\chi}(d_4,\x_4,...,d_1,\x_1)=  \label{Ichi} \\
&& \int d^D\x
 \langle \mathfrak{V}(\x,\tilde{\chi} ,\x_4,\chi_4,\x_3, \chi_3),
 \mathfrak{V}(\x,\chi ,\x_2,\chi_2,\x_1, \chi_1)\rangle \label{Ipw} \nn \\
 &&= (2\pi)^{-D}\int d^D\x  N_l(c^{34}_+,c_-^{34},-c)N_l(c^{12}_+,c_-^{12},c)\langle f^\lambda_l, f^\mu_l\rangle
 \nn \\ 
&& \quad (\half \x_{12}^2)^{-\half ( h - c + l)-c^{12}_+} 
\  (\half \x_{34})^{-\half ( h + c + l)-c^{34}_+}\nn \\
&&
\quad (\half \x_{10}^2 )^{-\half(h+c-l)-c_-^{12}}\ (\half \x_{20}^2 )^{-\half (h+c-l)+c_-^{12}} \nn \\
&&  \quad
(\half \x_{30}^2)^{-\half(h-c-l)-c_-^{34}}\ (\half \x_{40}^2 )^{-\half (h-c-l)+c_-^{34}}  \nn   
\ea
with
\ba
\lambda &=& 2\left( \frac {\x_{10}}{\x_{10}^2} - \frac {\x_{20}} {\x_{20}^2} \right),  \\
\mu &=&2\left( \frac {\x_{40}}{\x_{40}^2} - \frac {\x_{30}} {\x_{30}^2} \right).
\ea
$f^\lambda_l $ is the vector in the representation space $V^l$ of the Euklidean Lorentz group $SO(D)$ associated with the homogeneous polynomial
 $f^\lambda_l(\mathfrak{z})=(\lambda \z )^l$. The bilinear form 
$$ 
\langle f^\lambda_l, f^\mu_l\rangle =
 (\lambda_{\nu_1}...\lambda_{\nu_l}- traces ) (\mu_{\nu_1}...\mu_{\nu_l}- traces )\ .$$
$f^\lambda_l $ is invariant under the $SO(D-1)$ subgroup of rotations which leave $\lambda$ invariant, and it is homogeneous in $\lambda $ of degree $l$.
The expression $\langle f^\lambda_l, f^\mu_l\rangle$ is therefore given by the zonal spherical function 
 of $SO(D)$ multiplied with $(|\lambda||\mu|)^l$ times the norm squared
 $|c_l|^2$ of $f^e_l$, $e$= unit vector. It depends on the angle $\theta $  
between
 $\lambda$ and $\mu$, i.e. $\cos  \theta = \lambda \mu /|\lambda||\mu|$ .
The zonal sperical function of a irreducible representation $l$ of 
$SO(D)$ is the matrix element of the corresponding representation operator 
$D^l(g)$ for $g\in SO(D)$
between normalized $SO(D-1)$-invariant states. It depends on $g$ only 
through one angle of rotation $\theta$. It is proportional to the Gegenbauer 
polynomial $C^{h-1}_l$. I choose to absorb the normalization factor 
$|c_l|^2$ into the definition of the zonal spherical function $Y^D_l$. Explicitly
\ba
\langle f^\lambda_l, f^\mu_l\rangle &=&  (|\lambda||\mu|)^lY^D_l(\cos \theta) \label{ff:scalarProduct} \\
&=& \frac {l!}{(h-1)_l} 2^{-l} (|\lambda||\mu|)^l
C_l^{h-1}(\cos \theta) \nn \\
|c_l|^2 &=&  2^{-l} \frac {(2h-2)_l}{(h-1)_l} 
\ea
Inserting this result we get
\ba 
I^\chi &=& (2\pi)^{-D} 2^{3h+c^{12}_++c^{34}_+ - l}N_l(c^{34}_+,c_-^{34},-c)N_l(c^{12}_+,c_-^{12},c) \nn \\
&&
(\x_{12}^2)^{-\half (h-c)-c^{12}_+}(\x_{34}^2)^{-\half(h+c)-c^{34}_+} \nn 
\\
&& \int d^D\x_0 (\x_{10}^2)^{-\half (h+c)-c_-^{12}}
(\x_{20}^2)^{-\half (h+c)+c_-^{12}} \nn \\
&& \quad (\x_{30}^2)^{-\half(h-c)-c_-^{34}}(\x_{40}^2)^{-\half (h-c)+c_-^{34}}
 Y^D_l(\cos \theta) \ .\nn
\ea
The integral can be evaluated with the help of the generalized Symanzik 4-star 
formula (\ref{Symanzik:4:generalized}) of section \ref{sec:Symanzik:generalized}
with
\ba
\delta_1 &=& \half (h+c) + c_-^{12}\ , \nn\\
\delta_2 &=& \half (h+c) - c_-^{12}\ , \nn \\
\delta_3 &=& \half (h-c) + c_-^{34}\ , \nn \\ 
\delta_4 &=& \half (h-c) - c_-^{34}\ , \label{delta:i}
\ea
The result involves variables which I denote by $\delta_{ij}^\prime$ subject to the constraints $\sum_j\delta_{ij}^\prime = \delta_i$, and an integration over imaginary parts $s_1$, $s_2$ of two independent ones among the so constrained variables. $I^\chi$ comes out to equal  
\ba 
 && \pi^{-h} 2^{c^{12}_+ + c^{34}_+ -l +h} N_l(c^{34}_+,c_-^{34},-c)
N_l(c^{12}_+,c_-^{12},+c) \nn \\
&& (2\pi i)^{-2} \int_{-i\infty}^{i\infty}\int_{-i\infty}^{i\infty} ds_1 ds_2 
\mathfrak{P}^D_l(\{ \delta^\prime_{ij}\}) \prod^\prime \Gamma(\delta^\prime_{ij})(\x_{ij}^2)^{-\delta^\prime_{ij}}
\nn \\
&& \Gamma(\delta^\prime_{12}- \half l)(\x_{12}^2)^{-\delta^\prime_{12}- \half(h-c) - c^{12}_+}  \nn \\ &&
\Gamma(\delta^\prime_{34}- \half l)(\x_{34}^2)^{-\delta^\prime_{34} - \half(h+c) - c^{34}_+}  \nn
\ea
The product $\prod^\prime $ runs over the four links $(13),(14),(23),(24)$. 

Introduce new variables 
\ba
\delta_{12}&=& \delta^\prime_{12}+ \half (h+c)+ c^{12}_+ \label{delta:shift} \\
\delta_{34} &=& \delta^\prime_{34}+ \half (h-c)+ c^{34}_+ \nn \\
\delta_{ij} &=& \delta^\prime_{ij} ,  \qquad (ij)\neq (12),(34) \ .\nn 
\ea
They satisfy the constraints $\sum_j\delta_{ij}=d_i$.

We may compare with the general form (\ref{MellinRep}) of the Mellin representation of a 4-point function to deduce the Mellin amplitude $\mathfrak{M}_\chi $ of a Euklidean partial wave $I^\chi$,
\ba &&
\mathfrak{M}_\chi(\{\delta_{ij}\}) =  N_l(c^{34}_+,c_-^{34},-c)N_l(c^{12}_+,c_-^{12},c)
\tilde{\mathfrak{M}}_\chi(\{\delta_{ij}\})\ , \nn \\
&& \tilde{\mathfrak{M}}_\chi(\{\delta_{ij}\}) =
(\pi / 2)^{-h}2^{-l}  
\mathfrak{P}^D_l(\{\delta_{ij}^\prime\})  \label{M:chi:1} \\
&& \frac {\Gamma(\delta_{12}-\half(h+c +l)-c^{12}_+)}{\Gamma(\delta_{12})}
 \frac {\Gamma(\delta_{34}-\half(h-c +l)-c^{34}_+)}{\Gamma(\delta_{34})}
\nn
\ea
where $\delta^\prime_{ij}$ is supposed  to be expressed in terms of 
$\delta_{ij}$ using eqs.(\ref{delta:shift})ff, and the polynomials $\mathfrak{P}^D_l$ are 
defined in Appendix \ref{sec:n-star}. 

Let us express the result in terms of the independent variables $\beta_{12}$ and  $\gamma_{12}=\beta_{23}-\beta_{13}$ which determine the variables $\delta_{ij}$ according to the rules of section \ref{sec:MellinRep4}, given the field dimensions $d_1,...,d_4$. It is helpful to use, in place of $\beta_{12}$ the shifted variable
$$  \delta^{(43)(21)}= \beta_{12}-\frac 1 6 (d_1+d_2+d_3+d_4) = -\frac 1 2 s_{12} \ . $$
It is related to  the Mandelstam variable $s_{12}$ of the introduction.  
Its significance transpires from the relations
\be 
 \delta_{12} -\half (d_1+d_2) =\delta^{(43)(21)}= \delta_{34}-\half(d_3+d_4)  \label{delta:4321}
\ee
One computes 
\ba
\delta^\prime_{12} &=& \delta^{(43)(12)} + \half(h+c) \ , \nn \\
\delta^\prime_{34} &=& \delta^{(43)(12)} + \half (h-c) \ , \nn \\
\delta^\prime_{13} &=& -\half \delta^{(43)(12)}+ 
\half(c_-^{12} + c_-^{34} - \gamma_{12})  \ , \nn \\
\delta^\prime_{14} &=& -\half \delta^{(43)(12)}
+ \half ( c_-^{12}-c_-^{34} + \gamma_{12})  \ , \nn \\
\delta^\prime_{23} &=& -\half \delta^{(43)(12)}
+\half (-c_-^{12}+c_-^{34} + \gamma_{12})  \ , \nn \\
\delta^\prime_{24} &=& -\half \delta^{(43)(12)} 
+\half (-c_-^{12}-c_-^{34} - \gamma_{12})  \ , \label{delta:prime} 
\ea
Define, for $\chi=[l,h+c]$  the polynomial in $\gamma_{12}$ which depends parametrically on the differences of dimensions $c^{12}_-=\half (d_1-d_2)$ and $c^{34}_-=\half (d_3-d_4)$, and on $\delta^{(43)(21)}$ by
\be
 \mathfrak{P}^D_\chi(c^{12}_-,c^{34}_-; \delta^{(43)(21)}|\gamma_{12}) =
\mathfrak{P}^D_l(\{\delta^\prime_{ij}\}) \ , \label{pol:DnChi} 
\ee
evaluated at $\{\delta^\prime_{ij}\}$ as given by eq.(\ref{delta:prime}). 
Then
\ba 
&&  \tilde{\mathfrak{M}}_\chi (\{\delta_{ij}\}) = 
(\pi/2)^{-h}2^{-l}
 \mathfrak{P}^D_\chi(c^{12}_-,c^{34}_-;\delta^{(43)(21)}|\gamma_{12}) 
  \label{M:chi} \\
&& \frac {\Gamma(\delta^{(43)(21)}+\half(h-l)-\half c)}{\Gamma(\delta_{12})}
 \frac {\Gamma(\delta^{(43)(21)}+\half(h-l)+\half c)}{\Gamma(\delta_{34})}
\ .\nn
\ea

We see that $\tilde{\mathfrak{M}}_\chi$ has two families of poles in
 $\delta_{12}$ 
which are related to each other by the substitution $c\mapsto -c$,
and it  is a polynomial of degree $l$ in the 
 other independent variable $\gamma_{12}$.

The factors $\Gamma(\delta_{12})^{-1}\Gamma(\delta_{34})^{-1}$ were left in this form in order not to obscure the cancellation which occurs in the Mellin representation (\ref{MellinRep}), and which ensure that the reduced Mellin amplitude $\hat{M}^c$, which is related to $M^c$ by eq. (\ref{reduce}), does not have extra poles at non-positive integer $\delta_{12}$ and $\delta_{34}$. 

After splitting the pairs 
of poles and shifting the path of the $c$-integration as described in section
\ref{sec:mellin:poles},
expression (\ref{M:chi}) will (only) produce  the poles in $M^c$ coming from fields of Lorentz spin 
$l$ in the OPE. It is nevertheless amusing to compare with the structure of the Veneziano Beta-function $B(-\alpha(s),-\alpha(t))$.  
Its poles in $s$ come from a quotient of $\Gamma$-functions
 $\Gamma(-\alpha(s))/\Gamma(-\alpha(s)-\alpha(t))$, so that their positions are
obtained by a shift from the positions of its zeroes.

It follows from the amputation identity
(\ref{Norm1Vertex}) that
$$ \mathfrak{M}_\chi = \mathfrak{M}_{\tilde{\chi}} \ . $$
This reflects the equivalence of representations $\chi $ and $\tilde{\chi}$.
This implies the same for $\tilde{\mathfrak{M}}_\chi$ up to a factor. 
One finds by inspection that
\be  \mathfrak{P}^D_\chi(c^{12}_-,c^{34}_-;\delta^{(43)(21)}|\gamma) =
\mathfrak{P}^D_{\tilde{\chi}}(c^{34}_-,c^{12}_-;\delta^{(43)(21)}|\gamma)\ .\label{pol:tilde}\ee
Therefore 
\be \tilde{\mathfrak{M}}_\chi = \tilde{\mathfrak{M}}_{\tilde{\chi}} \ 
\mbox{ if } c_-^{12}=c_-^{34}. \label{tildeM:symm}
\ee
To verify eq.(\ref{pol:tilde}) note that $c\leftrightarrow -c$, $c^{12}_-
\leftrightarrow c^{34}_-$ takes  $\delta_1\leftrightarrow \delta_3$, 
$\delta_2\leftrightarrow \delta_4$ and 
$\delta_{34}^\prime\leftrightarrow
\delta_{12}^\prime$. The result now follows from 
eqs.(\ref{P:Dl:l},\ref{P:m}) by a change of summation variables 
$k_{14}\leftrightarrow k_{23}$.


The coefficient of the leading power in $\gamma$ in the polynomial 
$\mathfrak{P}_\chi^D(c_-^{12},c_-^{34}; \delta^{(43)(21)}|\gamma)$ can be evaluated in closed form, starting  from the defining equations (\ref{P:Dl:l}) and (\ref{P:m}) and using expressions (\ref{delta:prime}) for $\delta^\prime_{ij}$.To leading order in $\gamma$,
 \mbox{$\prod^\prime(\delta_{ij})_{k_{ij}}=(-)^{k_{13}+k_{24}}
 \prod^\prime(\gamma/2)^{k_{ij}}$ }.
Inserting the standard integral representation (8.381) of \cite{GR}, the summations over the four $k_{ij}$ can be performed with the help of the multinomial theorem. The result reads
\ba  &&\mathfrak{P}_\chi^D(c_-^{12},c_-^{34}; \delta^{(43)(21)}|\gamma) = 
\label{leadingOrder:pol1}\\ 
&&  a^l_0(D) (h+c-1)_l(h-c-1)_l  
\mathfrak{p}_\chi^D(c_-^{12},c_-^{34}) ( -\gamma/ 4 )^l\nn  + ...
 \nn 
\ea
with
\ba
\ \mathfrak{p}_\chi^D(c_-^{12},c_-^{34}) &=& 
\Gamma(\half (h+c)+c_-^{12})^{-1}\Gamma(\half (h+c)-c_-^{12})^{-1} \nn \\
&& \Gamma(\half (h-c)+c_-^{34})^{-1}\Gamma(\half (h-c)-c_-^{34})^{-1}
\nn 
\ea
The important observation is that the dependence on $c_-^{12}$ and on $c_-^{34}$ in $\mathfrak{p}(c_-^{12},c^{34}_-)$ factorizes. 

The result (\ref{leadingOrder:pol1})f is in agreement with the relation (\ref{pol:tilde}). It permits
 to extract the constant factor which relates
$\mathfrak{P}_\chi^D(c_-^{12},c_-^{34}; \delta^{(43)(21)}|\gamma)$ and $\mathfrak{P}_{\tilde{\chi}}^D(c_-^{12},c_-^{34}; \delta^{(43)(21)}|\gamma).$ The same factor is obtained  from eq.
(\ref{M:chi:1}) 
and expressions in the literature \cite{mack:dobrevClebsch} for the normalization factors. 

In later applications of the result (\ref{leadingOrder:pol1}), $\chi_k$ will substitute for $\chi$; it will be given by spin and dimension of some field in the OPE. 
\section{Poles in the Mellin amplitude}\label{sec:mellin:poles}
\subsection{Dynamical poles from poles in the partial wave 
}\label{sec:dynPoles}

Inserting the Mellin representation (\ref{M:chi}) 
of the individual partial waves into the partial wave expansion (\ref{scalarEpwe}), one finds that the connected 4-point Green function admits a Mellin representation with 
Mellin amplitude
\be
M^c(\{\delta_{ij}\}) = 
 \int d\chi \  g^{(43)(21)}(\chi)\ 
\mathfrak{M}_\chi
 (\{\delta_{ij}\}). \label{Mc:pwe} 
\ee
$\chi=[l,h+c]$, and this formula involves a sum over $l$ and an 
integration over $c$ along the imaginary axis (or along another path if there
 is a Born term, see below).

OPE imply asymptotic expansions of the full Greens functions, {\em not} of their connected parts. Nonvanishing disconnected parts $G_{i_4 i_2}(\x_4,\x_2)G_{i_3 i_1}(\x_3,\x_1)$ and 
$G_{i_4 i_1}(\x_4,\x_1)G_{i_3 i_2}(\x_3,\x_2)$ require that $d_4=d_2, d_3=d_1$ and $d_4=d_1, d_3=d_2$, respectively. We discuss first the case when neither of these two disconnected parts is present. 

Both the amplitude $g^{(43)(21)}(\chi)$ and the normalization factors $N_l$ depend on normalization conventions, but the product 
\mbox{$N_l(c_+^{34},c_-^{34},-c)N_l(c_+^{12},c_-^{12},c)g^{(43)(21)}(\chi)$}
 does not. 
 We know
from section \ref{sec:OPE:PWE}
 that non-derivative Lorentz-irreducible fields $\phi^k$ with spin and 
dimension 
$\chi_k=[l_k,d_k=h+c_k]$
which appear in the OPE of both $\phi^{i_1}\phi^{i_2}$ and of 
$\phi^{i_3\ast}\phi^{i_4\ast}$ are in bijective correspondence with pairs of poles in $N_l(c^{34}_+,c_-^{34},c)N_l(c^{12}_+,c_-^{12},-c) \ g^{(43)(21)}(\chi)$ for $l=l_k$ in $c$ at $c=\pm c_k$. A complete derivation of the result was published in ref.\cite{dobrevEtAl:OPE} for even space time dimension $D$, and under some restrictions on the field dimensions.
It was stated that the result can be generalized.

We wish to avoid restrictions on differences $c_-^{12}$ and $c_-^{34}$ of dimensions of the external fields, because factorization of 4-point functions 
is a constraint on the dependence of residues on the external fields. 
For the canonical choice of normalization factors \cite{mack:dobrevClebsch} which is used in ref. \cite{dobrevEtAl:OPE}, the product of normalization factors 
$N_l(c^{34}_+,c_-^{34},-c)N_l(c^{12}_+,c_-^{12},c)$ has branch cuts, for general $c_-^{12}, c_-^{34}$. Therefore it is only $\tilde{\mathfrak{M}}_\chi $ rather than $\mathfrak{M}_\chi$ which is meromorphic. We need to extract the product of normalization factors from $\mathfrak{M}_\chi$. It is convenient to postpone this, however.

Our object is to show that to each of the aforementioned pairs of poles there corresponds a 
family of poles of $M^c(\{\delta_{ij}\})$ in the variable $\beta_{12}$, or equivalently, in 
$ \delta^{(43)(21)}=-\frac 1 2 s_{12}=\delta_{12}- \frac 1 2 (d_1+d_2)$ at positions
\ba  
\delta^{(43)(21)} &=& \delta^n_l(d_k)\ ,  \qquad n=0,1,2,... \nn \\
\delta^n_l(d) &=& -\half(d-l)-n \ .
 \label{mellin:Poles} 
\ea
 Poles of $M^c$ in the $\beta_{ij}$ can only arise when the path of the
 $c$-integration  
in the integral (\ref{Mc:pwe}) gets pinched between singularities of the integrand when $\beta_{ij}$ is moved in the complex plane. It follows then from the 
polynomial character of the factor $\mathfrak{P}^D_\chi$ in the expression (\ref{M:chi}) 
for $\tilde{\mathfrak{M}_\chi}$ that the only poles are in $\beta_{12}$ and their residues are polynomials of degree $l$ in the remaining independent variable $\gamma_{12}=\beta_{23}-\beta_{13}$.

To get more explicit results, the symmetry in $c\mapsto -c$ is exploited in the same way as in the derivation of OPE from Euklidean partial wave expansions.
The first step was to decompose the Euklidean partial wave into two parts such that the second part is obtained from the first by substitution $c\mapsto -c$, and the strength of the singularity of the first part at $x_{12}^2=0$ decreases as the real part of $c$ is increased; this is achieved by 
the split (\ref{kernel2ndKind}) of the 3-point function.
 It is analog to the split 
of Legendre functions $P_{-\half+i\sigma}$ of the first kind into Legendre functions of the second kind which appears in the group theoretical approach to Regge theory \cite{ruehl:book}. The corresponding decomposition of the Mellin representation ${\mathfrak{M}}_\chi$ of the Euklidean partial wave reads
\ba
{\mathfrak{M}}_\chi &=& \frac {\pi}{\sin \pi (c+l)}
 \left[     N_l(c^{34}_+,c_-^{34},-c)N_l(c^{12}_+,c_-^{12},c)\mathfrak{M}^+_\chi \right.
\nn \\ 
&& \qquad - \left.
N_l(c^{34}_+,c_-^{34},c)N_l(c^{12}_+,c_-^{12},-c)\mathfrak{M}^+_{\tilde{\chi}}\right]\label{M:split}
\ea
with the requirement on $\mathfrak{M}^+_\chi$ that the poles in $c$ of 
$\mathfrak{M}^+_\chi $ 
occur only at the positions of the poles of $\Gamma(\delta_{12}-\half(h-c+l)-c^{12}_+)$, i.e. when the condition (\ref{mellin:Poles}) is fulfilled.
For the canonical choice \cite{mack:dobrevClebsch}, the ratio of the two products of normalization factors in eq.(\ref{M:split}) is meromorphic; therefore 
$\mathfrak{M}^+_\chi$ shares the meromorphy properties of $\tilde{\mathfrak{M}}_\chi$. 

 Since we know that poles in $\beta_{12}$ arise only from pinches, it suffices to represent $\tilde{\mathfrak{M}}_\chi $ as a formal sum of pole terms in $c$ labeled by $n=0,1,2,...$. 
\footnote{This is true in spite of the fact that the formal sum of pole terms does not indicate the correct asymptotic behavior as the imaginary part of $c$
 tends to $\infty$. We know from the work in \cite{dobrevEtAl:OPE} that the 
path of the $c$-integration in the Euklidean partial wave expansion can be shifted, after the split. This results from the good asymptotic behavior of the 
$\mathfrak{Q}$-kernels}

Using that $\Gamma(-n-c)= \pi [\sin \pi(-c-n)\Gamma(1+n+c)]^{-1}$, we find from the requirements on its poles that
\ba && \Gamma(\delta_{12})\Gamma(\delta_{34})\mathfrak{M}^+_\chi = -(\pi/2)^{-h} \sum_{n=0}^\infty \frac {(-2)^{-l}}{n!\Gamma(n+c+1)}  \nn \\
 && \quad
(\delta^{(43)(21)} - \delta^n_l(h+c))^{-1}
 \mathfrak{P}^D_\chi(c_-^{12},c_-^{34};\delta^n_l(h+c)|\gamma_{12}) \nn \\ &&
+ \mbox{nonsingular}
\label{M:+}\ea

Remember that the factor $\Gamma(\delta_{12})\Gamma(\delta_{34})$ cancels in the Mellin representation (\ref{MellinRep}).

Inserting the split (\ref{M:split}) and making a change of variables $c\mapsto -c$ in the second term, the integral (\ref{Mc:pwe}) becomes
\ba
&& M^c(\{\delta_{ij}\})= \frac {2\pi}{\sin \pi(c+l) } \label{Mc:pwe:split}
\int d\chi 
\nn \\ &&   N_l(c^{34}_+,c^{34}_-,-c) N_l(c^{12}_+,c^{12}_-,c)
\ g^{(43)(21)}(\chi)
\mathfrak{M}^+_\chi
 (\{\delta_{ij}\}) \ .  \nn 
\ea
Using the pole structure of $\mathfrak{M}^+$ as given by eq.(\ref{M:+}) 
we can identify the pinches which lead to poles in $\delta^{(43)(21)}$.

Let us initially ignore the poles of $\sin\pi(c+l)$. It will turn out that 
they make no contribution. 

Consider fixed $n$ and start with  $\delta^{(43)(21)}$ large and positive.  
Then the pole of $ (\delta^{(43)(21)} + \half(h+c-l) + n)^{-1}$ in $c$ is at large negative $c$. As we move $\delta^{(43)(21)} $ in negative direction, the pole will eventually touch the path of $c$-integration. We 
can deform the path of $c$-integration to avoid the pole, until the path gets pinched between the pole of $ (\delta^{(43)(12)} + \half(h+c-l) + n)^{-1}$
and of $N_l(c^{34}_+,c_-,c)N_l(c^{12}_+,c_-,-c) \ g^{(43)(21)}([l,h+c])$ at $c=c_k$.

 Suppose now that the residue of the latter pole factorizes as in 
eq.(\ref{coupling:const})

The pinch will lead to a pole in $M^c$ at position 
$ \delta^{(43)(21)}=\delta^n_l(h+c_k)$ with residue given by
\ba
\Gamma(\delta_{12})\Gamma(\delta_{34}) res\   M^c &=& \bar{g}^{i_3i_4}_k g^{i_1i_2}_k \label{res:M} \\ 
&& \mathfrak{r}^n_{\chi_k}
\mathfrak{P}^D_{\chi_k}(c^{12}_-,c^{34}_-;\delta^n_l(h+c_k)|\gamma_{12}).
 \nn  \\ 
\mathfrak{r}^n_\chi &=& - (\pi/2)^{-h} \frac {(-2)^{-l}}{n!\Gamma(n+c+1)} \label{r:n}\\
\chi_k &=& [l,h+c_k] \ . \nn 
\ea

If the OPE contains fields $\phi^a, \phi^b$ of the same Lorentz spin $l$ which differ in
 dimension by an even integer $d_b-d_a=2N$, then a coincidence of the positions of poles $\delta^n_l(d_a)=\delta^{(n+N)}_l(d_b)$ occurs.

It remains to discuss possible contributions from pinches with poles of 
 $1/\sin\pi(c+l)$ at integer $c$. The same problem occurs in the derivation of OPE from Euklidean partial wave expansions and was discussed in ref.\cite{dobrevEtAl:OPE}, for even $D$. The result is that all such contributions cancel in a subtle way, partly because of zeroes of the Plancherel measure
and partly because of partial equivalences of elementary representations
$\chi$ at so-called exceptional integer points. 
For odd $D$, the cancellation follows from zeroes of the Plancherel weight 
$\rho_l$ at integer $c$. But in this case the Plancherel weight has zeroes 
at half odd integral values of $c$, whose contributions need to be cancelled 
either by zeroes of $g^{(43)(21)}(\chi)$, or by partial equivalences among exceptional integer points. [Discrete series representations cannot occur for scalar amplitudes].

Let us now turn to situations with disconnected parts. Consider the extreme case of four equal real fields $\phi$ of dimension $d_4=d_3=d_2=d_1= h+c_+$ where all three disconnected parts are present. One of them represents the 
contribution of 
the unit operator to OPE.  We have now $c^{34}_+=c^{12}_+=c_+$ and
$c_-=0$. We assume that normalization conventions are so chosen that the Euklidean partial wave expansion of the 4-point function involves
\be   N_l(c_+,0,c)
N_l(c_+,0,-c)[1+ g^{(43)(21)}([l,h+c])]\ ,\label{mellin:factorize} \ee
where $1$ is the contribution of the disconnected part. 
The appropriate product of normalization factors is given in
 eq.(\ref{ProdNorm}); it is a meromorphic function of $c$. 

In this situation, the fields in the OPE of $\phi\phi$ with Lorentz spin $l$ 
 and dimension $d_k=h+c_k$ are in bijective correspondence with pairs of poles 
of expression (\ref{mellin:factorize}) at $c=\pm c_k$. We distinguish two sets of poles of \\ $N_l(c_+,0,c)N_l(c_+,0,-c)  g^{(43)(21)}([l,h+c])$. 

i) poles of  \\
$N_l(c_+,0,c)
N_l(c_+,0,-c)[1+ g^{(43)(21)}([l,h+c])]$ \\ I call these 
dynamical poles

ii) poles of $N_l(c_+,0,c)
N_l(c_+,0,-c)$ at which $g([l,h+c])=-1.$ \\ I call these kinematical poles. 

The dynamical poles make contributions as discussed before; in the expression for the residue, $[1+g^{(43)(21)}([l,h+c])]$ should be substituted for 
$ g^{(43)(21)}([l,h+c])$.

\subsection{Kinematical poles in $M^c$}
 \label{kinPolesOfM}

In the generic case of anomalous dimensions, when the disconnected parts have
 asymptotic expansion which involve integer powers of $x_{12}^2$ which are incompatible with OPE, the purpose of the kinematical poles is to cancel the disconnected parts of the full amplitude. As will be shown right below,
it follows from this that it must be true that $g(\chi)=-1$ at all the poles
of $N_l(c_+,0,c)N_l(c_+,0,-c)$. Therefore, all the poles of 
$N_l(c_+,0,c)N_l(c_+,0,-c)[1+ g^{(43)(21)}([l,h+c])]$ will actually be poles of $[1+g(\chi)]$. Another argument for this, which is valid when $\phi$ is the fundamental field of a Lagrangean field theory, was given in section \ref{footnote:poles}.  

To see the cancellation, compare with the generalized free field theory which 
has the same
disconnected parts as the interacting theory, but zero connected part; its conformal partial wave expansion is given by eq.(\ref{scalarEpweDisc}) with 
$g(\chi)\equiv 0$. Inserting the Mellin representation $\mathfrak{M}_\chi$ of individual partial waves, we end up with a sum over $l$ and an integral over 
$c$. To avoid divergences, we imagine that the summation over $l$ is done at the end. Repeating the procedure above, we get poles in the Mellin representation of the contributions from individual $l$ from the poles of $N_l(c_+,0,c)N_l(c_+,0,-c)$ which are the negative of the contributions from the 
contribution of kinematical poles to the connected amplitude $M^c$ in the interacting theory.

Because positivity is a property of the full Wightman functions,
it follows from this cancellation that  the contributions from the kinematical
 poles to the connected amplitude do not violate positivity.
%
\subsection{Born terms}\label{sec:BornTerms}
A word should be added concerning the case, when a fundamental field $\phi^f$ of dimension $h+c_f$, $c_f<0$ contributes to the OPE. This happens in $\phi^4$-theory in 3 dimensions, for instance, with $\phi^f =:\phi^2:$. As discussed in section \ref{sec:Lagrangean:SD}
 the initial path of the $c$-integration is different in this case, 
see figure \ref{fig:paths} a), because it is possible to include the Born term 
by altering the path of the $c$-integration. 

 Because of the symmetry of the partial wave $g^{(43)(21)}([l,h+c])$ under
  $c\mapsto -c$, the situation is unchanged after the split (\ref{M:split}) and subsequent substitution $c\mapsto -c$ in the second term. Upon shifting the path of the
 $c$-integration, there will be a contribution from the pole at $c=c^f$  
 which reproduces the  contribution of $\phi^f$ to the OPE, but there will be
 no contribution from its shadow at $c=-c^f$, see figure \ref{fig:paths} b). 

\section{Factorization of scalar 4-point functions} \label{sec:factorization}
In the following I consider the Mellin amplitudes of 4-point functions of 
scalar fields $\phi^{i_4},...,\phi^{i_1}$. They may be regarded as functions of two independent variables $\beta_{12}$ and $\gamma_{12}= \beta_{23}-\beta_{13}$. Fields $\phi_k$ which appear in the OPE of $\phi^{i_2}\phi^{i_1}$  and $\phi^{i_3\ast }\phi^{i_4 \ast}$ are symmetric tensor fields of some rank $l_k$ and some dimension $d_k$. As we saw,  they determine  integer spaced families of poles in $\beta_{12}$ at positions of $\beta_{12}=\delta^{(43)(21)} +\frac 1 6 (d_1+d_2+d_3+d_4)$
 given by $\delta^{(43)(21)} = \delta_{l_k}^n(h+c_k)= -\frac 1 2 (h+c_k-l_k)-n$. Their 
  residues are polynomials of degree $l_k$ in $\gamma_{12}$. Factorization is a statement about the $i_4,...,i_1$ dependence of these residues which follows from OPE. 

First I explain how the spin and dimension $\chi_k=[l_k,d_k]$ of all the fields is read off from the poles of $M^c_{i_4...i_1}$. In doing so, the amplitude 
$M^c_{i_4...i_1}$, regarded as formal sums of pole terms, will be rewritten so that the integer spaced families of poles which correspond to individual fields $\phi_k$ will be exhibited, as in eq.(\ref{pole:sum}) below. I write $c_-^{12}=d_{i_1}-d_{i_2}$ etc.  for differences of dimensions of the external fields. 
\ba
 M^c_{i_4...i_1}&=& \Gamma(\delta_{12})^{-1}\Gamma(\delta_{34})^{-1} \sum_k  \sum_{n=0}^\infty g_k^{i_4...i_1}
\mathfrak{r}^n_k  \label{pole:sum} \\
&&  \mathfrak{P}^D_{\chi_k}(c^{12}_-,c^{34}_-;\delta^n_{l_k} |\gamma_{12})
 \frac{1}{\delta^{(43)(21)} - \delta^n_{l_k}(d_k)-n } \nn 
\ea

Herein, $\mathfrak{P} $ are the kinematically determined polynomials in $\gamma_{12}$ defined by eq.(\ref{pol:DnChi}), using the Appendix (section \ref{sec:n-star}), and $\mathfrak{r}$ are kinematically determined factors which govern the $n$-dependence of the residues of satellite poles $n=1,2,...$, see eq.(\ref{r:n}). 
There is a dependence on differences of external dimensions in these quantities.

I assume that the normalization of fields is so chosen that the normalization factor $\c_k$ of their two-point function is $+1$ for physical fields, and $-1$ for ghosts. 

Factorization now requires that the coefficients $g_k^{i_4...i_1} $ factorize
in products 
\be g_k^{i_4...i_1}=\bar{g}_k^{i_4i_3}g_k^{i_1i_2} \label{facCoupl} \ee
for physical fields $\phi^k$. In the case of a ghost, there would be an overall $-$-sign. 

Let me explain how the spin and dimension $\chi_k=[l_k,d_k]$ of the fields and the coefficients $g_k^{i_4...i_1}$ are determined from the poles of 
$M^c_{i_4...i_1}$. 

I assume again that field dimensions are anomalous, so that there is at most a finite number of fields with given twist $d_k-l_k$. Nevertheless there can be limit points of poles, as mentioned in the introduction. The procedure outlined below must then be carried out in an appropriate order, as specified below, in order to find the fields in all trajectories associated with  the different limit points.  

Consider what remains of  $M^c_{i_4...i_1}$, considered as a sum of pole terms, after the contributions of zero or some fields $\phi^j$ in the OPE are subtracted. These contributions have the form of individual terms in the sum over $k$ in the right hand side of eq.(\ref{pole:sum}). Suppose that the remainder has
a leading pole in $\beta_{12}$ (i.e. the pole with largest $\beta_{12}$)  whose  residue is a polynomial in $\gamma_{12}$ of degree $l_k$, at a position which is such that $\delta^{(43)(21)}= \delta^0_{l_k}(d_k)\equiv - \frac 1 2 (d_k-l_k)$; this determines $d_k$.  Thereby spin $l_k$ and dimension $d_k$ of a field $\phi_k$ are determined which has to be introduced in order to reproduce this pole. Abbreviate $\chi_k=[l_k,d_k]$ as usual. 
The coefficient of the highest power $-(\gamma_{12}/4)^{l_k}$ of the polynomial residue can be written in the form 
\be    g_k^{i_4...i_1}\mathfrak{r}_{\chi_k}
a^{l_k}_0(D)
 (d_k-1)_{l_k}(D-d_k-1)_{l_k}\mathfrak{p}_{\chi_k}^D(c_-^{12},c_-^{34})\nn 
\ee
with some suitable coefficient $g_k^{i_4...i_1}$ which depends on the labels 
$i_4...i_1$ of the external fields.[The quantities $a^l_0$ are the coefficients in the power series expansion (\ref{power:Y}) of the zonal spherical function, while $\mathfrak{r}$ and $\mathfrak{p}$ are kinematically determined quantities defined in eqs. (\ref{r:n}), and (\ref{leadingOrder:pol1})f.]
Now subtract the contribution of this field in the OPE from the remainder of $M^c_{i_4...i_1}$. It follows from the definition (\ref{leadingOrder:pol1})  of 
$\mathfrak{p}_{\chi_k}^D(c_-^{12},c_-^{34})$ as coefficient of the leading power $(-\gamma_{12}/4)^{l_k}$ in the polynomial $\mathfrak{P}_{\chi_k}(c^{12}_-,c^{34}_-; \delta^{(43)(21)}|\gamma_{12})$ that the leading pole of what remains afterwards will either have a smaller position $\beta_{12}$ or a residue of lower order.

If there are no limit points of poles, all poles can be subtracted recursively, typically in infinitely many steps. 
In the case with limit points, one must proceed in a slightly different order, removing successively the finitely many poles outside smaller and smaller neighborhoods of 
the limit points and of $\beta_{12}=\infty$. 

\section{Asymptotic expansions from the Mellin representation}
Like the Mellin representation itself, the formulas of this section are valid for arguments throughout the axiomatic domain of analyticity (the permuted extended tube.) If $x_i$ are in Minkowski space, appropriate $i\epsilon$-prescriptions are to be imposed to obtain the Wightman function with a particular ordering of the fields. If $x_i$ are Euklidean, $-x_{ij}^2= \x_{ij}^2$. 

Here I examine the asymptotic expansion of a 4-point function with Mellin representation (\ref{MellinRep}) as $x_{12}\mapsto 0$. It is determined by the poles of $M^c$ in $\delta_{12}$.   

Remember that the exponents $\delta_{ij}$ in the Mellin representation of a scalar 4-point-function  have the form $\delta_{ij}=\delta^0_{ij} + \beta_{ij}$,
 where 
$\delta_{ij}^0$ are kinematically determined by the dimensions $d_i$, $i=1,\dots d_4$ of the fields whose 4-point function is considered, and $\beta_{ij}$ are integration variables, two of which are independent. 
As independent variables we may take $\beta_{12}$ and $\gamma_{12}$, 
\be \gamma_{12} = \beta_{23}-\beta_{13} \label{def:gamma12}\ee 
The exponents $\beta_{ij} $ are expressed through the independent ones as
\ba
\beta_{13}&=& - \half \beta_{12}- \half \gamma_{12} \quad = \quad \beta_{24} \nn \\
\beta_{23} &=& - \half \beta_{12} + \half \gamma_{12}\quad = \quad
\beta_{14}
 \label{beta:indep} 
\ea
The measure in the Mellin representation is 
$d\delta \equiv d^2\beta=\half d\beta_{12}d\gamma_{12}$.

We examine here the particular channel $(43)(21)$, and for this purpose it is 
again convenient to use in place of $\beta_{12}$ the shifted variable 
$\delta^{(43)(21)}=\beta_{12}-\frac 1 6(d_1+d_2+d_3+d_4)$ ($=-\frac 1 2 s_{12}$) so that 
\be \delta_{12}-\frac 1 2 (d_1+d_2)=\delta^{(43)(21)}=
\delta_{34}-\frac 1 2 (d_3+d_4) \ . \label{delta:4321:again} \ee
Suppose the Mellin amplitude $M^c_a(\{ \delta_{ij}\})$ of a connected scalar 
4-point function $G^c_a$ has a pole in $\beta_{12}$ at position 
$\delta^{(43)(21)}=\hat{\delta}$ 
with residue
$P(\hat{\delta}|\gamma_{12})$ which is a polynomial in $\gamma_{12}$. In applications it will depend on the position $\hat{\delta}$ of the pole.

When the path of the $\beta_{12}$-integration in the Mellin representation is shifted past such a pole, a contribution from the residue is obtained which 
is equal to 
\ba
R(\hat{\delta}| x_4,\dots,x_1) &=& 
\Gamma(\hat{\delta}_{12})\left( - \half x_{12}^2 \right)^{-\hat{\delta}_{12}}
\nn \\ &&
\Gamma(\hat{\delta}_{34})\left( - \half x_{34}^2 \right)^{-\hat{\delta}_{34}}
  \label{def:R} \\
&&
(-2\pi i)^{-1} \int_{-i\infty}^{i\infty}\half 
d\gamma_{12}  
P(\hat{\delta}|\gamma_{12})\X \nn
\ea
with 
\be
\X = \prod_{(ij)}{}^\prime \Gamma (\delta_{ij})
\left( -\half x_{ij}^2 \right)^{-\delta_{ij}} 
\label{def:X}
\ee
Herein, $\hat{\delta}_{12}$ and $\hat{\delta}_{34}$ are determined by 
$\hat{\delta}=\delta^{(43)(21)}$ through eq.(\ref{delta:4321:again}), and 
the product $\prod^\prime$ runs over the four unordered pairs 
$(13),(23),(14),(24)$, i.e. all except $(12)$ and $(34)$.
In eq.(\ref{def:X}), $\delta_{ij}=\delta^0_{ij} + \beta_{ij}$ 
 with $\beta_{ij}$ as in 
eq.(\ref{beta:indep}).

The factors in the first line may become 
singular when $x_{12}\mapsto 0$ and/or $x_{34}\mapsto 0$, while the 
factor $\X$ is not singular when $x_1,x_2$ stay away from $x_3,x_4$.

It remains to perform the $\gamma_{12}$-integration.  

The following well known identity can be used to combine the product of four
$x$-dependent  factors into a single factor, at the cost of introducing some
variables of integration.
\ba
&& \Gamma(\nu)A^{-\nu}\Gamma(\mu)B^{-\mu} = 
\int_0^1 du u^{\nu -1} (1-u)^{\mu -1} \Gamma(\mu + \nu) \nn \\
&& \qquad \left(uA+(1-u)B\right)^{-\nu -\mu}\ .
\label{id:usual}
\ea
We combine first the $x_{13}$-dependent factor with the $x_{14}$-dependent factor, using integration variable $u_1$, and the remaining two factors using integration variable $u_2$. 
As a result, there appears a product of two propagators,

\ba 
&& \X = \nn \\
&& \int\int du_1 du_2 u_1^{\delta_{13}-1}
(1-u_1)^{\delta_{14}-1}u_2^{\delta_{24}-1}(1-u_2)^{\delta_{23}-1}\nn \\
&& \Gamma(\delta_{13} + \delta_{14}) 
\left( - \half u_1 x_{13}^2 - \half (1-u_1) x_{14}^2\right)^{-\delta_{13} - \delta_{14}} \nn  \\
&&
\Gamma(\delta_{23} + \delta_{24})
\left( - \half u_2 x_{24}^2 - \half (1-u_2) x_{23}^2\right)^{-\delta_{23}-\delta_{24}} \label{two:propagators}  .
\ea
Let us introduce
\be x_5(u)= ux_3 + (1-u)x_4 \label{x5} \ee
 As $u$ varies from $0$ to $1$, the point $x_5$ runs along the straight line 
from $x_3$ to $x_4$. Defining $x_{15}(u)=x_1-x_5(u) $ and $x_{25}(u)=x_2-x_5(u)$, one can  rewrite the arguments of the above two propagators by using 
identities like
\be 
ux_{i3}^2 + (1-u)x_{i4}^2 = x_{i5}(u)^2 + u(1-u)x_{34}^2
\label{x:interpolate} 
\ee
for $i=1,2$. 
This shows that expression (\ref{two:propagators})  is a product of two factors
 of the form $\Gamma(\nu)\left(-\half x_{i5}^2 - \half  m_{i5}^2 \right)^{-\nu} $, which we can interpret as propagators from $x_1$ to $ x_5(u_1) $ and from 
$x_2$ to $x_5(1-u_2)$. These ``propagators'' do not have good positivity properties, though. 

In the special case $u_1=1-u_2$ the two points $x_5(u_1)$ and $x_5(1-u_2)$ coincide, and another application of identity (\ref{id:usual}) (with integration variable denoted $v$) would produce a formula with a single propagator connecting
\be 
x_0(v)=vx_1+ (1-v)x_2 
\ee to $x_5(u_1)$. The point $x_0(v)$ runs along the straight line connecting $x_1$ and $x_2$. 

In general $u_1\neq 1-u_2$. But it turns out that performing the $\gamma_{12}$-integration in expression eq.(\ref{def:R}) produces a derivative of a 
Dirac $\delta$-function $\delta(u_1+u_2-1)$. 

Expecting this, the two $x$-dependent factors in expression 
(\ref{two:propagators}) are combined using identity (\ref{id:usual}), with integration variable $v$. Noting that the expression (\ref{x:interpolate}) is linear in $u$, and using identity (\ref{x:interpolate}) appropriately, the 
result for $\X$ can be written as follows, using eq.(\ref{beta:indep})
 
\ba
\X &=& \int \int du_1  du_2
 \left( \frac {u_1}{1-u_1}\frac {u_2}{1-u_2} \right)^{-\half \gamma_{12}}\nn \\
&& \quad \Y
\nn \ea 
with $\hat{\beta}_{12}= \hat{\delta}_{12}- \delta^{0}_{12}$ and 
\ba
&&\Y =  \label{expr:X} \\
 &&\quad  u_1^{\delta_{13}^0 - \half \hat{\beta}_{12} -1}
(1-u_1)^{\delta_{14}^0 - \half \hat{\beta}_{12}-1} \nn \\  
&& \quad u_2^{\delta_{24}^0 - \half \hat{\beta}_{12} -1 }
(1-u_2)^{\delta_{23}^0 - \half \hat{\beta}_{12}-1} \nn \\
 && \int dv v^{\delta_{13}^0 + \delta_{14}^0 - \hat{\beta}_{12}-1}
(1-v)^{\delta_{23}^0 + \delta_{24}^0 - \hat{\beta}_{12}-1} \nn \\
&&  \Gamma(\Delta_{12})
\left(
-\half x_{05}(v,1-u_2) -\half v(1-v)x_{12}^2 - \right.
\nn  \\ && 
\left. - \half u_2(1-u_2)x_{34}^2 -\half v(u_1+u_2-1)(x_{13}^2 - x_{14}^2) \right)^{-\Delta_{12}}\nn \ea
with 
\ba \Delta_{12} &=& \delta_{13}^0+\delta_{14}^0 +\delta_{23}^0 + \delta_{24}^0 -2\hat{\beta}_{12} \nn \\
&=& d_1+d_2- 2 \hat{\delta}_{12}  \nn \\
&=& d_3 + d_4 - 2 \hat{\delta}_{34} \ . \label{expr:Delta}
\ea
All integrations over $u_1,u_2$ and $v$ run from $0$ to $1$. It is understood that  $x_{05}(v, 1-u_2) = x_0(v)-x_5(1-u_2)$ and $\hat{\beta}_{12}=\hat{\delta}_{12}-\delta^0_{12}$. 

The factor $\Y$ does not depend on $\gamma_{12}$.  

The expression (\ref{def:R}) for $R$ involves 
\ba
&& (2\pi i)^{-1} \int_{-i\infty}^{i\infty}
d\gamma_{12} P(\hat{\delta}_{12}| \gamma_{12}) \X 
= (2\pi i)^{-1} \int_{-i\infty}^{i\infty}
d\gamma_{12} \nn \\
&&
 \int \int du_1  du_2 P(\hat{\delta}_{12}| \gamma_{12})
 \left( \frac {u_1}{1-u_1}\frac {u_2}{1-u_2} \right)^{-\half \gamma_{12}}
\nn \\ && \qquad \Y \nn
\ea
The $\gamma_{12}$-integration can be performed. $P$ being a polynomial in $\gamma_{12}$, we get a derivative of a Dirac $\delta$-function
$\delta(y_1+y_2)$ with $y_1=-\half \ln (u_1/(1-u_1)) $ and 
$y_2=-\half \ln (u_2/(1-u_2)).$. 

The argument of the $\delta$-function becomes $0$ if and only if 
$u_2=1-u_1$. One may use the formula
$$\delta(f(x))= |f^\prime(x_0)|^{-1}\delta(x-x_0) \ ,$$
 which is valid when 
$f(x)$ has a single zero at $x=x_0$. In our case the variable is $u_1$, and 
$f(u_1)= y_1+y_2$, so that $f^\prime(u_1)^{-1}= -2u_1(1-u_1)$. 
The  result is
\ba
&& (2\pi i)^{-1} \int_{-i\infty}^{i\infty} \half
d\gamma_{12} P(\hat{\delta}_{12}| \gamma_{12}) \X   \label{int:Y} \\
&& =
\int\int du_1du_2\ u_1(1-u_1) \Y \nn \\
&&
\qquad  P\left( \hat{\delta}_{12}| 2 
u_1(1-u_1)\frac {\partial}{\partial u_1}\right)\delta(u_1+u_2-1)  \nn 
\ea
This can be inserted into the expression (\ref{def:R}) for $R$.
Thanks to the prescription $u_2=1-u_1$, which results from the $\delta$-function,  the $x$-dependent last factor in 
$Y$ becomes a propagator. But before the prescription can be applied, the 
derivatives with respect to $u_1$ must be performed. This gives a sum of terms involving propagators whose exponents differs from $-\Delta_{12}$ 
by non-positive integers, and extra factors of $x_{13}^2-x_{14}^2$, which tend
 to zero when $x_3-x_4\mapsto 0$.

Now we wish to obtain an asymptotic expansion as $x_{12}\equiv x_1-x_2\mapsto 0$ and 
$x_{13}\equiv x_3-x_4\mapsto 0$, keeping the two pairs of points $x_1,x_2$ and $x_3$, $x_4$ well separated. In this limit, $x_{05}$ tends to a nonzero limit which is independent of $v$ and $u_2$. 

To obtain an asymptotic expansion for future use expand the 
$x$-dependent last factor in $Y$ in powers of $x_{12}^2$ and $x_{34}^2$, using the expansion formula
\be \Gamma(\alpha)\left( A - \epsilon\right)^{-\alpha} = 
\sum_k \frac {1}{k!} \epsilon^k \Gamma(\alpha+k) A^{-\alpha - k}
 \label{eps:expand}\ee
We use this first for $\epsilon = \half v(1-v)x_{12}^2$ and then again for 
$\epsilon= \half u_2(1-u_2) x_{34}^2$. 
The result is
\ba
&& \Y =  \nn \\ 
&&
\sum_{p=0}^\infty \sum_{q=0}^\infty \frac {(-)^{p+q}}{p!q!}
\left( -\half x_{12}^2 \right)^p \left( - \half x_{34}^2\right)^q  
 \label{exp:Y} \\
&& \quad u_1^{\delta_{13}^0- \half \hat{\beta}_{12} + q -1 }
(1-u_1)^{\delta_{14}^0 - \half \hat{\beta}_{12} +q -1} \nn \\ 
&&
u_2^{\delta_{24}^0 - \half \hat{\beta}_{12}-1} 
(1-u_2)^{\delta_{23}^0 - \half \hat{\beta}_{12} -1} \nn \\
&& \quad \int dv  \
 v^{\delta_{13}^0 + \delta_{14}^0 - \hat{\beta}_{12}+p-1 }
(1-v)^{\delta_{23}^0 + \delta_{24}^0 - \hat{\beta}_{12} + p -1} \nn   \\
&& \quad \Gamma(\Delta_{12} + p + q) 
\left( - \half x_{05}(v, 1-u_2)^2 \right. \nn \\
&& \qquad \left.
 - \half v(u_1+ u_2 -1) (x_{13}^2 - x_{14}^2)\right)^{-\Delta_{12} - p - q}   
\nn
\ea 
$-R$ is obtained by multiplying expression (\ref{int:Y}) with  the two 
propagators $\Gamma(\hat{\delta}_{12})\left( - \half x_{12}^2 \right)^{-\hat{\delta}_{12}}
\Gamma(\hat{\delta}_{34})\left( - \half x_{34}^2 \right)^{-\hat{\delta}_{34}}$, inserting eq.(\ref{exp:Y}) for $Y$, and performing the differentiations in $P$.

For future reference let us specialize to the case when the polynomial $P$
is constant in $\gamma_{12}$, 
$$ P(\hat{\delta}|\gamma_{12}) =
 r(\hat{\delta}) \in {\mathbf R}\ ,$$
and for equal dimensions $d_1=d_2=d_3=d_4=d$, whence 
$$\hat{\delta}_{34}=\hat{\delta}_{12}= \hat{\delta}+d$$
 by eq.(\ref{delta:4321:again}).

The resulting asymptotic expansion for $R$ is 

\ba
&& \,\,\, R(\hat{\delta}| x_4,\dots,x_1) 
= - r(\hat{\delta}) \sum_{p=0}^\infty \sum_{q=0}^\infty 
\frac {(-)^{p+q} }{p!q!} \nn \\
&&
 \Gamma(\hat{\delta}_{12})\left
(-\half x_{12}^2\right)^{-\hat{\delta}_{12}+p}
 \Gamma(\hat{\delta}_{12})\left(-\half x_{34}^2\right)^{-\hat{\delta}_{12}+q} 
\nn \\
&& \quad \int du\int dv [u(1-u)]^{d-\hat{\delta}_{12} + q -1}
[v(1-v)]^{d-\hat{\delta}_{12} + p -1} \nn \\
&& \quad \Gamma(2d-2\hat{\delta}_{12} + p + q)
\left( -\half x_{05}^2(v,u)\right)^{-2d+2\hat{\delta}_{12} -p -q} \ ,
\label{exp:R:scalar}
\ea
where $x_0(v)=vx_1+(1-v)x_2$, $x_5(u)=ux_3 + (1-u)x_4$ and $x_{05}(v,u)=x_0(v)-x_5(u)$. 

\section{Comparison of contributions to asymptotic expansions as $x_{12}\mapsto 0 $ from OPE and from families of poles in the Mellin amplitude}\label{comparison}
%
%
\label{sec:comparison}
We compare the contribution to the asymptotic expansion of the scalar 
4-point Green function $G_{0000}$ as $\x_{12}\mapsto 0$ and/or $\x_{34}\mapsto 0$
which come  from a scalar field in the OPE with the contribution of the
 corresponding family of dynamical poles in the Mellin amplitude. 
The kinematical poles need not be considered since they just cancel 
disconnected parts, by their definition. 
\subsection{Summation of the contributions from a family of poles in the Mellin amplitude which corrresponds to a scalar field in the OPE}\label{sec:summation}
Consider the four-point function $G_{0000}$  of four real fields $\phi^0$ of dimension $d$.
 
As we have seen in section \ref{sec:dynPoles}, a scalar field $
\phi^k$ of dimension $\delta=h+c$ which contributes to the operator product expansion in the $(12)$-channel  produces a family of poles in
$\delta_{12}$ in the Mellin amplitude at positions $\delta^{(43)(21)}=\delta^n_0(\delta)$, whence
\be \hat{\delta}^n_{12}= d -\half \delta -n , \qquad, n=0,1,... \label{pole:position} \ee
with residues  as follows: For the scalar case, $\chi=[0,h+c]$, the 
polynomial $\mathfrak{P}^D_\chi $ is of degree $0$ and equals 
$\prod_{n=1}^4 \Gamma(\delta_n)^{-1}$, with $\delta_i$ given by eqs.(\ref{delta:i}). Therefore the residue $r_n$ of the $n$-th pole obtains from eq.(\ref{res:M})f as
\be
 r_n =\frac {const}{\Gamma(\delta_{12})\Gamma(\delta_{34})} 
 \ 
\frac{(-)^n}{n!} \Gamma(-n-c) \ . \label{Mellin:scalar:residue}
\ee
with 
\ba const&=& - \frac{\sin \pi c}{\pi} |g^{00}_k|^2 (\pi/2)^{-h} \label{const} \\
&&  \Gamma(\half(h+c))^{-2}\Gamma(\half(h+c))^{-2} \nn 
\ea
In the expression for $r_n$, the arguments $\delta_{12}=\delta_{34}$ are determined by the position of the pole and the  two inverse $\Gamma$-functions cancel against the first two $\Gamma$-functions in expression (\ref{exp:R:scalar}) for $R$. 

 The residues have the same sign for all $n$.
\footnote{This implies that the contribution from individual poles is not positive. Canceling a pole with $n >0$ would require adding a ghost field of dimension $\delta+2n$ which contributes to the OPE}

The $n$-th pole makes a contribution $R_n$ to the asymptotic expansion of the 
four-point function, which is given by eq.(\ref{exp:R:scalar}), with 
$\delta^n_0(\delta)$ substituted for $\hat{\delta}$, and
$$ r(\hat{\delta}) = r_n . $$
We wish to sum these contribution over $n$. 

To do so, we introduce new summation variables $N$,$K$ in place of $p$, $q$ by
\ba
n+p&=&N \nn \\
n+q &=&K
\ea
They run over $N\geq n, K\geq n$, but we may extend the summation from 0, because the extra contributions vanish, since $\frac {1}{p!}=\Gamma(N-n + 1)^{-1}=0$
when $N<n$, and similarly for $K$.
 Inserting expression (\ref{pole:position}),
the sum of the contributions $R_n$ comes out as 
\ba
&& \sum_n R(\delta^n_0(\delta)|x_4,...,x_1) 
 = - const \nn \\
&& \sum_{N,K}S_{N,K}(c)
 \left(-\half x_{12}^2 \right)^{-d+\half \delta +K }
 \left(-\half x_{34}^2 \right)^{-d+\half \delta +N }\nn \\
&& \quad \int\int du dv [ u(1-u)]^{\half \delta +K-1}
[v(1-v)]^{\half \delta +N-1}\nn \\
&& \Gamma(\delta + N+K) \left( -\half x_{05}^2(v,u)  \right)^{-\delta-N-K}
\label{scalar:pole:sum}
\ea
where $\delta=h+c $ is the dimension of the field $\phi^k$ in the OPE and 
\be S_{N,K} = \sum _n \frac {(-)^n}{n!}\Gamma(-c-n) 
\frac {(-)^{N+K}}{(N-n)!(K-n)!}
\ee
$S_{N,K}$ is manifestly symmetrical in $N$ and $K$. We may therefore assume 
without generality that $N\geq K$. The sum can be evaluated by 
writing
$$\frac {\Gamma(-c-n)}{\Gamma(N-n+1)} = \frac {B(-c-n, c+N+1)}{\Gamma(c+N+1)}\ ,$$
 inserting the standard integral representation for the Beta-function and 
 interchanging sum and integral.The result can be written in the form
\be
S_{N,K} = \frac {-\pi}{\sin\pi c} \frac {1}{K!N!} 
\frac {(-)^K }{\Gamma(c+N+1)} \frac {\Gamma(-c-K)}{\Gamma(-c-K-N)}
\label{sum:S} 
\ee
 for $N\geq K$. 
%
%
\subsection{Short distance behavior from the OPE}
Consider the Wightman 4-point function of four scalar fields $\phi^{i_4},...,\phi^{i_1}$ of dimensions $d_{i_4},...,d_{i_1}$.

  According to eq.(\ref{OPE:4ptfct}), 
the operator product expansion takes the form 
\ba
W_{\bar{i_4},\bar{i_3},i_2,i_1}(x_4,...,x_1)
&=& \sum_k c_k \bar{g}^{i_3 i_4}_k g^{i_2 i_1}_k  \label{OPE:4ptfct:again}
\\
&&  \mathfrak{W}_{\chi_k,\bar{i_4},\bar{i_3},i_2,i_1}(x_4,...,x_1)\ ; 
\nn \ea
where $\bar{g}$ is the complex conjugate of $g$. 

The term labeled by $k$ represents the contribution of field $\phi^k $ 
with Lorentz spin and dimension given by $\chi_k=[l_k,d_k]$; $c_k>0$ fixes the normalization of its two point functions. The $c_k$ and coupling constants $g_{..}^{.}$ are the dynamical quantities which need to be determined in order to construct a theory, and $\mathfrak{W}$ are kinematically determined quantities.

There is freedom of choosing normalization factors. 


According to eq.(\ref{QV}), 
\ba
\mathfrak{W}_{l,m,j,i}(x_4,...,x_1)
&=&  \mathfrak{Q}(\chi_k,p;\chi_l,x_4,\chi_m,x_3)\label{W=QV} \\
&& \quad \mathfrak{V}(\chi_k,z,\chi_j,x_2,\chi_i,x_1)|_{z=0}\ . \nn
\ea
with $p=-i\nabla_z$. 
It is understood that $c^{43}_- = d_l-d_m$ etc.


The kinematically determined quantity $\mathfrak{W}$ given by eq. (\ref{W=QV})
 is boundary value of an analytic function, whose domain of analyticity (the tube) includes Euklidean points. Upon substitution of $\mathfrak{Q}^u$ and $\mathfrak{V}^u$ for $\mathfrak{Q}$ and $\mathfrak{V}$, I denote this quantity by 
$(2\pi)^{-D}I_l^u$, where $l$ is the Lorentz spin of the field $\phi^k$ in the OPE. 
I will evaluate this quantity for $l=0$, thereby computing the contribution 
of a scalar field in the OPE to the short distance behavior. 

Inserting the formulas (\ref{Vscalar:u}) for $\mathfrak{V}^u$, and (\ref{Q})
for the partial  Fourier transform of $\mathfrak{Q}^u $ into eq.(\ref{W=QV}),
 the factor $e^{ip[ux_4+(1-u)x_3]}$ acts as a translation operator which shifts
$z$ form $0$ to
$ x_5=ux_4+ (1-u)x_3 $,
and $p^2$ acts as $-\Box_5$. Inserting the power series expansion of $J_c$, one obtains
\ba
& I_0^u& = \frac{-1}{\Gamma(\half(h-c)+c_-^{43}) \Gamma(\half(h-c)-c_-^{43})}
\nn \\ 
&&
\sum_{N=0}^\infty \frac {(-1)^N }{N!\Gamma(c+N+1)} 2^{-N} \nn \\
&& (-\half x_{43}^2)^{-\half(h-c)-c_+^{43}+N} 
(-\half x_{12}^2)^{-\half(h-c)-c_+^{12}}(-\Box_5)^N \nn \\
&& \int_0^1 du \ u^{\half (h+c) +c_-^{43}+N-1}
\ (1-u)^{\half (h+c) -c_-^{43}+N-1} \nn \\
&& (- \half x_{25}^2)^{-\half(h+c) +c_-^{12}}
(-\half x_{15}^2)^{-\half(h+c)-c_-^{12}}
\nn \\ 
&x_5&\equiv ux_4+(1-u)x_3 \nn 
\ea
Now the two propagators in the last line are combined into a single propagator 
by use of the standard identity (\ref{id:usual}), introducing an auxiliary variable of integration $v$. 
Defining 
\be x_0(v)= vx_2+ (1-v)x_1, \label{x_5}\ee
and  $ x_{05}=x_0-x_5$,
the new propagator can be rewritten with the help of the identity
\be vx_{20}^2 + (1-v)x_{10}^2 = x_{05}^2 + v(1-v)x_{12}^2 \ . \nn \ee 
and comes out proportional to 
\ba && \Gamma(h+c)(-\half x_{05}^2 -\half v(1-v)x_{12}^2)^{-h-c}  =
 \label{prop:expand} \\
 && \sum_{K=0}^\infty \frac {1}{K!}(\half v(1-v)x_{12})^2)^K 
\Gamma(h+c+K)(-\half x_{05}^2)^{-h-c-K}\nn 
\ea
Finally, the differentiation $\Box_5^N$ is carried out with the help of the 
formula
\ba
&& (\Box_5)^N \Gamma(\delta)\left( - \half x_{05}^2 \right)^{-\delta}=\label{Box:Prop} \\
&& \quad 2^N \frac{\Gamma(-\delta+h)}{\Gamma(-\delta + h -N)}
\Gamma(\delta + N)\left( - \half x_{05}^2\right)^{-\delta-N} \nn 
\ea 
Collecting everything and noting that $\delta_\chi=\half(h+c)$ 
produces the final result
\ba
I_0^u&=& \frac{-1}{\Gamma(\half(h-c) + c_-^{43})
\Gamma(\half(h-c) - c_-^{43})} \nn \\
&& \frac{1}{
\Gamma(\half(h+c)+ c_-^{12})
\Gamma(\half(h+c) - c_-^{12})} \nn \\
&& \sum_{N=0}^\infty \sum_{K=0}^\infty \frac {(-1)^K}{K!N!} 
\frac{\Gamma(-c-K)}{\Gamma(c+N+1)\Gamma(-c-K-N)} \nn \\
&& \left( -\half x_{43}^2\right)^{-\half(h-c)-c_+^{43}+N}
 \left( -\half x_{12}^2\right)^{-\half(h-c)-c_+^{12}+K}\nn \\
&&
 \int_0^1 du \ u^{\half(h+c)+c_-^{43}+N-1}(1-u)^{\half(h+c)-c_-^{43}+N-1}
\nn \\
&&
\int_0^1 dv \ v^{\half(h+c)-c_-^{12}+K-1}(1-v)^{\half(h+c)+c_-^{12}+K-1} \nn
\\
&& \Gamma(h+c+K+N)\left(-\half x_{05}^2 \right)^{-h-c-N-K} \ . 
\label{QV:result} \\
x_0&=&ux_1+(1-u)x_2 \nn \\
x_5&=& vx_3 - (1-v)x_4 \nn 
\ea
and $x_{05}=x_0-x_5$. According to eqs.(\ref{OPE:4ptfct},\ref{QvsV}),
the contribution of a real scalar field $\phi^k$ of dimension $d_k=h+c$ in the
 OPE to the 4-point function $W_{i_4,...,i_1}$ equals
\be (2\pi)^{-D}\bar{g}_k^{i_3i_4}g_k^{i_2i_1}\c_k I^u_0 , \label{OPE:4ptfct:scalar:eval}
 \ee
where $\c_k$ is determined by the normalization of the two-point function 
of $\phi^k.$

\subsection{Comparison with the expansion from the Mellin representation}
We now compare the contribution (\ref{OPE:4ptfct:scalar:eval} )
of a scalar field to the OPE (\ref{E:OPE:pSpace})
of the Euklidean 4-point function with the sum 
(\ref{scalar:pole:sum}) of the contribution
of the corresponding family of poles in the Mellin representation,
 with expressions (\ref{sum:S}) for $S_{N,K}$ and (\ref{const}) for $const$, for $c_-^{12}=c_-^{43}=0$. 

We see that both expressions agree, up to a constant factor which may be attributed to the normalization of two point functions.

The formulas obtained by summation of contributions from a family of Mellin poles  generalize to arbitrary 
dimensions of the scalar fields and agreement is again obtained.  

\section*{Acknowledgment}
I would like to thank Volker Schomerus and Yuri Pis'mak for stimulating discussions and for reading  preliminary versions of the manuscript. Ivan Todorov kindly informed me that Bakalov and Nikolov had independently studied dimensional induction \cite{bakalov:nikolov}.
\section{Appendix. Symanziks $n$-star integration formula in Euklidean space}
\label{sec:n-star}
In ref.\cite{symanzik:nstar}, Symanzik presented a formula which gives 
a Mellin representation of the integral of $n \geq 3$ conformal invariant
 scalar propagators over D-dimensional Euklidean space under conditions 
which guarantee the conformal invariance of the result. 

In the following, all variables $\x_i$ will be in Euklidean space, and 
$$ \x_{ij}=\x_i-\x_j \ . $$
The formula reads as follows
\ba
&&(\pi)^{-\frac 1 2 D} \int d^D\x_0 \prod_{i=1}^n 
\left(   \x_{0i}^2 \right)^{-\delta_i}\Gamma(\delta_i) 
=  \label{nStar}  \\
&& (2\pi i)^{-\frac 1 2 n(n-3)}
\int_{-i\infty}^{i\infty}...\int_{-i\infty}^{i\infty}
ds_1...ds_{\frac 1 2 n(n-3)}\nn \\
&& \quad
\prod_{1\leq i < j\leq n} \Gamma(\delta_{ij}(s))(\x_{ij}^2)^{-\delta_{ij}(s)}
\nn
\ea
provided
\be  0< \Re e \delta_i < \frac 1 2 D \label{ineqCond}\ee
 for all $i$, and ,
\be  \sum_i\delta_i = D . \label{sumCondDeltai}
\ee
Integration is over the hyper-surface of all solutions (with fixed real part)
$\delta_{ij}=\delta_{ji}$, $(i,j=1...n, i\neq j)$ 
of the system of equations 
\be
\sum_{j\neq i} \delta_{ij}=\delta_i , \label{sumCond}
\ee 
parametrized by $\frac 1 2 n(n-3)$ imaginary variables 
$s_1,...,s_{\frac 1 2 n(n-3)}$ as follows.

 Choose a particular 
solution $\delta^0_{ij}$ of eqs.(\ref{sumCond}) subject to the  
inequalities (\ref{ineqCond}), and real coefficients $c_{ij,k}=c_{ji,k}$
which obey
\be c_{ii,k}=0, \quad \sum_{j\neq i}c_{ij,k}=0 . \ee 
Then
\be
\delta_{ij}=\delta_{ij}^0+\sum_{k=1}^{\frac 1 2 n(n-3)}c_{ij,k}s_k .  
\ee
To obtain the proper normalization of the measure on the hyper-surface it is
demanded that the $(\frac 1 2 n(n-3))^2$ coefficients $c_{ij,k}$ with
$2\leq i <j\leq n$, excepting $c_{23,k}$, which may be taken as the independent
ones, must satisfy
\be
|det \ c_{ij,k}|=1 . \label{det} 
\ee
If $n=3$ there is no integration over $s$ to be performed, and eq.(\ref{nStar})
is a star-triangle relation. 

Our main interest will be in the case $n=4$ and its generalization with spin.
The general solution of eqs.(\ref{sumCond}) takes the form
\be
\delta_{ij}= \delta^0_{ij} + \beta_{ij} ,
\ee
with $\beta_{ij}=\beta_{ji} $ and
\ba && \beta_{34}=\beta_{12}, \ \beta_{24}=\beta_{13}, \ \beta_{14}=\beta_{23}
\label{betaIds}\\
&& \beta_{12}+\beta_{13}+\beta_{23}=0 \label{sumBeta}
\ea
Choosing $s_1=\beta_{12}, s_2=\beta_{13}$ the above conditions on the 
coefficients $c_{ij,k}$ are fulfilled, and similarly for any other choice
of two out of the three variables $\beta_{12}, \beta_{13}, \beta_{23}$. 
Thus we may write
\be
ds_1ds_2=d^2\beta=d\beta_{12}d\beta_{13}=... \label{d2beta}
\ee
\subsection{Generalization of the 4-star formula}\label{sec:Symanzik:generalized}
In the computation of the Mellin representation $\mathfrak{M}_\chi $ of a Euklidean partial wave, a generalization of the Symanzik 4-star formula is needed
which includes a factor $Y^D_l(\cos \theta)$ where 
\be \cos \theta = \frac {\lambda  \mu}{|\lambda ||\mu |} \label{cos:theta}\ee
is the cosine of the angle between the Euklidean $D$-vectors
\be
\lambda = 2\left( \frac {\x_{10}}{\x_{10}^2}-\frac {\x_{20}}{\x_{20}^2}\right)
\label{lambda:again}
\ee
and 
\be
\mu = 2\left( \frac {\x_{40}}{\x_{40}^2}-\frac {\x_{30}}{\x_{30}^2}\right)\ .
\label{mu}
\ee
 
In applications, $Y^D_l(\cos \theta)$ is the zonal spherical function in 
$D$ dimensions

 Its presence reflects the fact that the quantum numbers of a
field of Lorentz spin $l$ are exchanged between legs $1,2$ and $3,4$.
But the result holds for arbitrary polynomials of the form 
\be Y^D_l(t)= \sum_{k=0}^{E[l/2]} a^l_k(D)t^{l-2k}\ , \label{power:Y} \ee
i.e. which involves even powers of $t$ for even $l$, and odd powers for odd
 $l$. $E[l/2]$ is the largest integer $\leq l/2$,.

Given the external dimensions $\delta_i$, the result of the Symanzik 
integration formula involves an integration over two independent variables 
$s_1,s_2$ - e.g. $\delta_{12}$ and $\delta_{13}$ - 
which determine the six $\delta_{ij}=\delta_{ji}$ subject to the constraint
\be \sum_{j}\delta_{ij}=\delta_i \ . \label{delta:i:sum} \ee
We may equally well consider functions of $\delta_i$ and these variables
 as functions of the 
six $\delta_{ij}$. Because validity of the Symanzik integration formula presupposes $\sum \delta_i = D$, they obey $\sum_i\sum_j \delta_{ij} =D$, but the 
 definition of $\mathfrak{P}^D_l$ will make sense for arbitrary 
$\delta_{ij}$.

The result will involve  functions $\mathfrak{P}^D_l(\{\delta_{ij}\})$ which are polynomials of degree $l$ in the independent variables among the $\delta_{ij}$, for given $\delta_i$, and such that 
$\mathfrak{P}^D_{l}(\{\delta_{ij}\})\prod_{i=1}^4 \Gamma(\delta_i+\frac l 2)$ is a polynomial also in the $\delta_i$. The only explicit dependence on $D$, for given 
$\{\delta_{ij}\}$, is through the expansion coefficients $a^l_k(D)$ of 
$Y^D_l(t)$. Explicitly, the polynomials are defined as follows
\be
\mathfrak{P}^D_l(\{\delta_{ij}\}) = \sum_{k=0}^{E[l/2]} a^l_k(D) \
(\delta_{12}-\half l)_k(\delta_{34}-\half l)_k
  \mathfrak{P}_{l-2k}(\{\delta_{ij}\}) \label{P:Dl:l}
\ee  
where $\mathfrak{P}_m$ depends neither explicitly on $D$ nor on $l$
 and is explicitly defined as follows.

The definition involves a sum over nonnegative integers $k_{ij}=k_{ji}$ attached to the four links $(ij)=(13),(14),(23),(24)$. 
A sum or product over these four links will be denoted by $\sum^\prime$ and $\prod^\prime$ respectively. In the definition of $\mathfrak{P}_m$, the variables are subject to the constraint $\sum^\prime k_{ij}=m$. 

The definition of $\mathfrak{P}_m$ is as follows.
\ba 
&&
\mathfrak{P}_m(\{\delta_{ij}\})=2^{-m}m!\sum_{\{k_{ij}\}:\sum^\prime k_{ij}=m}
\frac {(-)^{k_{14}+k_{23}}}{\prod^\prime k_{ij}!}
\prod^\prime (\delta_{ij})_{k_{ij}}\nn \\
&& \qquad \qquad \prod_{n=1}^4 
\Gamma\left(\delta_n - \frac m 2 +\sum_j k_{jn}\right)^{-1} \ . \label{P:m}
\ea
Here as everywhere we use the notation $$(\alpha)_k= \Gamma(\alpha+k)/\Gamma(\alpha)= \alpha (\alpha+1)... (\alpha+k-1)\ ,$$
 and $\delta_n$ is to be read as $\sum_j\delta_{nj}$. 

For fixed $\delta_i$, $\mathfrak{P}_m$ is a polynomial in the remaining variables, and
$\prod_{n=1}^4 \Gamma(\delta_n + \frac l 2 )\mathfrak{P}_{l-2k}(\{\delta_{ij}\})$ is a 
polynomial also in the $\delta_n$. 

The generalization of the Symanzik 4-star formula reads as follows.
Given $\delta_i$, $ i=1...4$ subject to the constraint $\sum_i\delta_i=D$,
let $\cos \theta $ be defined through eqs.(\ref{cos:theta})ff.  Then   
\ba
S_l&\equiv& \pi^{-\half D} \int d^D\x_0 Y^D_l(\cos \theta )\prod_{i=1}^4
 (\x_{i0}^2)^{-\delta_i} \\
&=& (2\pi  i)^{-2}\int \int\  ds_1 ds_2 \mathfrak{P}^D_l(\{\delta_{ij}\})
 \label{Symanzik:4:generalized} \\
 &&  \Gamma\left(\delta_{12}- \half l \right) (\x_{12}^2)^{-\delta_{12}}
 \Gamma\left(\delta_{34}- \half l\right) (\x_{34}^2)^{-\delta_{34}}\nn \\
&& \prod^\prime \Gamma(\delta_{ij})(\x_{ij}^2)^{-\delta_{ij}} \ . 
\nn
\ea
The integrations are as in the ordinary Symanzik 4-star formula. It corresponds to the special case $l=0$, with 
$$\mathfrak{P}^D_0(\{\delta_{ij}\})=\prod_{i=1}^4 \Gamma(\delta_i)^{-1}. $$
\subsection{Derivation of the generalized 4-star formula}
One computes
\ba
\lambda^2 &=& 4 \frac {\x_{12}^2}{\x_{10}^2 \x_{20}^2} \label{lambda:sq}\\
\mu^2 &=& 4 \frac {\x_{34}^2}{\x_{30}^2 \x_{40}^2} \label{mu:sq}\\
\lambda \mu &=& 2\left(\frac {\x_{13}^2}{\x_{10}^2\x_{30}^2} + \frac {\x_{24}^2}{\x_{20}^2\x_{40}^2} - \frac {\x_{14}^2}{\x_{10}^2\x_{40}^2} - \frac {\x_{23}^2}{\x_{20}^2\x_{30}^2}\right) \ .
\label{lambda:mu}
\ea
The multinomial theorem asserts 
\ba
(\lambda\mu)^{l-2k} &=& 2^{l-2k} \sum_{\{k_{ij}\}:\sum^\prime k_{ij}=l-2k}
\frac {(l-2k)!\ (-)^{k_{14}+k_{23}} }{\prod^\prime k_{ij}!}\nn \\
&& \qquad \prod^\prime
(\x_{ij}^2)^{k_{ij}}(\x_{i0}^2 \x_{j0}^2)^{-k_{ij}}\nn
\ea 
Inserting the power series expansion (\ref{power:Y}) of $Y^D_l$, one obtains
\ba
&&\pi^{\half D}S_l=\nn \\
&& \int d^D\x_0 \sum_{k=0}^{E[l/2]} a^l_k(D) 
(\lambda^2 \mu^2)^{-\half l + k} (\lambda\mu)^{l-2k}
\prod (\x_{i0})^{-\delta_i}\nn \\
&=&  \int d^D\x_0 \sum_{k=0}^{E[l/2]} a^l_k(D)2^{-l+2k} (l-2k)!
(\x_{12}^2\x_{34}^2)^{-\half l +k}
\nn \\
 &&   \sum_{\{k_{ij}\}:\sum^\prime k_{ij}=l-2k}
\frac { (-)^{k_{14}+k_{23}} }{\prod^\prime k_{ij}!}
\prod^\prime (\x_{ij}^2)^{k_{ij}}
\prod_i (\x_{0i}^2)^{-\delta_i^\prime} \ ,
\ea
with 
\be
 \delta_{i}^\prime = \delta_i +\sum_j k_{ij} - \half l + k \ . 
\ee
One verifies that $\sum \delta_i^\prime = D$. Therefore the integral can be performed with the help of the Symanzik 4-star formula. It will involve 
exponents $\delta_{ij}^\prime $ subject to the constraint $\sum_j \delta_{ij}^\prime = \delta_i^\prime$, and produce factors 
$$\prod_{n=1}^4\Gamma(\delta_n^\prime)^{-1}
\prod_{ij}\Gamma(\delta_{ij}^\prime)(\x_{ij}^2)^{-\delta_{ij}^\prime} $$
which multiply the preexisting factors 
$(\x_{12}^2 \x_{34}^2)^{-\half l +k}\prod^\prime (\x_{ij}^2)^{k_{ij}}.$
Transform to new variables
\ba
\delta_{ij} &=& \delta_{ij}^\prime - k_{ij}\ , \quad (ij)\neq (12),(34), \nn
\\
\delta_{12} &=& \delta_{12}^\prime + \half l -k \ , \nn \\
\delta_{34} &=& \delta_{34}^\prime  + \half l -k \ .
\label{delta:ij}
\ea
As a consequence of the constraint on the $\delta_{ij}^\prime$, they satisfy the constraint $\sum_j \delta_{ij} = \delta_i \ .$

It follows that 
\ba
\Gamma(\delta_{ij}^\prime) &=&\Gamma(\delta_{ij}) (\delta_{ij})_{k_{ij}} \ , \quad (ij)\neq (12),(34) \nn \\
\Gamma(\delta_{12}^\prime) &=& \Gamma(\delta_{12}- \half l)
 (\delta_{12}-\half l )_k \nn \\
\Gamma(\delta_{34}^\prime) &=& \Gamma(\delta_{34}- \half l)
 (\delta_{34}-\half l))_k \nn 
\ea
Inserting this and carrying out the sum over the $k_{ij}$ with the help of the definition of $\mathfrak{P}_m$, and the sum over $k$ with the help of the definition 
of $\mathfrak{P}^D_l$ one obtains the final result, eq.(\ref{Symanzik:4:generalized}). 

The polynomial properties of $\mathfrak{P}_m$ and therefore of $\mathfrak{P}^D_l$ for fixed $\delta_i$ are obvious from the fact 
that 
$(\alpha)_k= \Gamma(\alpha+k)/\Gamma(\alpha)$ is a polynomial in $\alpha$ of degree $k$.

$\prod_{n=1}^4 \Gamma(\delta_n+ \frac l 2 )\mathfrak{P}_{l-2k}(\{\delta_{ij}\})$ is a
polynomial also in $\delta_n$ because $\Gamma(\delta_n + \frac l 2)
\Gamma(\delta_n -\frac l 2 +k + \sum_j k_{jn})^{-1}$ is a polynomial. This is
 so because $-l+k+\sum_jk_{jn}$ is an integer $\leq -k$ since 
$\sum_j k_{jn}\leq l-2k$.  
\section{Appendix: 3-point vertex functions and coefficients in the operator
 product expansion}
\subsection{Notations}\label{sec:notations}
In $D=2h$ space time dimensions, let 
$\chi=[l,\delta=h+c]$. Introduce
\ba
\tilde{\chi}&=&[\bar{l}, h-c], \label{tilde:chi}\\
\delta_\chi &=& \half  (h+c-l) \label{delta:chi}\\ 
\delta_{\tilde{\chi}}&=& \half (h-c-l)
\ea
$\bar{l}$ is the conjugate representation to $l$ which is the same as $l$ for symmetric tensor representations. 

I also use the shorthand 
$$ (\alpha)_k = \frac {\Gamma (\alpha + k)}{\Gamma(\alpha)}
= \alpha(\alpha+1)\dots (\alpha+ k -1) \ . $$
\subsection{The differential operators $D_l$}
Following ref.\cite{dobrevEtAl:OPE}, a convenient alternative expression 
for the conformal invariant 3-point function $\mathfrak{V}$ can be written down which uses a $l$-th order differential operator $D_l$. 
It acts on functions of  $x_1$ and $x_2$. It is homogeneous in the complex 
light-like $D$-vector  $\mathfrak{z}$ of degree $l$. 

\mbox{$
D_l(\delta_\chi + c_-, \mathfrak{z}\nabla_1, \delta_\chi - c_-, \mathfrak{z}\nabla_2) $}
is defined by
\ba
&& D_l (a, \alpha; b, \beta) = \label{D_l:def}\\
&& 
\sum_{k=0}^l \left(\begin{array}{l} l \\ k\end{array} \right)(a+k)_{l-k}(b+l -k)_k (-\alpha)^k\beta^{l-k}\nn \\
&=& (a)_l (\alpha+\beta)^l F\left(a+b+l-1, -l; a; \frac {\alpha}{\alpha+\beta}\right) 
\nn
\ea
where $F={}_2F_1$ is the hypergeometric function.  \\

The following identity is valid, and easily checked by use of the binomial theorem 
\ba
&&(a)_l(b)_l \left( \frac {2}{x_{13}^2} \right)^a \left( \frac {2}{x_{23}^2}\right)^b
(\lambda\cdot \z)^l \label{D:identity} \\
&&= \qquad D_l(a,\z\cdot\nabla_1; b, \z\cdot \nabla_2) \left( \frac {2}{x_{13}^2} \right)^a \left( \frac {2}{x_{23}^2}\right)^b \nn
\ea 
with $\lambda = 2\left( \frac {x_{13}}{x_{13}^2} - \frac {x_{23}}{x_{23}^2}\right)$
\subsection{The un-amputated 3-point vertex function with one spinning leg}
Let $\chi_1 = [0,d_1=h+c_1]$, $\chi_2=[0,d_2 = h+c_2]$, $\chi=[l, h+c]$, and
$c_+=\frac 1 2 (c_1+c_2)$, $c_-= \frac 1 2 (c_1-c_2)$, and retain the definition  $\delta_\chi=\frac 1 2 (h+c-l)$ of subsection \ref{sec:notations} %
Using  the identity (\ref{D:identity}), expression (\ref{Vscalar:u}) for the three-point function can be written in the alternative form
\ba
&& \mathfrak{V}^u(\x_0,\chi , \x_1,\chi_1, \x_2, \chi_2) = \label{V:D_l} \\
&& (2\pi)^{-h} \frac{1}{(\delta_\chi+c_-)_l (\delta_\chi - c_-)_l }
\left(-\frac {2}{x_{12}^2}\right)^{\frac 1 2 (h-c+l)+c_+} \nn \\
&&  D_l(\delta_\chi + c_-, \z\nabla_1, \delta_\chi - c_- , \z\nabla_2)
 \nn \\ &&
\left[ \left( \frac {-2}{x_{10}^2} \right)^{\frac 1 2 (h+c-l) + c_-}\left(\frac {-2}{x_{20}^2} \right)^{\frac 1 2 (h+c-l)-c_-} \right]
\nn \ea
This involves a Minkowski space scalar product so that  
$-x_{12}^2 > 0$ when $x_{12}$ is space-like or Euklidean.
 The 3-point functions in 
Euklidean space and in Minkowski space are obtained from each other by analytic continuation. $i\epsilon$-prescriptions as appropriate for Wightman functions are understood in Minkowski space. They follow from the spectrum condition. 
%
\subsection{The coefficient in the operator product expansion}
From the formula (2.34) of \cite{dobrevEtAl:OPE} for the \emph{scalar-leg-amputated} coefficient $Q_-$ the un-amputeted coefficient is obtained by changing 
$c_1,c_2\mapsto -c_1,-c_2$, assuming normalization conditions so that the scalar two-point function in momentum space for a field of dimension $h+c$ is 
$\left( \half p^2\right)^c$, and rewriting the normalization factor in terms 
of $N_l(c_+,c_-,-c)$ by use of eq.(3.42) of ref. \cite{mack:dobrevClebsch} . 
Using the notation of subsection \ref{sec:notations}, the  result is as follows, for the partial Fourier transform 
%
\ba 
&& \mathfrak{Q}^u(p,\tilde{\chi}; x_1, \chi_1, x_2, \chi_2) = \label{Q} \\
&&
  - (2\pi)^{-h}\frac {1}{\Gamma(\half(h-c+l)+c_-)\Gamma(\half(h-c+l)-c_-)}
 \nn \\
 &&  \left( - \frac 1 2 x_{12}^2\right)^{-\half(h+c+l) - c_+} 
D_l(\delta_{\tilde{\chi}}+c_-, \mathfrak{z}\nabla_1, \delta_{\tilde{\chi}}-c_-, \mathfrak{z}\nabla_2)
\nn\\
&& \quad \int du[u(1-u)]^{\frac h 2 - 1} \left( \frac {1-u}{u}\right)^{-c_-}
\left( \frac {-x_{12}^2}{p^2}\right)^{\frac l 2 + \frac c 2 }  \nn \\
&& \qquad J_{l+c} \left( [- x_{12}^2 p^2 u(1-u)]^{\frac 1 2 } \right)
\ e^{ip[ux_1 + (1-u)x_2]}\nn 
\ea
This is an entire function of $p$. In Euklidean space, the sign of the scalar product is reversed, so that 
$\p^2 = -\p^2 <0$ and $x_{12}^2= -\x_{12}^2 < 0$. One can use that 
$z^{-\nu}I_\nu(z\rho) = (e^{i\pi/2}z)^{-\nu}J_\nu(e^{i\pi/2}z\rho)$ for 
real $\rho$ in order to express the result in Euklidean space in terms of the modified Bessel function $I_{l+c}$.

\bibliography{cftMyPapersS,cftOtherLiteratureS,dualResonanceLiteratureS,dualityLiteratureS,AdSsusyYMRefs24APR08S,XFT_bibSPIRES,volkersReferences}

\begin{thebibliography}{10}

\bibitem{Aharony:1999ti}
Ofer Aharony, Steven~S. Gubser, Juan~Martin Maldacena, Hirosi Ooguri, and Yaron
  Oz.
\newblock {Large N field theories, string theory and gravity}.
\newblock {\em Phys. Rept.}, 323:183--386, 2000.

\bibitem{wilczek:anyons}
D.P. Arovas, J.R. Schrieffer, F.~Wilczek, and A.~Zee.
\newblock Statistical mechanics of anyons.
\newblock {\em Nucl. Phys. B}, 251:117--126, 1985.

\bibitem{bakalov:nikolov}
B.~Bakalov and N~Nikolov.
\newblock {\em (in preparation)}, 2009.

\bibitem{GCI:infinite:algebra}
B.~Bakalov, N.M. Nikolov, K.H. Rehren, and I.~Todorov.
\newblock Infinite dimensional {L}ie algebra in 4d conformal quantum field
  theory.
\newblock {\em J. Phys. A}, 41:194002, 2008.

\bibitem{todorov:bargmann}
V.~Bargmann and I.T. Todorov.
\newblock Spaces of analytic functions on a complex cone as carriers for the
  symmetric tensor representations of ${S}{O}(n)$.
\newblock {\em J. Math. Phys.}, pages 1141--1148, 1977.
\newblock reviewed in \cite{mack:dobrevBook:copy}, Appendix A.5.

\bibitem{bassoEtAl:cusp}
B.~Basso, G.P. Korchemsky, and J.~Kota\'{n}ski.
\newblock Cusp anomalous dimension in maximally supersymmetric {Y}ang {M}ills
  theory.
\newblock {\em Phys. Rev. Letters}, 100:091601, 2008.
\newblock eprint hep-th/07083933.

\bibitem{Beisert:2004ry}
Niklas Beisert.
\newblock {The dilatation operator of N = 4 super Yang-Mills theory and
  integrability}.
\newblock {\em Phys. Rept.}, 405:1--202, 2005.

\bibitem{Callan:1973pu}
Curtis~G. Callan and David~J. Gross.
\newblock {Bjorken scaling in quantum field theory}.
\newblock {\em Phys. Rev.}, D8:4383--4394, 1973.

\bibitem{DHoker:2002aw}
Eric D'Hoker and Daniel~Z. Freedman.
\newblock {Supersymmetric gauge theories and the AdS/CFT correspondence}.
\newblock In {\em {TASI-lectures Boulder 2001: Strings,Branes and EXTRA
  dimensions}}, pages 3--158, 2002.

\bibitem{Dirac36}
P.A.M. Dirac.
\newblock {\em Ann. Math.}, 37:429, 1936.

\bibitem{mack:dobrevClebsch}
V.K. Dobrev, G.~Mack, V.B. Petkova, S.G. Petrova, and I.T. Todorov.
\newblock On {C}lebsch {G}ordan expansions for the {L}orentz group in $n$
  dimensions.
\newblock {\em Reports on Mathematical Physics}, 9:219, 1976.

\bibitem{mack:dobrevBook}
V.K. Dobrev, G.~Mack, V.B. Petkova, S.G. Petrova, and I.T. Todorov.
\newblock {\em Harmonic analysis on the n-dimensional {L}orentz group and its
  applications to conformal quantum field theory}, volume~63 of {\em Lecture
  notes in physics}.
\newblock Springer Verlag, 1977.

\bibitem{mack:dobrevBook:copy}
V.K. Dobrev, G.~Mack, V.B. Petkova, S.G. Petrova, and I.T. Todorov.
\newblock {\em Harmonic analysis on the n-dimensional {L}orentz group and its
  applications to conformal quantum field theory}, volume~63 of {\em Lecture
  notes in physics}.
\newblock Springer Verlag, 1977.

\bibitem{dobrevEtAl:OPE}
V.K. Dobrev, V.B. Petkova, S.G. Petrova, and I.T. Todorov.
\newblock Dynamical derivation of operator-product expansions in {E}uclidean
  conformal quantum field theory.
\newblock {\em Phys. Rev. D}, 13:887--912, 1976.

\bibitem{erdmenger:osborn:currentsEnergy}
J.~Erdmenger and H.~Osborn.
\newblock Conserved currents and the energy momentum tensor in conformally
  invariant theories for general dimensions.
\newblock {\em Nucl. Phys. B}, 483:431--474, 1997.

\bibitem{ferraraEtAl:shadow}
S.~Ferrara, R.~Gatto, A.F. Grillo, and G.~Parisi.
\newblock The shadow operator formalism for conformal algebra. vacuum
  expectation values and operator products.
\newblock {\em Lettere Nuovo Cimento}, 4:115, 1972.

\bibitem{ferraraEtAl:conformalAlgebraOPE}
S.~Ferrara, A.F. Grillo, and R.~Gatto.
\newblock {\em Conformal algebra in space-time and operator product
  expansions}, volume~67 of {\em Springer Tracts in Modern Physics}.
\newblock Springer Verlag, 1973.

\bibitem{ferraraEtAl:OPE}
S.~Ferrara, A.F. Grillo, and R.~Gatto.
\newblock Tensor representations of conformal algebra amd conformally covariant
  operator product expansions.
\newblock {\em Ann.Phys.(N.Y.)}, 76:161, 1973.

\bibitem{ferraraEtAl:NPB:OPE}
S.~Ferrara, A.F. Grillo, G.~Parisi, and R.~Gatto.
\newblock Covariant expansion of the conformal four-point function.
\newblock {\em Nucl. Phys. B}, 49:77, 1972.

\bibitem{fradkin:1}
E.S. Fradkin.
\newblock The quantum theory of fields i.
\newblock {\em Zh. Exp. Teor. Fis.}, 29:121, 1955.
\newblock in Russian. Engl. Translation: JETP 2 (1956) 148.

\bibitem{fradkin:2}
E.S. Fradkin.
\newblock Green functions in a quantum field theory and quantum statistics.
\newblock In {\em Quantum field theory and hydrodynamics}, volume~29 of {\em
  Trudy FIAN}, Moscow, 1965.
\newblock in Russian. Engl. translation: Plenum Press, New York 1967.

\bibitem{fradkin:palchik:3}
E.S. Fradkin and M.Ya. Palchik.
\newblock Method of solving conformal models in d-dimensional space. secondary
  fields in $d>2$ and the solution of two-dimensional models.
\newblock {\em Int. J. Mod. Phys. A}, 13:4837--4888, 1998.

\bibitem{frampton:duality:reprint}
P.A. Frampton.
\newblock {\em Dual Resonance Models and Superstrings}.
\newblock World Scientific, 1986.

\bibitem{frs:CMP}
K.~Fredenhagen, B.~Schroer, and K.H. Rehren.
\newblock Superselection sectors with braid group statistics and exchange
  algebras, i: General theory.
\newblock {\em Commun. Math. Phys.}, 125:201--226, 1989.

\bibitem{freund:26D}
P.G.O. Freund.
\newblock Superstrings from 26 dimensions?
\newblock {\em Physics Letters}, 151B:387--390, 1985.

\bibitem{fubini:veneziano:NC.67A}
S.~Fubini and G.~Veneziano.
\newblock Duality in operator formalism.
\newblock {\em Nuovo Cimento}, 67A:29, 1970.

\bibitem{GelfandGindikin:CM-78}
M.~Gel'fand and S.~G. Gindikin.
\newblock Complex manifolds whose skeletons are semisimple real {L}ie groups,
  and analytic discrete series of representations.
\newblock {\em Funct. Anal. Appl.}, 7:258--265, 1978.
\newblock Translation of {\it Funkts. Anal. Prilozh.}, 11(4):19-27, 1977.

\bibitem{georgi:unparticles}
H.~Georgi.
\newblock Unparticle physics.
\newblock {\em Phys. Rev. Letters.}, 98:221601, 2007.

\bibitem{glimm:jaffe}
J.~Glimm and A.~Jaffe.
\newblock {\em Quantum Physics. A Functional Integral Point of View}.
\newblock Springer Verlag, 1981.

\bibitem{GR}
I.S. Gradshteyn and I.M. Ryzhik.
\newblock {\em Table of integrals, series, and products}.
\newblock Academic Press, 1965.

\bibitem{green:schwarz}
M.B. Green and J.H. Schwarz.
\newblock Anomaly cancellation in supersymmetric $d=10$ gauge theory and
  superstring theory.
\newblock {\em Phys. Letters}, 149B:117--122, 1984.

\bibitem{heterotic:string}
D.J. Gross, J.A. Harvey, E.~Martinec, and R.~Rohm.
\newblock Heterotic string.
\newblock {\em Phys. Rev. Letters}, 54:502--505, 1985.

\bibitem{gross:neveu:scherk:schwartz1970}
D.J. Gross, A.~Neveu, J.~Scherk, and J.H. Schwartz.
\newblock Renormalization and unitarity in dual resonance models.
\newblock {\em Phys. Rev. D}, 2:697--710, 1970.

\bibitem{GAdsCFT1}
Murat G{\"u}naydin.
\newblock Ads/cft dualities and the unitary representations of non-compact
  groups and supergroups: {W}igner versus {D}irac.
\newblock In {\em Proc. International Wigner Symposium}, Istanbul, August 1999.

\bibitem{johnson:SDequations}
R.W. Johnson.
\newblock Finite integral equations for green's functions for $:\phi^4:$
  coupling. {I}.
\newblock {\em J. Math. Phys.}, 11:2161, 1970.

\bibitem{pismak:wegner:kehrein:epsilon}
S.~Kehrein, F.~Wegner, and Yu.M. Pis'mak.
\newblock Conformal symmetry and the spectrum of anomalous dimensions in the
  n-vector model in four - epsilon dimensions.
\newblock {\em Nucl.Phys. B}, 402:669--692, 1993.

\bibitem{reuter:UVfixpoint}
O.~Lauscher and M.~Reuter.
\newblock Ultraviolet fixed point and generalized flow equation of quantum
  gravity.
\newblock {\em Phys. Rev. D}, 65:025013, 2002.

\bibitem{Lawson:semigroups}
J.D. Lawson.
\newblock Polar and {O}l'shanskii decompositions.
\newblock {\em J. Reine Angewandte Mathematik}, 448:191--219, 1994.

\bibitem{Luescher:OPE}
M.~L{\"{u}}scher.
\newblock Operator product expansions on the vacuum in conformal quantum field
  theory in two space time dimensions.
\newblock {\em Commun. Math. Phys.}, 50:23--52, 1976.

\bibitem{luescher:phd}
M.~L{\"{u}}scher.
\newblock {\em Theorie holomorpher {L}iehalbgruppen und deren Anwendung in
  konform kovarianten quantenfeldtheorien}.
\newblock PhD thesis, Universit{\"{a}}t Hamburg, Germany, 1976.

\bibitem{mack:shortDistance}
G.~Mack.
\newblock Conformal invariance and short distance expansions in quantum field
  theory.
\newblock In {\em Strong interaction physics}, volume~17 of {\em Lecture Notes
  in Physics}, page 300, Heidelberg, 1972. Springer Verlag.

\bibitem{mack:groupth:1}
G.~Mack.
\newblock Group theoretical approach to conformal invariant quantum field
  theory.
\newblock {\em Journal de physique}, 34(C1 (Suppl. No. 10)):99, 1973.

\bibitem{mack:groupth:2}
G.~Mack.
\newblock Group theoretical approach to conformal invariant quantum field
  theory.
\newblock In E.R. Caianello, editor, {\em Renormalization and invariance in
  quantum field theory}, New York, 1974. Plenum Press.

\bibitem{mack:OS}
G.~Mack.
\newblock {O}sterwalder {S}chrader positivity in conformal invariant quantum
  field theory.
\newblock In H.~Rollnik and K.~Dietz, editors, {\em Trends in elementary
  particle theory}, volume~37 of {\em Lecture Notes in Physics}, page~66.
  Springer Verlag, 1975.

\bibitem{mack:irreps}
G.~Mack.
\newblock All unitary ray representations of the conformal group with positive
  energy.
\newblock {\em Commun. Math. Phys.}, 55:1, 1977.

\bibitem{mack:OPE}
G.~Mack.
\newblock Convergence of operator product expansions on the vacuum in conformal
  invariant quantum field theory.
\newblock {\em Commun. Math. Phys.}, 53:155, 1977.

\bibitem{mack:duality}
G.~Mack.
\newblock Duality in quantum field theory.
\newblock {\em Nuclear Physics}, B118:445, 1977.

\bibitem{mack:deRiese}
G.~Mack and M.~{de Riese}.
\newblock Simple space time symmetries: generalizing conformal field theory.
\newblock {\em J. Math. Phys.}, 48:052304, 2007.

\bibitem{mack:luescher:global}
G.~Mack and Martin L{\"{u}}scher.
\newblock Global conformal invariance.
\newblock {\em Commun. Math. Phys.}, 41:203, 1975.

\bibitem{mack:pruestel:GR}
G.~Mack and T.~Pr{\"{u}}stel.
\newblock Generalized gauge theories with nonunitary parallel transporters:
  General relativity with a cosmological constant as an example.
\newblock {\em Eur. Phys. J. C}, 46:255--267, 2006.

\bibitem{mack:symanzik}
G.~Mack and K.~Symanzik.
\newblock Currents, stress energy tensor and generalized unitarity in conformal
  invariant quantum field theory.
\newblock {\em Commun. Math. Phys.}, 27:247, 1972.

\bibitem{mack:todorov:UV}
G.~Mack and I.T. Todorov.
\newblock Conformal invariant {G}reen functions without ultraviolet
  divergences.
\newblock {\em Physical Review}, D8:1764, 1971.

\bibitem{migdal:2}
A.A. Migdal.
\newblock Conformal invariance and bootstrap.
\newblock {\em Phys. Letters}, 37B:386, 1971.

\bibitem{migdal:1}
A.A. Migdal.
\newblock On hadronic interactions at small distances.
\newblock {\em Phys. Letters}, 37B:98, 1971.

\bibitem{GCI:bilocal}
N.M. Nikolov, K.H. Rehren, and I.~Todorov.
\newblock Harmonic bilocal fields generated by globally conformal invariant
  scalar fields.
\newblock {\em Commun. Math. Phys.}, 279:225--250, 2008.

\bibitem{GCI:todorov:gauge}
N.M. Nikolov, Y.S. Stanley, and I.T. Todorov.
\newblock Globaly conformal invariant gauge field theory and rational
  correlation functions.
\newblock {\em Nucl. Phys. B}, 670:373--400, 2003.

\bibitem{OS:I}
K.~Osterwalder and R~Schrader.
\newblock Axioms for euclidean green's functions, i.
\newblock {\em Commun. Math. Phys.}, 31:83--112, 1973.

\bibitem{OS:II}
K.~Osterwalder and R~Schrader.
\newblock Axioms for euclidean green's functions, ii.
\newblock {\em Commun. Math. Phys.}, 42:281--305, 1975.

\bibitem{palchik:fradkin:secondary}
N.Ya. Palchik and E.S. Fradkin.
\newblock A family of secondary fields in the $d$-dimensional conformal
  quantumk field theory.
\newblock {\em Dokl. Phys.}, 44:757--759, 1999.

\bibitem{palchik:secondary}
N.Ya. Palchik and V.N. Zaikov.
\newblock Secondary field in $d\geq 3$ conformal field theory: The existence
  condition for d-dimensional generalization of two-dimensional models in
  quantum field theory.
\newblock In {\em Quantization, gauge theory and strings}, pages 375--381,
  Moscow, 2000.

\bibitem{parisi:peliti:bootstrap}
G.~Parisi and L.~Peliti.
\newblock Calculation of critical indices.
\newblock {\em Lettere Nuovo Cimento}, 2:386, 1971.

\bibitem{pis'mak:2005me}
Yu.~M. Pis'mak.
\newblock {Renormalization group and infinite algebraic structure in
  D-dimensional conformal field theory}.
\newblock {\em J. Phys.}, A39:8157--8172, 2006.

\bibitem{holography:rehren}
K.H. Rehren.
\newblock Algebraic holography.
\newblock {\em Annales Henri Poincare}, 1:607--623, 2000.

\bibitem{ruehl:book}
W~R{\"{u}}hl.
\newblock {\em The {L}orentz group and harmonic analysis}.
\newblock Benjamin Inc., 1970.

\bibitem{ruehl:lifing:AdS}
W.~R{\"{u}}hl.
\newblock Lifting a conformal field theory from $d$-dimensional flat space to
  $d+1$ dimensional ads-space.
\newblock {\em Nucl.Phys. B}, 705:437--456, 2005.

\bibitem{speer}
E.R. Speer.
\newblock {\em Generalized Feynman amplitudes}.
\newblock Princeton University Press, 1969.

\bibitem{PCT}
R.F. Streater and A.S. Wightman.
\newblock {\em PCT, Spin and Statistics, and All That}.
\newblock Addison-Wesley Publishing Company, 1964, 1980.

\bibitem{holography:susskind}
L.~Susskind and J.~Lindesay.
\newblock {\em An introduction to black holes, information and the string
  theory revolution: The holographic universe}.
\newblock World Scientific, 2005.

\bibitem{symanzik:hercegnovi}
K.~Symanzik.
\newblock Green's functions method and renormalization of renormalizable
  quantum field theories.
\newblock In B.~Jaksic, editor, {\em Lectures on High Energy Physics}, pages
  485--517, New York, 1966. Gordon and Breach.
\newblock Proceedings Sixth Summer Meeting of Nuclear Physicists, Hercegnovi
  1961.

\bibitem{symanzik:nstar}
K.~Symanzik.
\newblock On calculations in conformal invariant field theories.
\newblock {\em Nuovo Cimento}, 3:734--738, 1972.

\bibitem{pismak:1overNexpansion:bootstrap}
A.N. Vasiliev, Yu.M. Pis'mak, and Yu.R. Khonkonen.
\newblock 1/n expansion. calculation of the exponent eta in the order 1/n**3 by
  the conformal bootstrap.
\newblock {\em Theor. Math. Phys.}, pages 127--134, 1982.

\bibitem{vasiliev:RG}
N.N. Vasiliev.
\newblock {\em Quantum field renormalization group in the theory of critical
  behavior and stochastic dynamics}.
\newblock Chapmann and Hall, London, 2004.

\bibitem{Veneziano:model}
G.~Veneziano.
\newblock Construction of crossing-symmetric, {R}egge behaved amplitude for
  linearly rising trajectories.
\newblock {\em Nuovo Cimento}, 57:190, 1968.

\bibitem{veneziano:erice1970}
G.~Veneziano.
\newblock Narrow-resonance models compatible with duality and their
  developments.
\newblock In A.~Zichichi, editor, {\em Elementary processes at high energy},
  pages 94--158. Academic Press, 1971.
\newblock Proc. Erice summer school 1970.

\bibitem{wilson:OPE}
K.~Wilson.
\newblock Non-{L}agrangean models of current algebra.
\newblock {\em Phys.Rev.}, 179:1499, 1969.

\bibitem{wilson:OPE:Thirring}
K.~Wilson.
\newblock Operator product expansion and anomalous dimensions in the {T}hirring
  model.
\newblock {\em Phys.Rev. D}, 2:1473, 1970.

\bibitem{Wilson:1973jj}
K.~G. Wilson and John~B. Kogut.
\newblock {The Renormalization group and the epsilon expansion}.
\newblock {\em Phys. Rept.}, 12:75--200, 1974.

\bibitem{witten:GR}
E.~Witten.
\newblock (2+1)-dimensional gravity as a exactly soluble system.
\newblock {\em Nucl. Phys. B}, 311:46, 1988.

\end{thebibliography}

\end{document}